\documentstyle[aps,prd,preprint]{revtex}
\tightenlines

\newcounter{th}
\newcounter{proof}
\begin{document}

\newcommand{\qed}{\hfill$\Box$}
\newcommand{\qedd}{\qed\vspace{4mm}}
\newcommand{\beq}[1]{\begin{equation}\label{#1}}
\newcommand{\eeq}{\end{equation}}
\newcommand{\bea}{\begin{eqnarray}}
\newcommand{\eea}{\end{eqnarray}}
\newcommand{\req}[1]{(\ref{#1})}
\newcommand{\diag}{{\rm diag}}
\newcommand{\vp}{\varepsilon}
\newcommand{\R}{{\bf R}}
\newcommand{\E}{{\cal E }}
\newcommand{\A}{{\cal A }}
\newcommand{\F}{{\cal F }}
\newcommand{\G}{{\cal G }}
\newcommand{\B}{{\cal B }}
\newcommand{\D}{{\cal D }}
\newcommand{\xx}{{\bf r }}
\newcommand{\uu}{{\bf u }}
\newcommand{\vv}{{\bf v }}
\newcommand{\ww}{{\bf w }}
\newcommand{\bb}{{\bf b }}
\newcommand{\ee}{{\bf e }}
\newcommand{\sss}{{\bf s }}
\newcommand{\0}{{\bf0}}
\newcommand{\g}{\gamma}
\newcommand{\la}{\lambda}
\newcommand{\al}{\alpha}
\newcommand{\aaa}{ {\bf a} }
\newcommand{\ff}{ {\bf f} }
\newcommand{\ggg}{ {\bf g} }
\newcommand{\hh}{ {\bf h} }
\newcommand{\circUp}{\mathop{\raise 1.2pt\hbox{${\scriptstyle \circ}$} } }
\newcommand{\sign}{{\rm sign}}
\newcommand{\dfrac}[2]{\displaystyle{\frac{#1}{#2}} }
\newcommand{\pg}[1]{page \pageref{#1}}

\newcommand{\paragraphh}[1]{\vspace{7mm} \refstepcounter{th} {\bf\arabic{th}.\quad} {\em #1} }
\newcommand{\aparagraphh}[1]{\vspace{7mm} \refstepcounter{proof} 
{\bf A\arabic{proof}.\quad} {\em #1} }
\draft

\title{Generalized Lorentzian Adjustment of Reference Frames\\
\small{\bf and Waves of Transformation of Spacetime}}
\author{Iosif Pinelis}
\address{Michigan Technological University, Department of Mathematical Sciences, Houghton, MI 49931}
\date{\today}
\maketitle

\begin{abstract} It is demonstrated that {\em any} two reference frames (RFs),
which are  uniformly and rectilinearly moving relative to each other, can be
adjusted via  (possibly anisotropic) rescaling and re-synchronization  so that
the resulting pair of RFs is Lorentzian; this statement {\em remains} true if
the word ``Lorentzian" is replaced by ``Galilean" or ``Riemannian", i.e., if a
finite positive value of $c^2$ is replaced by $\infty$ or by a negative real
number.  In this particular sense, the Lorentzian, as well as Galilean or 
Riemannian,
phenomenon turns out to be merely a matter of an arbitrary choice of
appropriate rescaling and re-synchronization of any given pair of RFs. 
Generalizations and refinements of this result are obtained, including
universal generalized Lorentzian adjustment via rescaling and
re-synchronization of arbitrarily large families of RFs.  Alternatively, the
generalized Lorentzian property of a pair of RFs is shown to be a consequence
of reciprocity and isotropy, with no adjustment needed in this case. The
universality of light and of the corresponding Lorentzian property of the
spacetime is questioned. Waves of transformation of spacetime are introduced,
which have in a certain sense a more universal character than electromagnetic
or gravitational waves.  \end{abstract}


\tableofcontents

\widetext

\paragraph*{}

\section{Introduction}\label{intro}

\noindent\P\quad This paper is peculiar in more aspects than one. 
Here are some indications as to
what this paper is and does, and what it is not and does not.

\begin{itemize}
\item This paper is an attempt at a careful critical
	reading of Einstein's paper on the special theory of relativity
	\cite{einstein-special}. In particular, we rigorously examine the physical
	procedures required to establish an appropriate correspondence between
	space- and time-measuring devices in two reference frames moving uniformly
	and rectilinearly relative to each other; we refer to such physical
	procedures as adjustment of reference frames.  
\item As such, this paper is most definitely not in the mainstream.  To the best
	of my knowledge, it 
	bears little relation with any work which has followed Einstein's 
	\cite{einstein-special}. Therefore, ``reading" this paper by way of
	associations with existing literature will most probably result in
	misunderstanding. 
\item This paper is entirely self-contained, except for the purely
	mathematical references \cite{alexandrov,fund-th,bor-heger}.
\item In no way does this paper use the notion of the metric tensor or any 
	terms
	based on that notion. The consideration is local throughout the paper,
	except for Section \ref{waves}, where a condition of differentiability 
	in an
	entire region of spacetime is imposed. Therefore, an attempt to interpret 
	the results of this
	paper in 
	metric-based terms will most probably lead to misunderstanding. 
\item Neither any properties of electromagnetic waves nor even their
	existence are used in this paper. 
\item Thus, in the usual sense, the theories presented here pertain neither to
	the ``special" theory of relativity (since we do not unilize the notion of
	light) nor to the ``general" one (since we do not unilize the notion of
	the metric).
\item No specific properties such as inertiality, group properties, properties
	of specific measuring devices as rigid or elastic bodies are used.  
\item No properties of spacetime such as isotropy and reciprocity are fixed
	throughout this paper. Rather, an entire spectrum of possible physical 
	scenarios is considered, ranging from scenarios with no assumptions
	whatsoever on isotropy or reciprocity to ones where both isotropy and
	reciprocity are assumed to be fully present. 
\item In each such scenario, it is shown that there exists an adjustment of 
	the pair
	of reference frames in question which makes the pair Lorentzian; the
	``amount" of the required adjustment is the less the more isotropy or
	reciprocity is there.
\item The treatment is completely rigorous, which is convenient and almost
unavoidable for a careful logical examination and especially when dealing
	with the multitude of the possible scenarios. 
	Yet, the mathematics involved is most elementary, even though at times
	tedious.
\item One should note that the notion of reference frames (RFs) and that of 
	RF change 
	transformations (RFCTs) are defined in this paper to allow the most general
	consideration: namely, so that {\em any} non-singular $4\times4$ real 
	matrix is a matrix
	of some RFCT.
\end{itemize}
Since the first days of the theory of relativity, the common belief has been 
that the Lorentzian transformations can appear only as a consequence of special 
physical conditions, which the RFs under consideration must satisfy, such as 
inertiality, constancy (in various senses) of the speed of light, isotropy, 
reciprocity, the principle of relativity, properties of rigid or elastic 
bodies, various group conditions, etc.

Quite contrary to this belief,
we demonstrate in this paper that the Lorentzian transformations can arise 
simply 
as the result of an appropriate (possibly anisotropic) rescaling and 
re-synchronization 
of {\em any} given pair of mutually uniformly and rectilinearly moving 
(URMoving) RFs.

This is the first main result of this paper, Theorem \ref{TH1-cor}, 
page \pageref{TH1-cor},
stated also in the first sentence of the above Abstract. 
The import of this statement depends foremost on the definition of an RF and on 
that of an RF URMoving relative to another RF.  
At this point, suffice it to say that our definitions will be such that 
{\em any} non-singular linear --- or, even more generally, affine -- 
transformation of $\R^4$ 
serves as the RF change transformation (RFCT) for some pair of RFs, URMoving 
relative to each other. (An affine transformation is any composition of a 
linear transformation and a parallel translation.) 
Thus, our notion of an RF URMoving relative to another RF is as wide as it can 
possibly be. 

We call an RF $\tilde f$ an adjustment of another RF $f$ if $\tilde f$ is at 
rest relative to $f$. 
It is easy to see -- refer to Proposition \ref{adjust-structure}, 
\pg{adjust-structure} --
that any adjustment of an RF may be obtained as a composition of the following 
four elementary types of adjustment: the (trivial) space-time origin 
adjustment, temporal adjustment, spatial adjustment, and re-synchronization.

In the fundamental paper by Einstein \cite {einstein-special} and in most 
texts, the special theory of relativity is derived based on the principle of 
relativity and on the postulate of the constancy of the speed of light.

Even at the first attempt to examine these two cornerstones of the theory, it 
becomes clear that their possible meaning crucially depends on the adjustment 
procedures employed in order to put into correspondence spacetime measurements 
in the two given RFs, which includes adjustment of the rates of the clocks, 
of the directions of the 
spatial axes, of the units along them, and 
synchronization of the clocks as a function of the spatial position of the 
clock. 
We shall refer to such procedures as (mutual) adjustment of (the pair of) RFs. 

Most of the existing accounts of the theory of relativity do not emphasize the import of the
choice of adjustment procedures. 
However, the original paper by Einstein \cite {einstein-special} treats the 
matter of adjustment quite explicitly.  
In particular, beams of light are used for synchronization of clocks; Einstein 
assumes, in addition to the principles of relativity and of the constancy of 
the speed of light, that the relation ``the clock at point $A$ synchronizes 
with 
the clock at point $B$" that he defines will be symmetric and transitive, 
i.e., it will be a relation of equivalence.
The spatial geometry in each of the two given RFs is assumed to be Euclidian.
Physically, this means the existence of rigid bodies with their usually 
assumed properties.

The correspondence between the spatial units is established according to 
Einstein 
\cite {einstein-special} by transporting rods from one RF into another RF, 
URMoving relative to the first one.
With such an approach, one could ask whether the logical foundations of the 
{\em special} theory of relativity are not thus compromised, since the rods to 
be so transported must be {\em accelerated} if the relative speed of the two 
RFs is nonzero. 

To avoid this difficulty, some authors just require -- tacitly or, less often, 
explicitly -- only the {\em existence} of a universal adjustment of all, say 
inertial, RFs such that the principles of relativity and that of the constancy 
of the speed of light are satisfied. 
However, this approach not only needs the additional, certainly not trivial, 
requirement of the existence of a universal adjustment but also leaves open 
the question as to how such a universal adjustment can be physically achieved.

To overcome all these difficulties, we engage into a comprehensive study of 
adjustment of RFs in relation with the generalized Lorentzian property. 
Surprisingly, this appears to be the first systematic study of adjustment of 
RFs.

We introduce generalized Lorentzian -- i.e., $C$-Lorentzian for some real 
$C$ -- pairs of RFs and the corresponding RFCTs;
we refer to an RFCT (and to a corresponding pair of RFs) as $C$-Lorentzian if 
the RFCT preserves the $C$-metric 
$$(t_2-t_1)^2-C[(x_2-x_1)^2+(y_2-y_1)^2+(z_2-z_1)^2]$$
in $\R^4$ -- cf. the Minkowski metric 
$c^2(t_2-t_1)^2-[(x_2-x_1)^2+(y_2-y_1)^2+(z_2-z_1)^2]$.
If $C>0$ and $c=1/\sqrt C$, then the two metrics are essentially the same, 
differing only by a constant factor. 
 
It is the sign of $C$ that is of utmost importance.
Let us refer to the $C$-metric as positive-Lorentzian or simply Lorentzian if 
$C>0$, 0-Lorentzian or Galilean if $C=0$ (which corresponds to $c=\infty$ and 
implies the preservation of the time interval $|t_2-t_1|$), and 
negative-Lorentzian or Riemannian if $C<0$.

An implication of Theorem \ref{TH1-cor}, \pg{TH1-cor}, is that any given pair 
of mutually URMoving RFs is Lorentzian up to adjustment. A significant feature 
of Theorem \ref{TH1-cor} is that, however wide or narrow definition of 
URMoving RFs is assumed, the word ``Lorentzian" in the last implication can be 
freely replaced by ``Galilean" or ``Riemannian".
This may seem highly surprising, since the Lorentzian property, as contrasted 
to the Galilean or Riemannian one, now looks merely as a matter of the choice 
of rescaling and re-synchronization of one and the same pair of RFs, rather 
than a fundamental property of a physical spacetime. 

Such a impression would be true only in part. 
In Sections \ref{recip-iso-natur} and \ref{level-2}, we show that -- if 
certain verifiable
physical conditions of isotropy and reciprocity take place -- the sign of $C$ 
can be
described as a natural local property of the physical spacetime. We 
propose critical experiments which could discriminate, again locally, 
between 
the three possible types of spacetime geometry: Lorentzian, Galilean, or 
Riemannian.
We shall refer to any of these three types of spacetime geometry as generalized Lorentzian.
We do not assume that the entire spacetime is of any one of these three types.

Mathematically, Theorem \ref{TH1-cor} is very simple; it just means that any
non-singular
$4\times4$ real matrix $A$ (i.e., the matrix of any RFCT) can be represented as
\beq{adjust-intro}
A
=\pmatrix{\tau_1&\bb^T_1\cr \0&S_1\cr}B\pmatrix{\tau&\bb^T\cr \0&S\cr},
\eeq
where $B$ is a $C$-Lorentzian matrix and the matrix blocks $S$ and $S_1$ are
$3\times3$; note that $\pmatrix{\tau&\bb^T\cr \0&S\cr}$ is the general form of
the matrix of adjustment transformations.

Now a surprise possibly produced by the statement of Theorem \ref{TH1-cor}
should all but disappear. Indeed, the L.H.S. of \req{adjust-intro} can be 
described by $4\times4=16$
real parameters (``degrees of freedom"), while the R.H.S. of \req{adjust-intro} 
contains two times more,
32 parameters
in all:
$2(16-3)=26$ parameters of the two adjustment matrices plus 6 parameters of
the $C$-Lorentzian matrix $B$. 

Moreover, not only does representation \req{adjust-intro} exist, it is not
unique. A reason for this, as one can now see, is that a general adjustment
matrix contains ``too many", 13, ``degrees of freedom". One may therefore want 
to allow
only certain special forms of adjustment, rather than the
general one. Alternatively or concurrently, one may also want to choose
a standard form of the $C$-Lorentzian matrix $B$.

That is just one way to look at results of Subsections \ref{universal} through 
\ref{w-out-resynchro}, where we have certain uniqueness. In particular,
in Subsection \ref{parametrization}, one of the two adjustment matrices in
\req{adjust-intro} is required to be the identity matrix while the 
$C$-Lorentzian
matrix $B$ is required to be a $C$-boost; then the total on each side of
\req{adjust-intro} is $(16-3)+3=16$ 
``degrees of freedom",
which provides for a unique representation of any non-singular $4\times4$ real
matrix as the product of the matrix of an adjustment and that of a $C$-boost. 

One may want just to put up with such a matrix language, without delving into 
such
questions as what an RF itself is; then one may skip some material of Section 
\ref{definitions}.  
 
In the general theory of relativity (TR), since {\em all} locally linear (i.e. 
differentiable) RFCTs are allowed,
the qualitative distinction between time and space, rather strong in the 
special TR, seems to almost disappear. 
This almost complete elimination of the distinction between time and space may 
seem hardly reconcilable with experimental practice, in which time-measuring 
devices and processes are quite different from space-measuring ones.
Thus, a reasonable question is, How could it be substantiated that {\em all} 
linear RFCTs should be allowed, be it in a special or general TR?

Another result of this paper may serve to address this concern.
This result is Theorem \ref{TH1}, \pg{TH1}, which at the first glance and by
itself might seem even more 
surprising than Theorem \ref{TH1-cor}, since the latter is only an immediate 
corollary to the former.

Theorem \ref{TH1} may be stated as follows: 
Let $(f,g)$ be a pair of  RFs which are mutually URMoving with a nonzero 
velocity and let $(f_1,g_1)$ be {\em any} other such pair; then RFs $f$ and 
$g$ can be respectively adjusted to some RFs $\tilde f$ and $\tilde g$ so that 
the RFCT from $\tilde f$ to $\tilde g$ is the same as the RFCT from $f_1$ to 
$g_1$. 

In other words, {\em any} affine RFCT with the corresponding nonzero relative 
velocity is {\em reducible} by means of RF adjustment to {\em any other} such 
RFCT. In addition, it is easy to see that this statement remains true if one 
replaces here the nonzero relative velocity requirement by the requirement 
that for {\em both} RFCTs the corresponding relative velocity {\em is} zero; 
however, because of unavoidable measurement errors, {\em exactly}-zero 
velocities 
are obviously exceptions, which cannot possibly be 
experimentally detected.
In this sense, practically all affine RFCTs can be obtained from practically 
any other 
affine RFCT via RF adjustment. 
Thus, Theorem \ref{TH1} provides a reason as to why {\em all} linear RFCTs 
should be allowed, and not only in the general theory of relativity but in the 
special one as well.
More exactly, however, what Theorem \ref{TH1} says is that being relatively in motion 
or being relatively at rest is the only invariant of RF adjustment. 
We see that some degree of distinction between time and space must remain
so that the relations of being relatively at rest or not at rest can be defined.

Pauli \cite {pauli}, pg. 11, describing results by Ignatowsky, 
Frank and Rothe \cite {frank-roth}, wrote: ``Nothing can, naturally, be said 
about the sign, 
magnitude, and physical meaning of $\al$"; Pauli's $\al$ corresponds to $C$ in 
our notation. 
Contrary to Pauli's opinion, in Section \ref{experiment} we describe an 
experiment 
through which the sign and magnitude of $C$ can be measured, even though 
indirectly; the dimension of $C$ is naturally that of [velocity]${}^{-2}$. 
Moreover, we provide a physical interpretation of  $1/C$ as the product of the 
velocities of certain time and space waves -- see
\req{C-interpret}, page \pageref{C-interpret}.

The main distinction between time and space is that time is one-dimensional 
and space is three-dimensional. For in the cases when only one spatial 
dimension 
is of interest, time and space become exchangeable. 
This may be not very surprising in certain everyday situations or, more 
generally, whenever there is a standard velocity. 
We say, e.g, "the distance from point A to point B is a ten-minute walk". 

Much deeper insights are provided by results of Section \ref{waves} which 
demonstrate, in particular, that there exists a complete in a certain sense 
duality between time and {\em one-dimensional} space in terms of certain waves 
of transformation of the spacetime.  

(Incidentally, in a number of derivations of the Lorentzian property, the 
spatial component of the spacetime is 
assumed to
be easily reducible to 
one dimension. But with one spatial dimension, there is virtually no problem.
Indeed, assuming just the reciprocity of the RFCT, its
$2\times2$ matrix $A$ must satisfy the equation $A=A^{-1}$; now a few lines of 
most elementary algebra show that the RFCT is generalized
Lorentzian.)

One may argue that the privileged role of the positive-Lorentzian geometry (with
$C>0$), in 
contrast to the negative- or 0-Lorentzian ones, is related to the special role 
ascribed to light, and this is true.
In fact, a much stronger statement is true \cite {alexandrov}:
Suppose that, for a given pair of RFs, there is {\em some} signal whose speed 
$c$ is always the same in both of the RFs, i.e., the equality 
$c^2(t_2-t_1)^2-[(x_2-x_1)^2+(y_2-y_1)^2+(z_2-z_1)^2]=0$ 
in one of the RFs implies the same in the other RF; 
then the RFCT is a scalar multiple $\al{\cal L}$ of a Lorentzian 
transformation ${\cal L}$,  where $\al$ is a positive real. 

Hence, if a slight reciprocity requirement is also satisfied (in order for the 
factor $\al$ in $\al{\cal L}$ to be necessarily equal to $\pm1$), then the RFCT is 
simply Lorentzian. 
Such an additional reciprocity requirement can be considered as an extremely 
non-restrictive form of the principle of relativity. 
Thus,
the principle of the constancy of the speed of light is so strong that almost 
by itself, just with an addition of a very weak trace of the principle of 
relativity, it implies the Lorentzian property;
here one even does not need to assume that the two given RFs 
are mutually URMoving -- the latter is {\em already} implied by the only 
assumption of the constancy of the speed of the signal!  

Such extreme restrictiveness of the requirement of the constancy of the speed 
of light is obviously related with 
its being too counterintuitive as perceived by many researchers since the 
first days of the relativity theory.
Some of them have also thought that to give the {\em electrodynamic} notion of 
light any special role in a theory of {\em kinematics} means to reverse the 
natural 
order of ideas.
For how can one possibly define a theoretical concept of electromagnetic waves 
{\em before} such kinematic notions as the coordinates of events in time and
space and relations between them have been developed?  

At this point, one may further argue that the special role of light does not 
necessarily imply putting electrodynamics before kinematics but is merely 
justified by the agreement between the Lorentzian kinematics and the equations 
of electrodynamics. 

The latter objection would be theoretically justified if the conventional form 
of the equations of electrodynamics were the only one theoretically possible 
or at least logically preferable. 
However, it requires no effort to give a simple (and just as inherently 
consistent as the 
conventional form) extension of the equations of electrodynamics, comprising 
the negative-Lorentzian and 0-Lorentzian spacetimes (in addition to the 
positive-Lorentzian ones).
As could be expected, the so extended equations 
do {\em not admit electromagnetic waves at all} in negative-Lorentzian and 
0-Lorentzian spacetimes; we also briefly describe here the corresponding 
hypothetical mechanics of not charged particles as well.

The generalized Maxwell-Hertz equations for empty space, with $\dfrac1c$ 
replaced by $\sqrt C$, are
$$\sqrt C \,\dfrac{\partial{\bf E}}{\partial t}=\nabla\times{\bf H},\quad
\sqrt C \,\dfrac{\partial{\bf H}}{\partial t}=-\nabla\times{\bf E},\quad
\nabla\cdot{\bf E}=0,
\quad{\rm and}\quad
\nabla\cdot{\bf H}=0,
$$
where $\bf H$ is the magnetic field (whose components take on imaginary values 
when $C<0$) and $\bf E$ is the electric field.
This implies the system of 6 scalar equations
$$C\dfrac{\partial^2{\bf E}}{\partial t^2}-\Delta{\bf E}=\0
\quad{\rm and}\quad
C\dfrac{\partial^2{\bf H}}{\partial t^2}-\Delta{\bf H}=\0,
$$
which are the conventional hyperbolic (wave) equations if $C>0$ but elliptic 
ones if $C<0$.

Thus, electromagnetic waves cannot exist in the negative-Lorentzian domains, 
where $C<0$. 

Similarly, the conventional formula for the electromagnetic force is 
generalized as 
$${\bf F}=e{\bf E}+\sqrt C e\vv\times{\bf H},$$
and so, the components of $\bf F$ always take on real values, no matter what 
is the sign of $C$. 
The formulas for the mass and the energy become
$$m=\dfrac{m_0}{\sqrt{1-Cv^2}}\quad{\rm and}\quad
E=\dfrac m C=\dfrac{m_0}{C\sqrt{1-Cv^2}};$$
thus, the mass $m$ {\em decreases} as $v$ increases if $C<0$, whereas the 
energy $E$ always increases as $v$ increases, whether $C$ is positive or 
negative.

Electromagnetic waves, as well as the conventionally described gravitational 
ones, can exist only in positive-Lorentzian domains. 
In Section \ref{waves}, we describe certain transformation waves that can 
exist in domains of {\em any} of the three types of spacetime geometry. 
We see this as another argument against the necessity of the 
positive-Lorentzian geometry and against that of the universal existence and 
character of light throughout the entire universe.

Yet one more potential objection can be seen here -- that so far all 
experiments have been in agreement with the positive-Lorentzian structure 
of 
spacetime. 
By itself, this statement can hardly be doubted; furthermore, we believe that 
if the aforementioned critical experiment proposed in Section \ref{experiment} 
of this paper, were conducted in a vicinity of the Earth, it only would once 
again positively confirm the positive-Lorentzian property.

However, what any experiment or at least any experiment dealing with signals 
with bounded velocities
can really test is a property of only a bounded spacetime domain in which we 
happened to be situated, rather than that of the entire universe.
The latter may nevertheless have negative-Lorentzian domains as well, even if 
very remote from us, plus three-dimensional 0-Lorentzian hypersurfaces 
between the positive-Lorentzian and negative-Lorentzian domains.
At least, no substantial logic is seen which would exclude the possibility of 
such an intermittent structure of the universe at large.

Actually, if there are any purely theoretical reasons to discriminate between 
the three types of spacetime, some general preference should be given to the 
negative-Lorentzian and not to the positive-Lorentzian type.
Indeed, in this paper we also describe certain {\em universal} adjustments of 
however large families of mutually \mbox{URMoving} RFs.
We show that {\em any} family of mutually URMoving RFs possesses a universal 
negative-Lorentzian adjustment.
However, to possess a universal positive-Lorentzian or 0-Lorentzian 
adjustment, a family of RFs must satisfy certain restrictions -- see 
Theorem \ref{univ}, \pg{univ}.
In this sense, the negative-Lorentzian adjustment is more universal than the 
positive-Lorentzian and 0-Lorentzian ones.

An intriguing question is, If there are negative-Lorentzian domains in the 
universe, how could they be experimentally detected? 
At this point, we are far from being able to fully describe the nature of 
signals that 
may originate in negative-Lorentzian domains and to say in what manner can 
such signals get transformed upon entering our, doubtlessly 
positive-Lorentzian, part of the universe.
Note that in negative-Lorentzian domains there may exist signals of any finite 
or infinite speed; if such infinite-speed or too-high-speed signals can 
penetrate into Lorentzian domains at 
all, they must at least appropriately decrease their speed upon such 
penetration. 

What we can also say with certainty is that those hypothetical signals cannot 
originally exist in the negative-Lorentzian domains as either electromagnetic 
or conventionally described gravitational waves.
It might therefore seem to be not a very good idea to try to detect 
negative-Lorentzian domains via electromagnetic or gravitational waves.
Instead, it makes sense to try to detect waves of transformation of spacetime, 
described in Section \ref{waves}, using methods of Section \ref{experiment}.

In this paper, almost no assumptions are made in general; instead, a number of 
possible scenarios are proposed;
these scenarios depend on the number and nature of assumptions.  
If anything is being assumed, it is explicitly stated.
To make such an approach effective, it is important to verify every time that 
the assumptions made are necessary or essential, and we adhere to this maxim.

By the already mentioned Theorem \ref{TH1-cor}, \pg{TH1-cor}, {\em any} pair 
of mutually URMoving RFs can be adjusted to a $C$-Lorentzian pair via 
(possibly anisotropic) rescaling and re-synchronization. 

On the other side of this spectrum of results is Theorem \ref{RECIP-AND-ISO}, 
\pg{RECIP-AND-ISO}, which says that certain reciprocity and isotropy 
properties of a pair of mutually URMoving RFs
imply that the given pair of RFs is {\em already} generalized Lorentzian. 
Moreover, by Theorem \ref{RECIP-AND-ISO-cor}, the generalized Lorentzian pairs 
can be fully characterized by reciprocity and isotropy.

It should therefore be clear that generally, the more is the extent to which 
reciprocity and isotropy 
conditions are satisfied by a given pair of mutually URMoving RFs, the less 
adjustment is needed in order to adjust such a pair to a 
generalized Lorentzian one.

We consider three main levels of assumptions regarding a given pair of 
mutually \\
URMoving RFs 
and identify the three corresponding levels of adjustment needed in order to 
reduce the given pair of RFs to a generalized Lorentzian pair: 
\begin{description}
\item[0.]
\label{levels}
no assumptions at all -- then both (possibly anisotropic) rescaling {\em and} 
re-synchronization may be needed; see Theorem \ref{TH1-cor};
\item[1.]
only reciprocity {\em or} isotropy is assumed -- then, respectively, only 
(possibly anisotropic) rescaling {\em or} isotropic rescaling with re-synchronization 
may be needed; see Theorem \ref{RECIP}, \pg{RECIP}, and Theorem 
\ref{skew-free-charact}, \pg{skew-free-charact};
\item[2.]
both reciprocity {\em and} isotropy are assumed -- then no adjustment is needed, 
the pair of RFs is then already generalized Lorentzian; see Theorem 
\ref{RECIP-AND-ISO}, \pg{RECIP-AND-ISO}.
\end{description} 

This is a physical paper of a rather infrequently encountered style.
We try to make explicit the distinction between the two modes of 
consideration:
(i) when we are discussing relations {\em between} the reality and the model 
and (ii) when we are acting {\em within} a rigorous mathematical model.
Thus, we first build various models -- see Section \ref{definitions}; then we 
work within the model under consideration by means of purely mathematical 
methods, without using nearly impossible to rigorously define -- at least
before kinematics is developed -- notions such as 
clocks, rods, light, inertiality, etc. -- see Sections \ref{results} and 
\ref{waves} and the Appendixes; finally, we go back to reality to consider 
methods of testing of the results and to interpret the predictions of the 
theory -- see Section \ref{experiment}.

The theorems in this paper can each be considered as a mini-theory of 
relativity; the assumptions of a theorem correspond to a possible real-world 
scenario; each of the assumptions corresponds to a postulate, i.e., to a 
statement about properties of physical objects. 

We find this style to be especially appropriate for the subject of this paper; 
it helps to organize the multitude of different scenarios, to clearly 
distinguish and, on the other hand, to show the correspondence between 
physical objects and relations and their counterparts in the model(s) used. 
This approach is also effective in that it provides maximum generality, since 
one and the same model notion may, and often does, admit many different 
physical realizations.

In particular, we need not restrict ourselves to inertial RFs, i.e. ones 
usually considered as ``freely falling" far away from large masses, where the divergence 
of the gravitational field is negligible. 
Instead, for the most part of this paper, we theoretically consider pairs of 
RFs which are only assumed to be URMoving relative each other, neither of them 
having to be inertial or otherwise distinguished by itself; obviously, this is 
in perfect correspondence with the spirit of the theory of {\em relativity}. 
On the other hand, pairs of inertial RFs can be considered as special, even 
if 
most common, physical realizations of the notion of pairs of mutually URMoving 
RFs.

This paper is devoted foremost to establishing as much order and clarity as 
possible in {\em kinematic} foundations of physics. For is any unifying, 
thoroughly penetrating physical theory possible other than one which is based 
on a firm and free of contradiction or vicious circles kinematic foundation?

One may argue that as soon as the Lorentzian transformations are derived and 
as long as the corresponding predictions are all well confirmed by all 
experiments, there is no need to question the basis on which the theory is 
built. It is however always an advantage to have a theory, based on less
contradictory 
and more general logical foundations, which would be more flexible and more 
easily 
adaptable whenever new experimental data appear.

The most basic notion in all the models introduced in this paper is that of 
the RF. An RF is understood, in accordance with Einstein 
\cite {einstein-general}, as any 1-to-1 correspondence between the space of 
all events and the space of all four-tuples of their temporal and spatial 
coordinates; the latter space may be either $\R^4$ or a subset of $\R^4$.   
The other notions are all built on the notion of the RF using logic, which 
parallels the corresponding relations between physical objects. 

This correspondence between physical objects and relations and their model 
counterparts is indicated by using, after necessary discussion, the same term 
both for the physical object and the respective model notion. 
The confusion between the two will hardly be possible because of the context; 
in particular, Section \ref{results} is explicitly devoted to the statement 
and discussion of theoretical results within different rigorous models, while 
in Section \ref{definitions} all the basic notions to be subsequently used are 
introduced and discussed.

Let us emphasize that we need not rigorously define notions like those of 
observers, clocks, rods, inertiality, light; none of them is among our basic 
notions. 
Therefore, our theoretical results do not depend on the concrete physical 
realizations of these notions.
Of course, these physical notions are important, but we refer to them only at the initial stage of building of the models and at the concluding stages of testing of the models and of interpreting of the results.

\section{Basic notions: Reality--Model connection}
\label{definitions}

\paragraph*{}


\subsection{   Events, reference frames (RFs), RF change transformations (RFCTs),  and relabeling of events}
\label{RF-RFCT}

Essentially, our notion of an RF coincides with the one proposed by Einstein \cite {einstein-general} in his general theory:
``We allot to the universe four space-time variables $x_1,x_2,x_3,x_4$ in such a way that for every point-event there is a corresponding system of values of the variables $x_1,x_2,x_3,x_4$. To two coincident point-events there corresponds one system of values of the variables $x_1,x_2,x_3,x_4$ ...". 
Thus, two events with same space-time coordinates in an appropriate RF are considered to be the same.  

We use the term {\em events}, rather than ``point-events", and denote events by $e,e_0,e_1,\dots$ . 
The set of all events is called the {\em event space} and is denoted by $\E$.
Let us stress that in this paper the nature of events is irrelevant; no structure on the event space $\E$ is assumed. 
The only assumption made about $\E$ is that it can be put into a 1-to-1 correspondence with $\R^4$. 
Any such correspondence is referred to as an RF.  
More exactly, let us define an {\em RF} as an arbitrary 1-to-1 mapping of $\E$ onto $\R^4$.
This definition will be used throughout this paper, except only for Section \ref{waves}; see further details there.

Thus, any RF $f$ takes every event $e$ in $\E$ to the corresponding 4-tuple $f(e)=\pmatrix{t\cr\xx}=:\pmatrix{t^f(e)\cr\xx^f(e)}$ in $\R^4$ of the time-space coordinates of event $e$ in RF $f$, so that the real number $t^f(e)$ and the vector $\xx^f(e)$ in $\R^3$ represent, respectively, the one temporal and the three spatial coordinates of event $e$ in RF $f$.
Following the common practice, we identify vectors in $\R^d$ with the corresponding $d\times1$ real column matrices and 
let small boldfaced Roman letters stand for vectors in $\R^3$ and the corresponding italicized letters, for their length: 
$r:=|\xx|$, $v:=|\vv|$, etc. 
We also let $X$, $Y$, etc., or $\pmatrix{t\cr\xx}$, where $t$ is a real number, denote vectors in $\R^4$. 
Whenever speaking of pairs or any other families of RFs, we shall always assume that the RFs in question are defined on the same event space, unless otherwise specified.

For any two RFs $f$ and $g$, the {\em RFCT} $\A^{g,f}$ from $g$ to $f$ is then defined as the mapping that carries the vector $g(e)\in\R^4$ of the temporal and spatial coordinates of every event $e$ in RF $g$ to the vector $f(e)\in\R^4$ of the coordinates of the same event $e$ in RF $f$, so that the following diagram is commutative:
\newcommand{\stk}[1]{\stackrel{#1}{\longrightarrow}}
\[
	\begin{array}{lcl}
	\E	& \stk{g} & \R^4\\  
	\downarrow\mbox{id}_\E&	& \downarrow\A^{g,f}\\  
	\E	& \stk{f} & \R^4
	\end{array}
\]  
Here, as usual, id${}_\Sigma$ denotes the identity mapping of a set $\Sigma$, i.e. the mapping that does not move any element of $\Sigma$. 
The diagram being commutative means here that 
$\A^{g,f}\circUp \,g=f\circUp \,\mbox{id}_\E(=f)$.
Thus, RFCT $\A^{g,f}$ is a 1-to-1 transformation of $\R^4$ onto itself, and is the composition of mapping $f$ and the inverse $g^{-1}$ of $g$, i.e., 
$$\A^{g,f}=f\circUp \,g^{-1}.$$

It is the relative motion of given RFs as represented by the RFCT only, rather the nature of the RFs themselves, that matters in a theory of relativity; 
in the rest of this subsection, this thesis is clarified in terms of re-labeling.   

Let us call any 1-to-1 mapping $\ell$ of the event space $\E$ onto itself or another set $\E^\ell$ a {\em re-labeling} of $\E$; let us then call $\E^\ell$ a {\em re-labeled} event space. 
Let $e^\ell:=\ell(e)\in\E^\ell$ denote the re-labeling of event $e$ in $\E$ under mapping $\ell$. 

Then the formula $f^\ell(e^\ell)=f(e)$ for all $e$ in $\E$ -- so that 
$f=f^\ell\circUp\ell$ and $f^\ell=f\circUp\ell^{-1}$ -- determines an obvious 1-to-1 correspondence between the RFs $f\colon\ \E\to\R^4$ defined on the ``original" event space $\E$ and their ``re-labeled" versions $f^\ell\colon\ \E^\ell\to\R^4$ defined on the re-labeled event space $\E^\ell$. 

This is illustrated by another commutative diagram:
\[
	\begin{array}{lcl}
	\E	& \stk{f} & \R^4\\  
	\downarrow\ell&	& \downarrow\mbox{id}_{\R^4}\\  
	\E^\ell	& \stk{f^\ell} & \R^4
	\end{array}
\]

Let us say that two pairs of RFs $(f,g)$ and $(f_1,g_1)$ are {\em the same up to re-labeling of events} if $f_1=f^\ell$ and $g_1=g\,^\ell$ for some re-labeling $\ell$, one and the same for $f$ and $g$. 


\paragraphh{Proposition: Identical RFCTs and re-labeling of events}
\label{re-label}
\ \\
Two pairs of RFs $(f,g)$ and $(f_1,g_1)$ are the same up to re-labeling of events if and only if the pairs have the same RFCTs: $\A^{g_1,f_1}=\A^{g,f}$.  
\qedd

It is very easy to verify this statement; see Appendix \ref{re-label-proof}, \pg{re-label-proof}.    


\subsection{   Uniform rectilinear motions (URMotions) and their velocities relative to RFs}
\label{URMotion}

Let $M$ be any {\em motion}, that is, any subset of the event space $\E$. 
Let $f$ be any RF. 
For any two different events $e_1$ and $e_2$, belonging to motion $M$, let us define the {\em (average) velocity of motion $M$ relative to RF $f$ between the two events} as
$$\vv^{M,f}(e_1,e_2)
:=\dfrac{\xx^f(e_2)-\xx^f(e_1)}{t^f(e_2)-t^f(e_1)}$$
provided that $t^f(e_2)\ne t^f(e_1)$; otherwise, $\vv^{M,f}(e_1,e_2)$ or, more exactly, the relative speed $|\vv^{M,f}(e_1,e_2)|$ is considered infinite, and  the direction of the line through the origin carrying the vector $\xx^f(e_2)-\xx^f(e_1)$ is assigned to velocity $\vv^{M,f}(e_1,e_2)$; the direction of such a line is defined by the unordered set $\{\ee,-\ee\}$ of unit vectors with $\ee:=\dfrac{\xx^f(e_2)-\xx^f(e_1)}{|\xx^f(e_2)-\xx^f(e_1)|}$ (note that necessarily $\xx^f(e_2)\ne\xx^f(e_1)$, since $e_2\ne e_1$ while $t^f(e_2)=t^f(e_1)$\ ). 

Thus, the scope of this paper is not restricted to finite velocities only.
Physically, though, this particular point does not represent a significant advantage, since, obviously, infinite velocities cannot possibly be experimentally detected, as well as any finite velocity cannot be measured precisely. 
However, it is certainly more convenient not to restrict modeling by the exclusion of infinite velocities; 
one reason for this is that the velocity $\vv^{M,f}(e_1,e_2)$ between two different events can always be {\em made} infinite simply by using re-synchronization (refer to Subsection \ref{adjustment}, \pg{adjustment}) in order to make events $e_1$ and $e_2$ synchronous relative to RF $f$, so that $t^f(e_2)=t^f(e_1)$. 

A {\em URMotion} relative to an RF $f$ 
is then defined as any motion $U$, containing at least two different events and such that the average relative velocity $\vv^{U,f}(e_1,e_2)$ between two different events $e_1$ and $e_2$ belonging to $U$ does not depend on the choice of such events $e_1$ and $e_2$. 

Let us denote the constant {\em velocity of a URMotion $U$ relative to RF} $f$ simply by 
$\vv^{U,f}$, so that $\vv^{U,f}=\vv^{U,f}(e_1,e_2)$ for any choice of two different events $e_1$ and $e_2$ belonging to $U$.

In other words, a subset $U$ of the event space $\E$ is a URMotion relative to an RF $f$ if and only if the image $f(U)$ (``world-line") of $U$ under mapping $f$ lies on a (straight) line in $\R^4$.

Thus, the so-defined model notion of the URMotion corresponds to the uniform and rectilinear motion of a negligibly small physical particle. 
The ``world-line" of such a particle does not have to be an {\em entire} (straight) line.
As follows from the {\em Fundamental Fact} cited below in Subsection \ref{URMoving}, the use of such a more general notion does not diminish the strength of the subsequent results;
on the other hand, such a model notion better corresponds to physical reality, since in practice only finitely many events can be observed, and so, the corresponding points in $\R^4$ can never fill a continuous line.

\subsection{    Mutually uniformly and rectilinearly moving (URMoving) RFs, their relative velocities, and linearity of RFCTs}
\label{URMoving}

We have modeled the uniform and rectilinear motion of a negligibly small physical particle.
A physical RF is thought of as consisting of a (usually very large) number of specially arranged small particles and hence cannot be considered to be same as just any one small physical particle. 
Therefore, another round of modeling is needed to introduce model notions of mutually URMoving RFs and their relative velocities.  

Let us say that an RF $g$ is {\em URMoving} relative to another RF $f$ if every URMotion $U$ relative to $g$ is a URMotion relative to $f$, too.
Geometrically, this simply means that the RFCT $\A^{g,f}=f\circUp \,(g)^{-1}$ maps every line in $\R^4$ into a line.

The {\em Fundamental Fact} in any special TR is that if an RF $g$ is URMoving relative to another $f$, then the RFCT $\A^{g,f}$ is affine; this fact is just a restatement of the fundamental theorem of affine geometry \cite {fund-th}. 
In particular, this fact implies that if RF $g$ is URMoving relative to RF $f$, then $f$ is URMoving relative to $g$; hence, one can refer to a pair $(f,g)$ of mutually URMoving RFs. 

Obviously, one could equivalently define URMoving RFs as follows. 
Let us call a motion $M$ accelerated relative to RF $f$ if there are three distinct events $e_1$, $e_2$, and $e_3$, belonging to $M$ and such that 
$\vv^{M,f}(e_2,e_3)\ne\vv^{M,f}(e_1,e_2)$. 

Then an RF $g$ is URMoving relative to another RF $f$ if and only if every motion which is accelerated relative to $f$ is accelerated relative to $g$ as well.
 
This statement may be considered as a weak form of the principle of relativity.


To test directly by the above definition whether two given RFs are mutually URMoving, one would have to examine whether ``{\em every} URMotion $U$ relative to $g$ is a URMotion relative to $f$".
It is therefore of importance for testing purposes that the latter requirement can be relaxed to the following \cite{bor-heger}: ``every URMotion $U$ relative to $g$ {\em with a small enough relative speed $|\vv^{U,g}|$}
is a URMotion relative to $f$", i.e.: ``there exists a real number $\delta:=\delta^{g,f}>0$ such that every URMotion $U$ relative to $g$ with $|\vv^{U,g}|<\delta$ is a URMotion relative to $f$".
The latter condition may be even further relaxed by replacing the inequality $|\vv^{U,g}|<\delta$ by $|\vv^{U,g}-\vv_0|<\delta$, for some fixed $\vv_0\in\R^3$, 
thus only requiring that every URMotion $U$ relative to $g$ with a relative velocity $\vv^{U,g}$ close enough to {\em some} given vector $\vv_0$
be a URMotion relative to $f$.  

If an RF $g$ is URMoving relative to another $f$, then, by the {\em Fundamental Fact}, the RFCT $\A^{g,f}$ is affine and its action on the vectors $X$ in $\R^4$
is therefore given by 
\beq{affine}
\A^{g,f}\colon\ X\longmapsto\A^{g,f}(X)=A^{g,f}X+s^{g,f},
\eeq
where $A^{g,f}$ is a $4\times4$ real matrix, which will be called the matrix of the RFCT $\A^{g,f}$, and $s^{g,f}$ is a vector in $\R^4$, which will be called the shift of $\A^{g,f}$.

Obviously, equation \req{affine} can be rewritten as 
\beq{affine1}
f(e)=A^{g,f}g(e)+s^{g,f},\quad \mbox{for all\ } e\in\E.
\eeq
The shift $s^{g,f}$ does not cause any essential difficulties, and so, will be assumed for simplicity to be zero, unless otherwise indicated, so that 
all for any mutually URMoving pair of RFs $(f,g)$, RFCT $\A^{g,f}$ will be assumed to be not just affine but linear. 
Dropping also, for brevity, the argument $e$ in eq. \req{affine1}, one may rewrite it simply as 
$$f=A^{g,f}g.$$

Any terms originally defined either for a pair $(f,g)$ of mutually URMoving RFs or for the affine RFCT $\A^{g,f}$ or for the matrix $A^{g,f}$ will apply interchangeably to all of these three notions. 
E.g., in Section \ref{lorentz}, \pg{lorentz}, we shall rigorously define $C$-Lorentzian matrices; then, a pair $(f,g)$ of mutually URMoving RFs or the RFCT $\A^{g,f}$ will be $C$-Lorentzian if and only if the matrix $A^{g,f}$ is $C$-Lorentzian.  

If every pair of a family of RFs possesses a certain property, then we shall refer to the family as possessing this property as well; e.g., a natural family of RFs is a family of RFs in which every pair is natural.

Next, if an RF $g$ is URMoving relative to another RF $f$, define the 
{\em velocity of $g$ relative to} $f$ as $\vv^{g,f}:=\vv^{U,f}$, where $U$ is any URMotion relative to $g$ with $\vv^{U,g}=\0$; it is not hard to see that
this definition is correct in the sense that $\vv^{U,f}$ does not depend on the choice of $U$ given $\vv^{U,g}=\0$.

Physically, the latter definition corresponds to the following. 
One fixes any small particle which is at rest relative to RF $g$; such a particle represents a URMotion relative to $g$ with the zero relative velocity, as though the particle was ``attached to" RF $g$.
Since RF $g$ is URMoving relative to RF $f$, the particle represents a URMotion relative to $f$ as well. 
Then, the constant velocity of this particle relative to $f$ will be the relative velocity of $g$ relative to $f$; 
this velocity does not depend on the choice of a particle at rest relative to $g$.  

The above definition of the relative velocity may be restated as follows. 
Let $e_1$ and $e_2$ be any two events whose spatial coordinates in RF $g$ are the same, i.e., $\xx^g(e_2)=\xx^g(e_1)$; then the velocity of $g$ relative  to $f$ is $\vv^{g,f}=(\xx^f(e_2)-\xx^f(e_1))/(t^f(e_2)-t^f(e_1))$ provided that $t^f(e_2)\ne t^f(e_1)$. If for any two events $e_1$ and $e_2$, $\xx^g(e_2)=\xx^g(e_1)$ implies $t^f(e_2)=t^f(e_1)$, then the relative velocity $\vv^{g,f}$ is infinite. It can be seen that all the vectors of the form $\xx^f(e_2)-\xx^f(e_1)$ for all pairs of events $e_1$ and $e_2$ satisfying the equality $\xx^g(e_1)=\xx^g(e_2)$ are directed along one line in $\R^3$; that line is the line of the direction of the vector of the relative velocity $\vv^{g,f}$, be it finite or infinite. 

For any $4\times4$ matrix $A$, we shall routinely use the block representation   
\beq{block}
A=\pmatrix{A_{00}&A_{01}\cr A_{10}&A_{11}\cr},
\eeq
where $A_{11}$ is $3\times3$. 

If an RF $g$ is URMoving relative to another RF $f$ and $A=A^{g,f}$, then it is easy to see that the velocity of $g$ relative to $f$ is 
\beq{vv-matrix}
\vv^{g,f}=\frac{A_{10}}{A_{00}}
\eeq
provided that $A_{00}\ne0$; otherwise, $|\vv^{g,f}|$ is infinite and $\vv^{g,f}$ has the direction of the line in $\R^3$ through $\0$ carrying the vector $A_{10}$. 

Any RF $\ell_*$, defined on an event space $\E$, may be considered as a re-labeling of $\E$ (see Subsection \ref{RF-RFCT}, \pg{RF-RFCT}). 
E.g., one may choose $\ell_*$ to describe a physical RF, which is stationary relative to ``remote stars". Then the re-labeled version $f^{\ell_*}$ of any RF $f$ coincides with the RFCT $\A^{\ell_*,f}$ from the ``stationary" RF $\ell_*$ to RF $f$, $f^{\ell_*}=\A^{\ell_*,f}$.

Next, an RF $f$ may be called inertial if it is URMoving relative to the ``stationary" RF $\ell_*$. i.e., if the re-labeled version $f^{\ell_*}$ of RF $f$ is an affine mapping of $\E^{\ell_*}=\R^4$ onto itself. 

Then any RF $g_1$ which is not inertial is moving with (possibly non-uniform) acceleration relative to the ``stationary" RF $\ell_*$. 
Let us define another RF $f_1$ by the formula $f_1(e)=Ag_1(e)$ for all events $e$ in $\E$, where $A$ is any non-singular $4\times4$ real matrix.
Then RF $f_1$, as well as $g_1$, is not inertial; it is accelerated relative to the ``stationary" RF $\ell_*$.

Nonetheless, RFs $f_1$ and $g_1$ are URMoving relative to each other, and so, pair $(f_1,g_1)$ belongs in the subsequent {\em special} theories of relativity given in this paper, even though RFs $f_1$ and $g_1$ are not inertial. 

Replacing the special and hard to rigorously define notion of the inertial RFs by the more general notion of relatively URMoving RFs is in better conformance with the spirit of relativity.

\subsection{    Notion and types of adjustment of reference frames}
\label{adjustment}

Given a physical RF, constructed using e.g. rods and clocks in the well-known manner, one can adjust it by changing the directions of the three coordinate rods or the spatial units along them. One can also adjust the RF by  changing the rates of the clocks or the directions of their hands' movement.
Finally, one can shift the readings of the clocks, possibly depending on their spatial locations; this latter kind of adjustment may be referred to as  re-synchronization.
Using any of these kinds of adjustment of the given RF, one obtains another RF;  
it is physically evident that the latter RF is at rest relative to the former one. 
This motivates the following model notion of adjustment of RFs.

Let us say that an RF $\tilde f$ is an {\em adjustment of another RF} $f$ -- or, equivalently, that RF $\tilde f$ is {\em at rest relative to RF} $f$ -- if RF $\tilde f$ is URMoving with a zero velocity $\vv^{\tilde f,f}$ relative to RF $f$. 

In this case, let us also say that the RFCT $\A^{f,\tilde f}$ is an {\em adjustment (transformation)}.

In view of \req{vv-matrix}, page \pageref{vv-matrix}, the matrix $A^{f,\tilde f}$ of an adjustment is any real $4\times4$ matrix $A$ with $A_{10}=\0$ and $A_{00}\ne0$, that is any one
of the form
\beq{rest}
A=\pmatrix{\tau&\bb^T\cr \0&S\cr},
\eeq
for a nonzero real number $\tau$ and a non-singular real $3\times3$ matrix $S$.

Here and in what follows, superscript ${}^T$ will denote matrix transposition, as usual.

It is easy to see that the adjustment transformations, as well as their matrices of form \req{rest}, constitute a group. 
This fact will not however play a significant role in this paper. 

Let us say that a pair of RFs $(\tilde f,\tilde g)$ is an {\em adjustment of another pair of RFs} $(f,g)$ if $\tilde f$ is an adjustment of $f$ and $\tilde g$ is an adjustment of $g$.

More generally, suppose that $\F$ is any family of RFs. 
For every RF $g$ in $\F$, let us take any adjustment $\tilde g$ of $g$. Let us call the resulting family $\tilde\F:=(\tilde g\colon\ g\in\F)$ a {\em universal adjustment} of the family $\F$. 

Along with expressions like ``$\tilde f$ (or $\tilde\F$) is an adjustment of $f$ (or $\F$)", we shall interchangeably use their self-explanatory paraphrases, such as ``$f$ (or $\F$) {\em can be adjusted (or is adjustable) to} $\tilde f$ (or $\tilde\F$)".

Let $\diag(A_1,\dots,A_n)$, where $A_1,\dots,A_n$ are real square matrices, stand for the block-diagonal matrix with the diagonal blocks $A_1,\dots,A_n$.  

Let $I_n$ denote the $n\times n$ identity matrix.

Adjustment transformations include the following four elementary model types or any composition thereof; in the listing below, $\pmatrix{t\cr\xx\cr}$ stands for an arbitrary vector in $\R^4$:   


\begin{enumerate}
\item
\label{types}
{\em space-time origin adjustment}:
$\pmatrix{t\cr\xx\cr}\longmapsto\pmatrix{t+t_0\cr\xx+\xx_0\cr}$, for some fixed $t_0\in\R$ and $\xx_0\in\R^3$;
the matrix of this transformation is $I_4$;
since we agreed to assume that the shift $s^{g,f}$ in \req{affine1}, \pg{affine1}, is zero, this trivial type of adjustment will not in fact be subsequently needed; 
\item
{\em temporal adjustment}:
$\pmatrix{t\cr\xx\cr}\longmapsto\pmatrix{\tau t\cr\xx\cr}$, for some fixed nonzero $\tau\in\R$;
the matrix of this transformation is $\diag(\tau,I_3)$; in particular, this type includes
	\begin{enumerate}
\item
{\em temporal re-orientation}, when $\tau=\pm1$; obviously, $\tau=-1$ corresponds to the change of the sign of the temporal coordinates of all events; in the case $\tau=1$, the given RF is left unchanged;  
\item
{\em temporal rescaling}, when $\tau$ is positive; physically, this corresponds to any proportional change of the rates of all the clocks in the given RF;
	\end{enumerate}  
\item
{\em spatial adjustment}:
$\pmatrix{t\cr\xx\cr}\longmapsto\pmatrix{t\cr S\xx\cr}$, for some fixed non-singular $3\times3$ real matrix $S$;
physically, this corresponds to any change of the rods determining the spatial basis in the given RF;
the matrix of this adjustment transformation is $\diag(1,S)$;
in particular, this type includes
\begin{enumerate}
\item
{\em spatial re-orientation}, when matrix $S$ is orthogonal;
\item
{\em (possibly anisotropic) spatial rescaling}, 
when matrix $S$ is symmetric and positive-definite; in other words, a spatial rescaling is a linear transformation of the form 
\vskip3pt
$\pmatrix{t\cr x\ee_1+y\ee_2+z\ee_3\cr}\longmapsto\pmatrix{t\cr \xi_1 x\ee_1+\xi_2 y\ee_2+\xi_3 z\ee_3\cr}$, for some fixed orthonormal 
\vskip3pt
basis $(\ee_1,\ee_2,\ee_3)$ of $\R^3$ and some fixed
{\em positive} real numbers $\xi_1$, $\xi_2$, and $\xi_3$, which may be called the coefficients of rescaling of the three mutually orthogonal axes along the spatial basis vectors $\ee_1$, $\ee_2$, and $\ee_3$; 
here, $x$, $y$, and $z$ are arbitrary real numbers;
the matrix of this rescaling transformation in the orthonormal basis of vectors $\pmatrix{1\cr\0\cr}$, $\pmatrix{0\cr\ee_1\cr}$, $\pmatrix{0\cr\ee_2\cr}$, $\pmatrix{0\cr\ee_3\cr}$ in $\R^4$ is $\diag(1,\xi_1,\xi_2,\xi_3)$;  
spatial rescaling
further includes
		\begin{enumerate}
	\item
{\em isotropic spatial rescaling}, when the rescaling coefficients $\xi_1$, $\xi_2$, and $\xi_3$ are 
\vskip3pt
equal to one another:
$\pmatrix{t\cr\xx\cr}\longmapsto\pmatrix{t\cr\xi\xx\cr}$, for some fixed {\em positive} $\xi$; the matrix 
\vskip3pt
of this transformation is $\diag(1,\xi I_3)$.
		\end{enumerate}   
	\end{enumerate}   
\item
{\em re-synchronization}:
$\pmatrix{t\cr\xx\cr}\longmapsto\pmatrix{t+\bb^T\xx\cr\xx\cr}$, for some fixed $\bb\in\R^3$; 
in other words, a re-synchronization is any linear transformation of $\R^4$ preserving the spatial coordinates of all events as well as the time intervals between any two events occurring at any one and the same point of space; if $\bb\ne\0$, then the temporal coordinates of all the events with the spatial coordinates $\xx$ are shifted by $\bb^T\xx$, proportionally to the projection of $\xx$ onto the axis through $\bb$.
The matrix of this transformation is $\pmatrix{1&\bb^T\cr \0&I_3\cr}$. 
Physically, a re-synchronization corresponds to a shift of the readings of all the clocks in the given RF without changing their rates and without changing the spatial coordinates of events; of course, the readings of the clocks must be shifted in such a way that all URMotions relative to the RF before re-synchronization remain so thereafter, so that the transformation of the spacetime coordinates is affine; it is also required that this transformation preserve the spacetime origin, so that the transformation is in fact linear.  
\end{enumerate}

It is evident from the above discussion that each of the listed model types of adjustment is physically realizable. 
The following simple proposition shows that the above listing of the types of adjustment is essentially complete, and so, the above-defined notion of adjustment is neither too general nor too narrow.


\paragraphh{Proposition: Adjustment structure}
\label{adjust-structure}\ \\
Any adjustment transformation can be represented as a composition of the listed above four elementary types of adjustment, in any order.
\qedd


This follows easily because any matrix of the form \req{rest} can be represented as the product of the matrices of adjustments of the four types, in any order. E.g.,
$$\pmatrix{\tau&\bb^T\cr \0&S\cr}=
I_4\pmatrix{\tau&\0^T\cr \0&I_3\cr}
\pmatrix{1&\0^T\cr \0&S\cr}
\pmatrix{1&(\bb/\tau)^T\cr \0&I_3\cr}.$$

Of the listed types of adjustment, re-synchronization and {\em anisotropic} spatial rescaling seem to be the least desirable. 
In Subsections \ref{RECIP-only}, \ref{skew-free}, and \ref{RECIP-iso} we shall see when it is possible to do without re-synchronization and when it is possible  to use isotropic rescaling rather than the anisotropic version.

Let us define an {\em adjustment without re-synchronization} as any adjustment which can be represented as a composition of the three elementary types of adjustment listed above other than re-synchronization. The matrix of an adjustment without re-synchronization is one of the form $\diag(\tau,S)$, where $\tau$ is a nonzero real and $S$ is non-singular.    

Let us define a {\em re-orientation} as any composition of a temporal re-orientation and a spatial re-orientation, that is, any composition of adjustment transformations 
of subtypes 2(a) and 3(a), listed above.
The matrix of a re-orientation is one of the form $\diag(\vp,Q)$, where $\vp=\pm1$ and $Q$ is an orthogonal matrix.

Let us define a {\em rescaling} as any composition of a temporal rescaling and a spatial rescaling, that is, any composition of adjustment transformations 
of subtypes 2(b) and 3(b).
The matrix of a rescaling is one of the form $\diag(\tau,S)$, where $\tau$ is a positive real and $S$ is symmetric and positive-definite. In other words, the matrix of any rescaling in an appropriate orthonormal basis of vectors $\pmatrix{1\cr\0\cr}$, $\pmatrix{0\cr\ee_1\cr}$, $\pmatrix{0\cr\ee_2\cr}$, and $\pmatrix{0\cr\ee_3\cr}$ in $\R^4$
has the form $\diag(\tau,\xi_1,\xi_2,\xi_3)$, where $\tau,\xi_1,\xi_2,\xi_3$ are positive reals -- the rescaling coefficients.

Let us define an {\em isotropic rescaling} as a composition of a temporal rescaling and an isotropic spatial rescaling, that is, any composition of transformations 
of subtypes 2(b) and 3(b)i.
The matrix of an isotropic rescaling is one of the form 
$\diag(\tau,\xi I_3)$, where $\tau$ and $\xi$ are positive reals. 

Note that any adjustment of any RF $g$ does not change 
its velocity $\vv^{g,f}$ relative to any other RF $f$, which is URMoving relative to $g$. 
This can be seen as another confirmation of consistency of the above modeling of adjustment of RFs.



\paragraphh{Remark: 
$C$-Lorentzian adjustment without re-synchronization means the same as $C$-Lorentzian rescaling}
\label{polar}
\ \\ 
(For a rigorous definition of a generalized Lorentzian pair refer to Section \ref{lorentz} below.)
A pair of RFs can be rescaled to a generalized Lorentzian pair if and only if it can adjusted without re-synchronization to a generalized Lorentzian pair.

Indeed, any spatial re-orientation obviously preserves $C$-Lorentzian pairs of RFs. 
On the other hand, any spatial adjustment can be represented as the composition of an anisotropic spatial rescaling and a spatial re-orientation, in either order, according to the polar decomposition of matrices. Hence, the statement of this remark follows. 
\qedd

If there is a mapping $f\mapsto\tilde f$ of a family $\F$ of RFs onto another family $\tilde\F$ of RFs so that for every RF $f$ in $\F$, $\tilde f$ is an adjustment of $f$ of a certain type, then we refer to family $\tilde\F$ as to that same type of (universal) adjustment of family $\F$.  
E.g., if for every $f$ in $\F$, $\tilde f$ is an isotropic rescaling of $f$, then we say that $\tilde\F$ is a (universal) isotropic rescaling of $\F$ or, in other words, $\F$ is isotropically rescalable to $\tilde\F$.

\subsection{    Reciprocal, isotropic, and natural pairs of RFs}
\label{recip-iso-natur} 

In this subsection, some rigorous model expressions for the principle of relativity will be given. 


Imagine two physical RFs located in the spacetime so that they can be considered completely symmetric to each other with respect to some center of symmetry. 
E.g., such a situation can be the case if the following conditions are fulfilled. 

(I) All the masses of the universe and their velocities are symmetric with respect to some point, which is thus the center of symmetry of the universe; an approximation to this ideal situation would be absence of large masses in a sufficiently large neighborhood of a comparatively small spacetime domain where the two RFs are located.
(II) The two RFs in question can be obtained only by means of physical processes which are symmetric with respect to the center of symmetry. 

Then, obviously, the central symmetry will coincide with the RFCT from one of the two RFs to the other. 

Clearly, instead of the central symmetry one consider here any (not necessarily orthogonal) symmetry with respect to any straight line or any two- or three-dimensional plane in the spacetime. 
Here, the spacetime is considered locally, so that it can be assumed to be approximately flat.

Instead of any of the described above kinds of symmetry of the coordinate space $\R^4$, one can consider any re-labeling $\ell\colon\ \E\to\E$ of events, which is involutive in the sense that $\ell\circUp\ell=\mbox{id}_\E$; in other words, if an event $\tilde e$ is the re-labeled version of another event $e$ under re-labeling $\ell$, i.e., $\tilde e=\ell(e)$, then event $e$ is the re-labeled version of event $\tilde e$ under the same re-labeling mapping $\ell$, i.e., $e=\ell(\tilde e)$;
for the definition of re-labeling of events, refer to Subsection \ref{RF-RFCT}.

One thus comes to the following definition.

Let us call a pair $(f,g)$ of mutually URMoving RFs {\em reciprocal} if $\A^{g,f}=\A^{f,g}$.

According to Proposition \ref{re-label}, a pair $(f,g)$ of mutually URMoving RFs is reciprocal if and only if the pairs $(f,g)$ and $(g,f)$ are the same up to a (necessarily involutive) re-labeling of events: $f^\ell=g$ and $g\,^\ell=f$, where the re-labeling is $\ell=g^{-1}\circUp f=f^{-1}\circUp g$. 


Note that a pair $(f,g)$ of mutually URMoving RF is reciprocal if and only if the RFCT matrix $A=A^{g,f}$ is involutive, i.e., $A^2=I_4$ or, equivalently, $A^{-1}=A$. (Remember that the shift $s^{g,f}$ in \req{affine1}, \pg{affine1}, is assumed to be zero throughout the paper.)

Hence, considering the Jordan canonical form of matrix $A$, it is easy to see
that in some basis in $\R^4$, the matrix of RFCT $\A^{g,f}$ for a reciprocal
pair  $(f,g)$ must be of the form $\diag(\vp_0,\vp_1,\vp_2,\vp_3)$, where
$\vp_0,\vp_1,\vp_2,\vp_3=\pm1$.   Thus indeed, the involutive transformation
$\A^{g,f}$ is any (not necessarily orthogonal) symmetry in $\R^4$ with respect
to any linear subspace of $\R^4$. 

Another important property a physical spacetime may have is isotropy. 
Let us assume for a moment that this is the case. 
Yet, from a viewpoint of at least one of any two physical RFs, URMoving relative each other with a nonzero velocity $\vv$, the inherent isotropy of the spacetime will necessarily appear violated because of the definite direction of the relative velocity $\vv$. 
However, if both of two appropriately constructed physical RFs are rotated around the vector of the relative velocity $\vv$ through one and the same angle, then one may expect that the pair of the RFs will remain essentially the same as before the rotation in the sense that the RFCT will not change.

One thus comes to the following definition.

We shall say that two mutually URMoving RFs $f$ and $g$ are  {\em mutually isotropically oriented} or, for brevity, that the pair $(f,g)$ is {\em isotropic} if for any $3\times3$ rotation matrix $Q$ such that $Q\vv^{g,f}=\vv^{g,f}$, the RFCT 
$\A^{\tilde g,\tilde f}$ coincides with $\A^{g,f}$, where 
$\tilde f:=\diag(1,Q)f$ and $\tilde g:=\diag(1,Q)g$.  

In other words, a pair of RFs $(f,g)$ is isotropic if the RFCT from $g$ to $f$ does not change when both RFs undergo adjustment of the spatial axes via one and the same rotation of $\R^3$ preserving the vector of the relative velocity $\vv^{g,f}$. 

By Proposition \ref{re-label}, \pg{re-label}, this can be also expressed as follows: a pair of RFs $(f,g)$ is isotropic if 
the pair of RFs $(\tilde f,\tilde g)$ obtained from $(f,g)$ via one and the same rotation of their spatial axes so that to preserve the vector of the relative velocity $\vv^{g,f}$ is the same as the original pair $(f,g)$ up to re-labeling of events.

The notion of isotropy remains meaningful even when the relative speed $|\vv^{g,f}|$ is infinite; in such a case, once again, the rotations verifying the isotropy are around the well-defined line of the direction of $\vv^{g,f}$. 


\paragraphh{Proposition: One rotation suffices to verify isotropy}
\label{one-angle}
\ \\
Let $(f,g)$ be  a pair of mutually URMoving RFs with $\vv:=\vv^{g,f}\ne\0$.
Then the following conditions are equivalent to one another:
\begin{enumerate}
	\item 
pair $(f,g)$ is isotropic;
	\item
for {\em some} $3\times3$ matrix $Q$ of rotation about $\vv$ through not a multiple of $180^\circ$, the RFCT $\A^{\tilde g,\tilde f}$, where $\tilde f:=\diag(1,Q)f$ and $\tilde g:=\diag(1,Q)g$, coincides with $\A^{g,f}$;
	\item
in any orthonormal basis of $\R^4$ of the form $\pmatrix{1\cr\0\cr}$, $\pmatrix{0\cr\vv/v\cr}$, $\pmatrix{0\cr\ee_2\cr}$, $\pmatrix{0\cr\ee_3\cr}$, the matrix of the RFCT $\A^{g,f}$ is of the form $B=\diag(B_0,\la P)$, where $\la$ is a positive real number and $P$ is a $2\times2$ rotation matrix.   
\end{enumerate}
\qedd

The equivalence of Conditions 1 and 2 of Proposition \ref{one-angle} means that in the definition of the isotropic pair, instead of the invariance of the RFCT with respect to all rotations about $\vv$, it suffices to require the invariance of the RFCT with respect to {\em only one} rotation through not a multiple of $180^\circ$; in particular, the angle of the rotation can be chosen to be arbitrarily small. 

Proposition \ref{one-angle} will be proved in Appendix \ref{Proof of Theorem RECIP-AND-ISO and of Remark 1Q}, \pg{Proof of Theorem RECIP-AND-ISO and of Remark 1Q}.

Let us say that a pair of mutually URMoving RFs $(f,g)$ is {\em natural\/} if it can be adjusted via re-orientation and isotropic rescaling to a reciprocal and isotropic pair of RFs. 


We suggest that the model notion of the natural pair of RFs generalizes the idea of the pair of specially constructed inertial RFs.
By an inertial RF we understand a physical RF, ``freely falling without rotation" and located in a small enough region of the physical spacetime, where the divergence of the gravitational field is negligible. 

The above-mentioned special construction consists in the following.
Let an inertial RF have three mutually perpedicular rigid coordinate axes realized as rods joined together at one point (the spatial origin), with the same scale unit along all the three axes; we thus assume that the 3-dimensional Euclidian geometry is an appropriate model for description of properties of rigid bodies. 
To synchronize the clocks, a sufficient number of completely identical clocks are prepared at the spatial origin, say. Then each clock is {\em slowly} transported to its designated spatial position so that a sufficiently dense network of clocks is obtained. 

The above special construction is applied to every inertial RF in question separately from any other RF. 
Let us refer to such a construction as {\em standard autonomous}. 

We may conjecture that any two inertial RFs, located in the same small region of the space-time and obtained via a standard autonomous construction, ``can be adjusted via re-orientation and isotropic rescaling to a reciprocal and isotropic pair of RFs"; the terms in the latter quoted phrase are to be understood as physical objects and relations corresponding to their model counterparts. 

Thus, the hypothesis is that all the pairs of inertial RFs obtained via a standard autonomous construction are adequately modeled by the notion of natural pairs. Hence, by Part 2 of Proposition \ref{C-F}, \pg{C-F} (cf. Section \ref{level-2}), the local sign of the constant $C$ is uniquely determined. Thereby, the most important local characteristic of the spacetime -- 
the local type of the spacetime geometry, whether positive-Lorentzian or negative-Lorentzian -- is determined by means of any pair of inertial RFs not at rest relative to each other, obtained via a standard autonomous construction and located in a spacetime neighborhood of the given point of the spacetime.     

Obviously, all the construction processes within a standard autonomous construction can be performed with however small accelerations as well as speeds. 
This allows one to avoid in principle the difficulty with the procedure described in Introduction, where measuring devices had to be {\em transported from one RF into another}, moving with a nonzero speed $v$ relative to the first one.







\subsection{Proper pairs of RFs}
\label{proper} 

Let $(f,g)$ be a pair of mutually URMoving RFs and let $A:=A^{g,f}$.  
Let
us call the pair $(f,g)$ {\em improper} if $A_{11}$ is non-singular and $A_{01}A_{11}^{-1}A_{10}=0$ (recall \req{block}, \pg{block}); otherwise, let us call the pair $(f,g)$ {\em proper}.

Let
us call the pair $(f,g)$ {\em strictly proper} if $A_{00}\ne0$, $A_{11}$ is non-singular, and $A_{01}A_{11}^{-1}A_{10}\ne0$.

It is easy to see that any adjustment without re-synchronization (as defined in Subsection \ref{adjustment}, \pg{adjustment})
does not turn a proper pair of RFs into an improper one, or vice versa. 
A similar statement is true regarding strictly proper pairs.

Note that improper, or even not strictly proper, pairs of RFs are exceptions, which cannot be possibly detected experimentally; indeed, no elements of the matrix $A:=A^{g,f}$ can be precisely determined because of random errors inherent in any physical measurement. 

At times, we exclude improper and not strictly proper pairs of RFs to avoid too many technicalities arising in the exceptional, inessential cases. 
Nevertheless, a reader who is interested in exploring the nature of these exceptions a little further may want to continue reading this subsection for such details.  

Given any two mutually URMoving RFs $f$ and $g$, let us write the matrix $A:=A^{g,f}$ as 
\beq{A-partial}
A=\pmatrix{A_{00}&A_{01}\cr A_{10}&A_{11}\cr}
=\pmatrix{
\dfrac{\partial t}{\partial t'}& \dfrac{\partial t}{\partial\xx'}\cr
\noalign{\phantom{a}\vskip-16pt}
\dfrac{\partial\xx}{\partial t'}& \dfrac{\partial\xx}{\partial\xx'}\cr},
\eeq
where
$t':=t^g(e)$, $\xx':=\xx^g(e)$, $t:=t^f(e)$, $\xx:=\xx^f(e)$, for any event $e$.

Then 
$$\vv^{g,f}=\dfrac{A_{10}}{A_{00}}
=\dfrac{\ \dfrac{\partial\xx}{\partial t'}\ } 
{\ \dfrac{\partial t}{\partial t'}\ }$$
and
\beq{vv-f-g}
\vv^{f,g}=-A_{11}^{-1}A_{10}.
\eeq
The row-matrix 
\beq{grad}
\mbox{grad}_{\xx'}t:=\dfrac{\partial t}{\partial\xx'}=A_{01}
\eeq
may be called {\em the gradient of the $f$-time $t$ relative to the $g$-space $\xx'$}, or {\em the spatial gradient of asynchrony of $f$ relative to $g$}. 
Since $t=A_{00}t'+A_{01}\xx'$, one can say that the $f$-time coordinate $t$ of an event $e$ depends only on the $g$-time coordinate $t'$ of $e$ and on the orthogonal projection of the $g$-space coordinate vector $\xx'$ of $e$ onto the gradient $\mbox{grad}_{\xx'}t$. 

In these terms, pair $(f,g)$ being improper means that the gradient $\mbox{grad}_{\xx'}t$ is orthogonal to the velocity 
$$\vv^{f,g}=\dfrac{\ \dfrac{\partial\xx'}{\partial t}\ } 
{\ \dfrac{\partial t'}{\partial t}\ }$$ 
of RF $f$ relative to RF $g$. 
Hence, for a improper pair $(f,g)$, the $f$-time coordinate $t^f(e)$ of an event $e$ depends -- in addition to 
$t^g(e)$ -- only on the component of the $g$-space coordinate vector $\xx^g(e)$ in a direction {\em perpendicular to} the velocity 
$\vv^{f,g}$ 
of RF $f$ relative to RF $g$. 
Such a situation would probably seem counterintuitive.

Recall that $A_{00}\ne0$ if and only if $|\vv^{g,f}|\ne\infty$.
Similarly, by a common algorithm of matrix inversion, $A_{11}$ is non-singular if and only if $|\vv^{f,g}|\ne\infty$.
Thus, a pair of RFs $(f,g)$ is strictly proper if and only if it is proper and the relative speeds $|\vv^{f,g}|$ and $|\vv^{g,f}|$ are both finite.

\section{Statements of results and discussion: Three levels of assumptions and the three corresponding levels of adjustment}
\label{results}

\subsection{Preliminary: $C$-Lorentzian transformations and their structure}
\label{lorentz}

Let $C$ be any real number. Let us say that a $4\times4$ real matrix $A$ is $C$-{\em Lorentzian} if for all real $t$ and $t'$ and all vectors $\xx$ and $\xx'$ in $\R^3$ the relation
$A\pmatrix{t'\cr\xx'\cr}=\pmatrix{t\cr\xx\cr}$ implies $t^2-Cr^2={t'}^2-C{r'}^2$.
This definition is equivalent to the following equation:
\beq{lor-matrix} 
A^T\diag(1,-CI_3)A=\diag(1,-CI_3).
\eeq 

In other words, a pair of mutually URMoving RFs is $C$-Lorentzian if and only if 
\beq{interval}
(g(e_2)-g(e_1))^T\diag(1,-CI_3)(g(e_2)-g(e_1))=
(f(e_2)-f(e_1))^T\diag(1,-CI_3)(f(e_2)-f(e_1))
\eeq
for all events $e_1$ and $e_2$. 
Actually, condition \req{interval} of the preservation of the ``$C$-interval" is so strong by itself that 
the restriction ``mutually URMoving" can be removed here without altering the meaning of the definition if $C\ne0$;
in the case $C>0$ this follows from the paper by Alexandrov \cite {alexandrov}; in the case $C<0$, from the fact that every isometry of $\R^n$ is affine.

(Since we assume throughout that the shift $s^{g,f}$ in \req{affine1}, \pg{affine1}, is zero, \req{interval} can be written simply as $g(e)^T\diag(1,-CI_3)g(e)=f(e)^T\diag(1,-CI_3)f(e)$, for all events $e$.)

Let us say that $A$ is {\em generalized Lorentzian} if $A$ is $C$-Lorentzian for some $C\in\R$.

The following theorem on the multiplicative parametrization of $C$-Lorentzian matrices will be a useful tool in the proofs of some of the main results of this paper. 
It may be also of interest by itself.


\paragraphh{Proposition: Multiplicative boost-orientation representation of $C$-Lorentzian transformations}
\label{LOR}
\ \\
Let $C$ be any {\em non-zero} real number. Let $A$ be a non-singular $4\times4$ real matrix. Then $A$ is $C$-Lorentzian if and only if 
one of the following two mutually exclusive cases takes place:
either (i) there exist some $\vp\in\{-1,1\}$, $\vv\in\R^3$, and orthogonal $3\times3$ matrix $Q$ such that $Cv^2<1$ and 
\beq{L}
A=B^{C,\vv}\diag(\vp,Q)=\pmatrix{\vp\g_v&-C\g_v\vv^T Q\cr \vp\g_v\vv &-S^\vv Q\cr}\quad 
\eeq
or (ii) there exist some unit vector $\ee\in\R^3$ and orthogonal $3\times3$ matrix $Q$ such that
\beq{L-infty}
A=B^{C,\ee}_\infty\diag(1,Q)
=\pmatrix{0&\sqrt{-C}\ee^T Q\cr \ee/\sqrt{-C} & (-I_3+P^{\ee})Q\cr}.\quad 
\eeq
Here
\beq{boost} 
B^{C,\vv}:=\pmatrix{\g_v & -C\g_v\vv^T\cr \g_v\vv & -S^{\vv}\cr};
\eeq
\beq{gamma} 
\g_v:=\g_{v,C}:={1\over\sqrt{1-Cv^2} };\quad 
\eeq
\beq{S-v}
S^{\vv}:=S^{\vv,C}:=I_3+(\g_v-1)P^{\vv},
\eeq
\beq{P}
P^{\vv}:={1\over v^2}\vv\vv^T \quad\mbox{if}\quad v\ne0;
\eeq
\beq{S-0}
S^{\0}:=I_3;
\eeq
\beq{boost-infty} 
B^{C,\ee}_\infty:=
\lim_{v\to\infty}B^{C,v\ee}
=\pmatrix{0&\sqrt{-C}\ee^T\cr \ee/\sqrt{-C} & -I_3+P^{\ee}\cr}.
\eeq
Note that \req{L-infty} may occur (but of course not necessarily does) only if $C<0$.

The parameters $\vp$, $\vv$, and $Q$ of representation \req{L}, as well as the parameters $\ee$ and $Q$ of representation \req{L-infty}, are uniquely determined by the matrix $A$. 
\qedd

This proposition is proved in Appendix \ref{Proof of Theorem LOR}, \pg{Proof of Theorem LOR}. 



\paragraphh{Remark: Interpretation of the boost-orientation representation}
\ \\
Let RFs $f$ and $g$ be such that $A^{g,f}=A$.
Then, according to \req{vv-matrix}, page \pageref{vv-matrix}, 
the unique $\vv$ in the representation \req{L} coincides with
$\vv^{g,f}$, the velocity of $g$ relative to $f$.  
Matrix $B^{C,\vv}$ may be called a {\em $C$-boost} matrix or, more exactly, 
the matrix of the $C$-boost in the direction of $\vv$. Respectively, $B^{C,\ee}_\infty$ may be called an {\em infinite $C$-boost} matrix or, more exactly, 
the matrix of an infinite $C$-boost in the direction of $\ee$; in case \req{L-infty} takes place, the velocity of $g$ relative to $f$ is infinite. 
Next, $\vp$ and $Q$ represent the mutual orientation of RFs $f$ and $g$ in time and space, respectively; 
indeed, consider RF $\tilde g:=\diag(\vp,Q)g$, which is a re-orientation of RF $g$; then \req{L} implies 
$f=B^{C,\vv}\tilde g$, so that the matrix $A^{\tilde g,f}$ coincides with $B^{C,\vv}$. 
Next,
$P^{\vv}$ is the matrix of the orthogonal projection of $\R^3$ onto the direction of $\vv$, and so, $S^{\vv}$ has a transparent geometrical interpretation: for any vector $\uu$ in $\R^3$, $S^{\vv}\uu$ is the vector obtained from $\uu$ by stretching $\g_v$ times the component of $\uu$ parallel to $\vv$ while leaving the component of $\uu$ perpendicular to $\vv$ unchanged; note that the stretch coefficient $\g_v$ tends to 1 and hence $S^{\vv}$ tends to $S^\0=I_3$ as $\vv$ tends to $\0$.
\qedd




\paragraphh{Remark:\ $0$-Lorentzian transformations}
\ \\
The structure of the $0$-Lorentzian transformations as defined above is trivial: a non-singular $4\times4$ real matrix $A$ is $0$-Lorentzian if and only if $A_{00}=\pm1$ and $A_{01}=\0^T$ (remember \req{block}, page \pageref{block}). This is immediate from relations \req{L1}--\req{L3} (with $C=0$) in the proof of Proposition \ref{LOR}, \pg{Proof of Theorem LOR}.

We see that there are ``too many" $0$-Lorentzian transformations; the cause is that the matrix $\diag(1,-CI_3)$ in the definition \req{lor-matrix} is triply degenerate if $C=0$, and so, the above definition of the $0$-Lorentzian transformations is insufficiently restrictive in this case. 

We shall therefore redefine the notion of the $0$-Lorentzian transformations by means of an additional requirement of continuity in $C$.
Namely, further on let us refer to a matrix as $0$-Lorentzian if it is a limiting point as $C\to0$ of both the set of all $C$-Lorentzian matrices with $C>0$ and the set of all $C$-Lorentzian matrices with $C<0$.

It is obvious that no sequence of matrices of the form \req{L-infty} has a limit as $C\to0$. 
Hence, by Proposition \ref{LOR}, a matrix $A$ is $0$-Lorentzian if and only if it has the form \req{L} with $C=0$, that is,
\beq{0-L}
A=B^{0,\vv}\diag(\vp,Q)=\pmatrix{1&\0^T\cr \vv &-I_3\cr}\diag(\vp,Q)
=\pmatrix{\vp&\0^T\cr \vp\vv &-Q\cr}
\quad 
\eeq
\qedd


\paragraphh{Remark:\ 
A pair of mutually URMoving RFs with a nonzero relative velocity can be $C$-Lorentzian for at most one $C$}
\label{C-uniq}
\ \\
It is easy to see that given $A=A^{g,f}$ satisfying \req{L} or \req{L-infty} and such that $\vv^{g,f}\ne\0$, the value of $C$ in \req{L} ot \req{L-infty} is uniquely determined -- namely, $C=(A_{00}^2-1)/|A_{10}|^2$ (recall \req{block}, page \pageref{block}). 

On the other hand, if $(f,g)$ is a generalized Lorentzian pair with $\vv^{g,f}=\0$, then, in view of \req{L}, \req{L-infty}, and \req{0-L}, $(f,g)$ is $C$-Lorentzian for {\em any} real $C$. 
\qedd




\paragraphh{Remark:\ Scalar $C$-boosts}
\ \\
Special cases of $C$-boost matrices $B^{C,\vv}$ and $B_\infty^{C,\ee}$ defined by \req{boost} and \req{boost-infty} are the {\em scalar} $C$-boost matrices
\beq{scalar-boost}
B^{C,v}=
\pmatrix{\g_v&-C\g_v v&0&0\cr
\g_v v&-\g_v&0&0\cr
0&0&-1&0\cr
0&0&0&-1\cr}
\eeq
or  
\beq{scalar-boost-infty}
B_\infty^{C}=\lim_{v\to\infty}B^{C,v}=
\pmatrix
{0&\sqrt{-C}&0&0\cr
1/\sqrt{-C}&0&0&0\cr
0&0&-1&0\cr
0&0&0&-1\cr},
\eeq
corresponding to $\vv=(v,0,0)^T$ and $\ee=(1,0,0)^T$.
One has 
\beq{scalar}
B^{C,\vv}=\diag(1,Q_\vv)B^{C,v}\diag(1,Q_\vv^T)
\eeq
and
\beq{scalar-infty}
B_\infty^{C,\ee}=\diag(1,Q_\ee)B_\infty^{C}\diag(1,Q_\ee^T),
\eeq
where $Q_\vv$ is any orthogonal matrix whose first column is $\vv/v$ if $v\ne0$ (if $v=0$, then $Q_\vv$ is any orthogonal matrix at all) and $Q_\ee$ is any orthogonal matrix whose first column is $\ee$. 
Hence, by Proposition \ref{LOR}, a non-singular $4\times4$ real matrix $A$ is $C$-Lorentzian if and only if 
either there exist orthogonal $3\times3$ matrices $Q_1$ and $Q_2$ such that either
\beq{L-scalar}
A=\diag(1,Q_1)B^{C,v}\diag(\vp,Q_2)
\eeq
for some $\vp\in\{-1,1\}$ and $v\in\R$
or
\beq{L-infty-scalar}
A=\diag(1,Q_1)B^{C}_\infty\diag(1,Q_2).
\eeq
In case $C>0$ representation \req{L-scalar} is well known. 
However, in contrast to the uniqueness of all the parameters in representations
\req{L} and \req{L-infty}, matrices $Q_1$ and $Q_2$ in 
\req{L-scalar} and \req{L-infty-scalar} are obviously not unique.
\qedd


\subsection{    Level 0: without any assumptions, any two mutually URMoving RFs are adjustable to a $C$-Lorentzian pair}
\label{2 RFs}



\paragraphh{Theorem:\ Any pair of RFs is $C$-Lorentzian up to rescaling and re-synchronization}
\label{TH1-cor}
\ \\
For {\em any} real $C$, {\em any} pair of mutually URMoving RFs can be adjusted to a $C$-Lorentzian pair. 
By Remark \ref{polar}, \pg{polar}, this can be done by rescaling and re-synchronization only.
\qedd 




\paragraphh{Remark:\ Scalar $C$-boost adjustment}
\label{boost-adj}
\ \\
Furthermore, any pair pair of mutually URMoving RFs can be adjusted to a scalar $C$-boost pair, for any given real $C$.
\qedd 

Since for any real $C$, there obviously exist both a $C$-boost pair of RFs not at rest relative to each other and a $C$-boost pair of RFs at rest relative to each other, Theorem \ref{TH1-cor} and Remark \ref{boost-adj} are immediate from the following general result.



\paragraphh{Theorem:\ Adjustment can turn almost any RFCT into almost any other RFCT}
\label{TH1}
\ \\
Suppose that an RF $g$ is URMoving relative to an RF $f$
and an RF $g_1$ is URMoving relative to an RF $f_1$.
Then the following two conditions are equivalent to each other:
\begin{enumerate}
	\item
there exists an adjustment $(\tilde f,\tilde g)$ of the pair $(f,g)$ such that the RFCT $\A^{\tilde g,\tilde f}$ is the same as $\A^{g_1,f_1}$;
	\item
Either (i) $\vv^{g,f}\ne\0$ and $\vv^{g_1,f_1}\ne\0$ or (ii) $\vv^{g,f}=\0$ and $\vv^{g_1,f_1}=\0$.
\end{enumerate}
\qedd

Thus, Theorem \ref{TH1} says that the only invariant of the RFCT under adjustment is whether or not the corresponding pair of RFs are at rest relative to each other. 


This can also be expressed as follows: The only invariant of the RFCT under adjustment of the pair of RFs is whether or not the two RFs are adjustments of each other. 
This latter restatement of Theorem \ref{TH1} may at first glance seem trivial but it certainly is not so -- the emphasis here is on the ``the only". 
Since the condition that two RFs are at rest relative to each other, i.e. that the relative velocity is precisely zero, cannot possibly be detected experimentally, one can also somewhat loosely restate Theorem \ref{TH1} as above: Adjustment can turn almost any RFCT into almost any other RFCT. 

Note also that the first of the two equivalent conditions in Theorem \ref{TH1}  can be restated as follows: $(f,g)$ can be adjusted to a pair $(\tilde f,\tilde g)$ which is the same as $(f_1,g_1)$ up to re-labeling of events (recall Proposition \ref{re-label}, \pg{re-label}).

Proof of Theorem \ref{TH1} is given in Appendix \ref{proof of TH1}, \pg{proof of TH1}.




\paragraphh{Remark:\ ``Symmetric" form of Theorem \ref{TH1}}
\ \\
It is easy to see, either from the proof of Theorem \ref{TH1} or directly, that the first of the two equivalent conditions of Theorem \ref{TH1} can be restated in the following symmetric manner, formally better reflecting the exchangeability of the roles of the pairs $(f,g)$ and $(f_1,g_1)$:  pairs of RFs $(f,g)$ and $(f_1,g_1)$ can be adjusted to some other two pairs of RFs $(\tilde f,\tilde g)$ and $(\tilde f_1,\tilde g_1)$, respectively, so that $\A^{\tilde g,\tilde f}=\A^{g_1,f_1}$;
in other words, pairs $(f,g)$ and $(f_1,g_1)$ can be adjusted to some other two pairs of RFs, which are the same up to re-labeling of events.
\qedd

Theorem \ref{TH1-cor} and Proposition \ref{adjust-structure}, \pg{adjust-structure}, imply that any pair of RFs can be adjusted, for any prescribed real $C$, to a $C$-Lorentzian pair by means of the four types of adjustment described in Subsection \ref{adjustment}. In this sense, the phenomenon of the RFCT being positive-Lorentzian (or 0-Lorentzian or negative-Lorentzian or any other) is seen merely as a matter of an appropriate adjustment, which may appear rather surprising. 
In particular, what may seem surprising is that
any positive-Lorentzian pair of RFs can be made just by a choice of adjustment, at one's will, into either a 0-Lorentzian or a negative-Lorentzian pair, any 0-Lorentzian pair -- into either a positive-Lorentzian or a negative-Lorentzian one, and any negative-Lorentzian pair of RFs -- into either a positive-Lorentzian or a 0-Lorentzian one.

In connection with Theorem \ref{TH1-cor},
one could ask, When is it possible to adjust only one of two URMoving RFs so that the resulting pair of RFs is $C$-Lorentzian?
The next theorem provides a complete answer to this question.



\paragraphh{Theorem:\ 
Unilateral $C$-Lorentzian adjustment}
\label{ONE-ADJUST} 
\ \\
Let $f$ and $g$ be two RFs, URMoving relative to each other. Let $\vv:=\vv^{g,f}$ and let $C$ be a real number such that $Cv^2<1$ (assuming that $0\cdot\infty^2:=\infty$).
Then 
RF $g$ can be adjusted via rescaling and re-synchronization to an RF $\tilde g$ such that the pair $(f,\tilde g)$ is $C$-Lorentzian. 
\qedd

Theorem \ref{ONE-ADJUST} is immediate from its more detailed version, Theorem \ref{BOOST-ADJUST}, \pg{BOOST-ADJUST}, taking also into account Remark \ref{polar}, \pg{polar}.



\paragraphh{Remark:\ 
Necessity of $Cv^2<1$ for unilateral $C$-Lorentzian adjustment}
\ \\
The condition $Cv^2<1$ is not only sufficient in Theorem \ref{ONE-ADJUST} but necessary as well. Indeed, if $\tilde g$ is an adjustment of $g$, i.e. if $\tilde g$ is at rest relative to $g$, then it is easy to see that $\vv^{\tilde g,f}=\vv^{g,f}=\vv$. 
Hence, the condition $Cv^2<1$ is necessary for the pair $(f,\tilde g)$ to be $C$-Lorentzian,
in view of \req{gamma}, \pg{gamma}.
\qedd

\subsection{Level 0: Universal $C$-Lorentzian adjustment}
\label{universal}

Given a $C$-Lorentzian pair of RFs $(f,g)$ with $\vv^{g,f}\ne\0$, $C$ is uniquely determined, according to Remark \ref{C-uniq}, \pg{0-L}. So, $C$ serves to relate RFs $f$ and $g$ for all events $e$. 
In this sense, $C$ is constant.

Suppose now that one has to deal with more than two RFs, so that there are at least three RFs $f_1$, $f_2$, and $f_3$ under consideration. Let us fix any real number $C$. By Theorem \ref{TH1-cor}, each of the pairs $p_1:=(f_2,f_3)$, $p_2:=(f_1,f_3)$, and $p_3:=(f_1,f_2)$ can be adjusted to a $C$-Lorenzian pair, to obtain $C$-Lorentzian pairs $\tilde p_1:=(\tilde f^1_2,\tilde f^1_3)$, $\tilde p_2:=(\tilde f^2_1,\tilde f^2_3)$, and $\tilde p_3:=(\tilde f^3_1,\tilde f^3_2)$, respectively; the superscripts here refer to the corresponding pair. Thus, for each of the three RFs $f_1$, $f_2$, and $f_3$, one has two adjustments, e.g. two adjustments $\tilde f^2_1$ and $\tilde f^3_1$ of $f_1$, depending into which of the two pairs the RF is included. One may now ask whether this dependence of the $C$-Lorentzian adjustment on the pair of RFs can be avoided. 
A positive and more general answer to this question will be given below in this section.

Suppose that $\F$ is any family of mutually URMoving RFs. 

Let $\tilde\F$ be a universal adjustment of $\F$, as defined at the end of Subsection \ref{adjustment}, \pg{adjustment}.
Let us refer to $\tilde\F$
as a {\em $C$-Lorentzian universal adjustment} of $\F$ if $\tilde\F$ is a $C$-Lorentzian family of RFs, i.e., if any pair of RFs in $\tilde\F$ is $C$-Lorentzian; let us call a $C$-Lorentzian universal adjustment  {\em positive-Lorentzian} if $C>0$, {\em 0-Lorentzian} if $C=0$, and {\em negative-Lorentzian} if $C<0$. 

Now, the more general question that we want to consider is the existence of a \\
$C$-Lorentzian universal adjustment of a given family of RFs. The next theorem shows that a $C$-Lorentzian universal adjustment always exists if $C<0$; for $C\ge0$, certain general conditions must be satisfied in order for a $C$-Lorentzian universal adjustment to exist. In other words, there always exists a negative-Lorentzian universal adjustment, and this is not so for either positive-Lorentzian or negative-Lorentzian adjustments. Thus, the negative-Lorentzian adjustment is more ``universal", so to speak, than either the positive-Lorentzian or 0-Lorentzian ones.


\paragraphh{Theorem:\ 
Existence of a $C$-Lorentzian universal adjustment}
\label{univ}
\ \\
Let $C$ be any given real number.
There exists a $C$-Lorentzian universal adjustment of $\F$ if and only if one of the following three conditions is satisfied:
	\begin{enumerate}
\item 
$C<0$;
\item 
$C>0$ and 
there exist an RF $f$ in $\F$ and an adjustment $\tilde f$ of $f$ such that  
the speeds of all RFs in $\F$ relative to $\tilde f$ are less than $1/\sqrt C$; 
\item 
$C=0$ and 
there exist an RF $f$ in $\F$ and an adjustment $\tilde f$ of $f$ such that the speeds of all RFs in $\F$ relative to $\tilde f$ are finite. 
	\end{enumerate}
In this statement, each of the two entries of the phrase ``there exist an RF $f$ in $\F$ and an adjustment $\tilde f$ of $f$" can be replaced by ``for any RF $f$ in $\F$ there exists an adjustment $\tilde f$ of $f$".
\qedd

Theorem \ref{univ} follows from Theorem \ref{BOOST-ADJUST}, \pg{BOOST-ADJUST};
under Condition 2 or 3 of Theorem \ref{univ}, apply Theorem \ref{BOOST-ADJUST} with $\tilde f$ in place of $f$ and with every $g$ in $\F$ other than $f$;
under Condition 1, before applying Theorem \ref{BOOST-ADJUST} in the same manner, choose arbitrarily 
and fix an RF $f$ in $\F$ and any adjustment $\tilde f$ of $f$.





\paragraphh{Remark:\ 
Uniqueness of a $C$-boost universal adjustment}
\ \\
Moreover, it follows from Theorem \ref{BOOST-ADJUST} that 
the universal $C$-Lorentzian adjustment in Theorem \ref{univ} can always be chosen so that all the RFCTs within the resulting family $\tilde\F$ are finite or infinite $C$-boosts. 
Let us call such an adjustment a {\em universal $C$-boost adjustment}.
It also follows from Theorem \ref{BOOST-ADJUST} that a universal $C$-boost adjustment is in a certain sense unique. 
E.g., given $f$ and $\tilde f$ such as in Theorem \ref{univ}, every adjustment $\tilde g$ within a universal $C$-boost adjustment is uniquely determined for each $g\in\F$ with a finite $\vv^{g,\tilde f}$; for each $g\in\F$ with an infinite $\vv^{g,\tilde f}$, there will be exactly two appropriate adjustments $\tilde g$; the latter duplicity can be eliminated if it is additionally required that $\tau$ in the matrix $\pmatrix{\tau&\bb^T\cr \0&S\cr}$ of the adjustment RFCT $\A^{g,\tilde g}$ is positive, say.
\qedd

Let $V^{\F,f}:=\{\vv^{g,f}\colon g\in\F\}$ denote the set of all the vectors (or, more exactly, the set of the terminal points of the vectors) of the velocities of all RFs in $\F$ relative to some RF $f$ in $\F$.




\paragraphh{Remark:\ 
Two-sheet hyperboloid condition for positive-Lorentzian universal adjustment}
\label{HYPERB}
\ \\
Theorem \ref{univ} shows that for {\em any given} $C<0$, there always exists a $C$-Lorentzian universal adjustment of any family $\F$. 
Thus, there always exists a negative-Lorentzian universal adjustment.
For the existence of a positive- or 0-Lorentzian universal adjustment, additional conditions on the family $\F$ are needed. 
The following statements hold, in which there is no mentioning of an adjustment $\tilde f$ of $f$. 
\begin{enumerate}
\item 
There exists a positive-Lorentzian universal adjustment of $\F$ if and only if  for some [or, equivalently, for any] $f\in\F$, the set $V^{\F,f}$ of relative velocities is either bounded or is
contained in the inside, say $H$, of a two-sheet hyperboloid in $\R^3$; the hyperboloid may have any center of symmetry and any orientation in $\R^3$; the inside $H$ of the hyperboloid is assumed here to also contain all the infinitely remote points in the directions contained in the asymptotic cone $\lim_{\alpha\downarrow0} \alpha  H$ of $H$; hence, some of the relative velocities in $V^{\F,f}$ may be {\em infinite}.
\item 
There exists a $0$-Lorentzian universal adjustment of $\F$ if and only if, for some [or, equivalently, for any] $f\in\F$, either the set $V^{\F,f}$ contains only finite relative velocities or 
is contained in the complement $\R^3\setminus P$ of a two-dimensional affine plane $P$ in $\R^3$ which does not pass through $\0$; the complement $\R^3\setminus P$ is assumed here to also contain all the infinitely remote points in the directions not contained in the plane passing through $\0$ and parallel to $P$; hence, some of the relative velocities in $V^{\F,f}$ may be {\em infinite}. 
Note that the set $\R^3\setminus P$ can be considered as a set-limit of the insides of a certain sequence of two-sheet hyperboloids, whose two sheets are getting closer to each other and flatter.
\end{enumerate}
Details on this remark are given in Appendix \ref{Detais on Remark HYPERB}, \pg{Detais on Remark HYPERB}.
\qedd

\subsection{Level 1: Given only reciprocity, only spatial adjustment may be needed} 
\label{RECIP-only}


Given two pairs $(f_1,g_1)$ and $(f_2,g_2)$ of mutually URMoving RFs, let us call the two pairs {\em spatially similar} 
if there exists 
a non-singular $3\times3$ real matrix $S$ such that 
\beq{space-similar}
f_2=\diag(1,S)f_1\quad{\rm and}\quad g_2=\diag(1,S)g_1.
\eeq 
In other words, two pairs $(f_1,g_1)$ and $(f_2,g_2)$ of RFs are spatially 
similar if $f_2$ and $g_2$ may be obtained from $f_1$ and $g_1$, respectively, 
by means of one and the {\em same} spatial adjustment.

Obviously, if two pairs of RFs are spatially similar, then they are adjustable to each other without re-synchronization.


Observe that two pairs $(f_1,g_1)$ and $(f_2,g_2)$ of RFs are spatially similar if and only if  
\beq{similar}
A^{g_1,f_1}=\diag(1,S^{-1})\,A^{g_2,f_2}\,\diag(1,S),
\eeq
for some non-singular $3\times3$ real matrix $S$. 
 


\paragraphh{Theorem:\ 
Reciprocity implies spatial similarity to a generalized Lorentzian pair}
\label{RECIP}
\ \\
If a proper pair of RFs is reciprocal, then it is spatially similar to a generalized Lorentzian pair. 
\qedd




\paragraphh{Remark:\ 
Improper reciprocal pairs are asymptotically spatially similar to $0$-Lorentzian pairs}
\label{IMPROPER}
\ \\
Any improper reciprocal pair of RFs $(f,g)$ is {\em asymptotically spatially similar to a \\
$0$-Lorentzian pair} in the sense that 
there exists a 
sequence of pairs of RFs $(f_k,g_k)$, which are spatially similar to $(f,g)$
and such that $\lim_{k\to\infty} A^{f_k,g_k}$ exists and is $0$-Lorentzian, i.e., Galilean. 
The relation of being spatially similar is carried here, as in 
\req{space-similar}, by spatial transformations whose matrices $S_k$ or their 
inverses $S_k^{-1}$ are nearly singular.  
\qedd

Proof of Theorem \ref{RECIP} and Remark \ref{IMPROPER}  is given in Appendix \ref{Proof of Theorem RECIP}, \pg{Proof of Theorem RECIP}. 




\paragraphh{Remark:\ 
Reciprocity of $C$-boosts}
\label{recip-lor}
\ \\
It is straighforward to check that any $C$-boost or infinite $C$-boost pair of RFs is reciprocal (recall definitions \req{boost} and \req{boost-infty}, page \pageref{boost-infty}). 
\qedd

The following theorem provides an interesting connection between reciprocity and \\
rescaling to a generalized Lorentzian pair. It is immediate from Theorem \ref{RECIP}, Proposition \ref{adjust-structure} (\pg{adjust-structure}), Remark \ref{polar} (\pg{polar}), Proposition \ref{LOR} (\pg{LOR}),
and Remark \ref{recip-lor}. 
 


\paragraphh{Theorem:\ 
Reciprocity and generalized Lorentzian rescaling}
\label{RECIP-cor}
\ \\
A proper pair $(f,g)$ of RFs can be adjusted without re-synchronization to a generalized Lorentzian pair of RFs
if and only if it 
can be adjusted without re-synchronization to a reciprocal pair of RFs.
\qedd

Note that by Remark \ref{polar}, \pg{polar}, the phrase ``adjusted without re-synchronization to a generalized Lorentzian pair" in the statement of Theorem \ref{RECIP-cor} can be replaced by ``rescaled to a generalized Lorentzian pair".

Some further details on adjustment without re-synchronization can be found in Subsection \ref{w-out-resynchro}, \pg{w-out-resynchro}.

\subsection{Another Level 1:
Given isotropy, only isotropic rescaling and re-synchronization may be needed}
\label{skew-free}


Euclidian geometry is usually assumed -- tacitly or explicitly -- as the model for the spatial component of the spacetime in accounts of the special theory of relativity. In reality, this assumption corresponds to certain assumed properties of rigid bodies. 
In this subsection, we establish a necessary and sufficient condition characterizing such an assumption. 

We begin with the following. 


\paragraphh{Theorem:\ 
Given isotropy, only isotropic rescaling and re-synchronization may be needed}
\label{skew-free-th}
\ \\
Let $C$ be any real number. 
Then any strictly proper isotropic pair of RFs can be adjusted via 
{\em isotropic} rescaling and re-synchronization to a $C$-Lorentzian pair.
\qedd

This theorem should be compared with Theorem \ref{TH1-cor}, \pg{TH1-cor}; without the isotropy assumption, anisotropic rescaling may be needed. 


The isotropy condition in Theorem \ref{skew-free-th} can be relaxed to the following weak isotropy version of it.

Let $(f,g)$ be a pair of mutually URMoving RFs with $\vv:=\vv^{g,f}\ne\0$. For any vector $\xx$ in $\R^3$, let $\xx^\perp:=(I_3-P^\vv)\xx$ denote the vector
component of $\xx$ perpendicular to $\vv$. 
Let us say that RFs $f$ and $g$ are {\em mutually weakly-isotropically oriented} or, for brevity, pair $(f,g)$ is {\em weakly-isotropic} if for any two events $e_1$ and $e_2$ which are simultaneous in RF $g$, the length of the component perpendicular to $\vv$ of the space interval between $e_1$ and $e_2$ in RF $f$ is proportional to that in $g$; 
in other words, $t^g(e_2)=t^g(e_1)$ implies $|\xx^f(e_2)^\perp-\xx^f(e_1)^\perp|=
\xi|\xx^g(e_2)^\perp-\xx^g(e_1)^\perp|$ for some real constant $\xi$.
Note that since matrix $A^{g,f}$ is non-singular, $\xi$ here must be nonzero, and so, $\xi>0$. 

It follows form Proposition \ref{one-angle}, \pg{one-angle}, that every isotropic pair of RFs is weakly-isotropic. 

The essential difference between the notions of isotropic and weakly-isotropic pairs of RFs is that for the latter, the space intervals are considered only for pairs of events simultaneous in RF $g$.

Theorem \ref{skew-free-th} is immediate from the following more detailed result. 


\paragraphh{Theorem:\ 
Characterization of anisotropy-free adjustment}
\label{skew-free-charact}
\ \\
Let $C$ be any real number. 
Let $(f,g)$ be a strictly proper pair of RFs. 
Then $(f,g)$ can be adjusted via isotropic rescaling and re-synchronization to a proper $C$-Lorentzian pair of RFs $(\tilde f,\tilde g)$ if and only if it can be adjusted via spatial re-orientation to a weakly-isotropic pair of RFs $(\hat f,\hat g)$.
\qedd



\paragraphh{Remark:\ 
Uniqueness}
\label{skew-free-uniqueness}
\vskip-15pt
\begin{enumerate}
\item
The proper $C$-Lorentzian adjustment $(\tilde f,\tilde g)$ of $(f,g)$ in Theorem \ref{skew-free-charact} can be chosen so that (i) $(\tilde f,\tilde g)$ is $C$-boost, i.e., $A^{\tilde g,\tilde f}=B^{C,\uu}$ for some $\uu$, 
(ii) $\tilde g$ is obtained from $g$ by isotropic rescaling and spatial re-orientation only, and (iii) $\tilde f$ is obtained from $f$ by re-synchronization and temporal rescaling only; if $C\ge0$, then $\tilde f$ may be taken to be just a re-synchronization of $f$ --
no temporal adjustment is then needed.
\item
Such a choice of $(\tilde f,\tilde g)$ is unique given 
$(f,g)$ and the (constant) value of 
$\dfrac{\partial t^{\tilde f}}{\partial t^f}$,
where $t^f:=t^f(e)$ and $t^{\tilde f}:=t^{\tilde f}(e)$, $e\in\E$.
\item
The weakly-isotropic pair of RFs $(\hat f,\hat g)$ can be chosen so that 
$\hat f=f$, and $\hat g$ is obtained from $g$ by spatial re-orientation only.
\end{enumerate}
\qedd

Proof of Theorem \ref{skew-free-charact} and Remark \ref{skew-free-uniqueness} is given in Appendix \ref{Details on skew-free}, \pg{Details on skew-free}.



\paragraphh{Remark:\ 
Weak isotropy vs. isotropy}
\label{weak iso vs iso}
\ \\
Let $(f,g)$ be a strictly proper reciprocal and weakly isotropic pair of RFs. 
Since $(f,g)$  is reciprocal, by Theorem \ref{RECIP}, \pg{RECIP}, $(f,g)$ can be rescaled, and hence adjusted without re-synchronization, to a generalized Lorentzian pair $(\hat f,\hat g)$.
On the other hand, since $(f,g)$ is weakly isotropic, by Theorem \ref{skew-free-charact} $(f,g)$ can be adjusted via re-synchronization and isotropic spatial rescaling to a generalized Lorentzian pair $(\check{f},\check{g})$, perhaps different from $(\hat f,\hat g)$.

The question is, Can one always choose $(\hat f,\hat g)$ and $(\check{f},\check{g})$ to be the same,
so that $(f,g)$ can be isotropically adjusted to a generalized Lorentzian pair? 
The answer is no; see a counterexample in Appendix \ref{weak-iso-and-iso}, \pg{weak-iso-and-iso}. 
\qedd

\subsection{Level 2: Reciprocity and isotropy already imply the generalized Lorentzian property}
\label{RECIP-iso}



\paragraphh{Theorem:\ 
Reciprocal and isotropic pairs are generalized Lorentzian}
\label{RECIP-AND-ISO}
\ \\
If a pair of RFs is reciprocal and isotropic, then it is generalized Lorentzian.
\qedd

This Theorem is proved in Appendix 
\ref{Proof of Theorem RECIP-AND-ISO and of Remark 1Q},
\pg{Proof of Theorem RECIP-AND-ISO and of Remark 1Q}.





\paragraphh{Theorem:\ 
Generalized Lorentzian characterization of natural pairs}
\label{RECIP-AND-ISO-cor}
\ \\
A pair of RFs is natural if and only if it can be isotropically rescaled to a generalized Lorentzian pair.
\qedd

This follows from Theorem \ref{RECIP-AND-ISO}, Proposition \ref{LOR} (\pg{LOR}), Remark \ref{recip-lor} (\pg{RECIP}), and the fact that any $C$-boost pair of RFs is isotropic.

\subsection{Level 2: Universal generalized Lorentzian isotropic rescaling}
\label{scale-orientation adjustment}

Let $\F$ be a family of mutually URMoving RFs. 
If $\F$ is natural, i.e. every pair of RFs in $\F$ is natural, then by Theorem \ref{RECIP-AND-ISO-cor}, every pair of RFs in $\F$ can be isotropically rescaled to a $C$-Lorentzian pair of RFs. Hence, the following question arises:
Is there always a \break
$C$-Lorentzian isotropic rescaling of the entire family $\F$? 
The following theorem answers yes to this question.




\paragraphh{Theorem:\ 
Existence of a universal $C$-Lorentzian isotropic rescaling}
\label{scale-ortho}
\ \\
Family $\F$ is natural if and only if $\F$ can be isotropically rescaled to a $C$-Lorentzian family for some real $C=:C_\F$.
\qedd

Proof of this result is given in Appendix \ref{proof of [universal rescaling]}, \pg{proof of [universal rescaling]}.

In view of Theorem \ref{RECIP-AND-ISO-cor}, Theorem \ref{scale-ortho} can be restated as follows. 


\paragraphh{Theorem:\ 
Existence of a universal $C$}
\label{universal rescaling}
\ \\
Suppose that every pair of RFs in $\F$ can be isotropically rescaled to a generalized Lorentzian pair.
Then $\F$ can be isotropically rescaled to a $C$-Lorentzian family for some real $C=:C_\F$.
\qedd

This theorem is immediate from Theorem \ref{scale-ortho} and Theorem \ref{RECIP-AND-ISO-cor}.



\paragraphh{Proposition:\ 
Choice of a universal $C$}
\label{C-F}
\ \\
Let us refer to the constant $C=C_\F$ mentioned in Theorems \ref{scale-ortho} and \ref{universal rescaling} as a {\em universal constant} of family $\F$, because in a $C$-Lorentzian family $\tilde\F$, every pair of RFs is $C$-Lorentzian for one and the same $C$, rather than $C$ depending on the choice of a pair in $\tilde\F$.
\begin{enumerate}
\item
Depending on the choice of the universal isotropic rescaling, the universal constant $C_{\F}$ can be chosen arbitrarily except for its sign, which may be 1, $-1$, or 0 (assuming that sign$\,(0)=0$). 
E.g., the universal constant $C_\F$ may be assumed without loss of generality to be 1, $-1$, or 0.
\item
The sign of the universal constant $C_{\F}$ is uniquely determined by $\F$ unless all RFs in $\F$ are at rest relative to one another; in the latter, exceptional case, the value of $C_{\F}$ is a completely arbitrary real number.
\item
For any fixed $f$ in $\F$, its isotropic rescaling $\tilde f$ 
as the part of a universal $C$-Lorentzian isotropic rescaling $\tilde\F$ of $\F$ can be chosen completely arbitrarily;
of course, the choice of the isotropic rescaling of RFs in $\F$ other than $f$ depends on the choice of $\tilde f$.
Moreover, given any $f$ in $\F$ and any isotropic rescaling $\tilde f$ of $f$, the entire isotropic rescaling $\tilde\F$ of $\F$ is uniquely determined.
\item
Given any fixed RF $f$ in $\F$ which is not at rest relative to at least one other RF $g$ in $\F$ and given
any fixed isotropic rescaling $\tilde f$ of $f$, the value of $C_{\F}$ is uniquely determined.
\end{enumerate}
\qedd  

Proof of this proposition is given in Appendix \ref{proof of C-F}, 
\pg{proof of C-F}.



\paragraphh{Remark:\ 
Isotropy is essential}
\ \\
Theorem \ref{universal rescaling} would no longer hold if the two entries of 
``isotropically rescaled" in its statement were replaced by  ``rescaled". --
See Remark \ref{UNIVERSAL-SYNCHRO-FREE} below.
\qedd



\paragraphh{Remark:\ 
Three spatial dimensions are essential}
\label{dimension}
\ \\
The analogue of Theorem \ref{universal rescaling} with less than three spatial dimensions would not hold, even if its conclusion 
``$\F$ can be isotropically rescaled to a $C$-Lorentzian family" for a universal $C$ were relaxed to merely ``$\F$ can be isotropically rescaled to a generalized Lorentzian family". -- See Appendix \ref{details on [dimension]},
\pg{details on [dimension]}.
\qedd 


\subsection{    Unilateral $C$-boost-adjustment and parametrization of affine transformations}
\label{parametrization}


\paragraphh{Theorem:\ 
Unilateral $C$-boost adjustment}
\label{BOOST-ADJUST}
\ \\
Let $f$ and $g$ be two RFs, URMoving relative to each other. Let $\vv:=\vv^{g,f}$ and let $C$ be a real number.
\begin{enumerate}
\item 
The following conditions are equivalent:
	\begin{enumerate}
	\item 
there exists an adjustment $\tilde g$ of $g$ such that the pair $(f,\tilde g)$ is $C$-boost;
	\item
$v<\infty$ and $Cv^2<1$. 
	\end{enumerate}
If either of these equivalent conditions takes place, then the appropriate adjustment $\tilde g$ of $g$ is uniquely determined, and $A^{\tilde g,f}=B^{C,\vv}$.   
\item 
Also, the following conditions are equivalent:
	\begin{enumerate}
	\item 
there exists an adjustment $\tilde g$ of $g$ such that the pair $(f,\tilde g)$ is infinite-$C$-boost;
	\item
$v=\infty$ and $C<0$. 
	\end{enumerate}
If either of the latter two equivalent conditions takes place, then there are exactly two appropriate adjustments $\tilde g$ of $g$, with $A^{\tilde g,f}$ equal to either $B_\infty^{C,\ee}$ or $B_\infty^{C,-\ee}$, where the pair of unit vectors $\{\ee,-\ee\}$ determines the direction of the infinite relative velocity $\vv$;
the appropriate adjustment $\tilde g$ of $g$ is determined completely uniquely if, in addition, the sign of $\dfrac{\partial t^g}{\partial t^{\tilde g} }$ is prescribed. 
(Loosely speaking, the sign of $\dfrac{\partial t^g}{\partial t^{\tilde g} }$ determines the relative orientation of the time axes in RFs $g$ and $\tilde g$.)   
\end{enumerate}
\qedd




\paragraphh{Remark:\ 
$C$-boost-adjustment parametrization of affine transformations}
\ \\
Obviously, any non-singular $4\times4$ real matrix $A$ is a matrix of some RFCT. 
Therefore, Theorem \ref{BOOST-ADJUST} means any such matrix $A$ possesses a unique multiplicative representation of the form 
\req{Cnot0Anot0}, \pg{Cnot0Anot0}, or, in the exceptional case $A_{00}=0$, of the form \req{Cnot0Ais0} with $\tau>0$.
One thus concludes that the $C$-boost transformations together with the adjustment transformations provide for a unique factorization representation of arbitrary affine transformations of $\R^4$. 
Now multiplicative representations \req{L}, \req{L-infty}, and \req{0-L}, \pg{0-L}, of the generalized Lorentzian transformations can be seen as special cases of \req{Cnot0Anot0} and \req{Cnot0Ais0}, with $\tau=\pm1$, $\bb=\0$, and $S=Q$ -- an orthogonal matrix.
\qedd

\subsection{    More on generalized Lorentzian adjustment without re-synchronization, or rescaling}
\label{w-out-resynchro}

Of the four types of adjustment, listed in Subsection \ref{adjustment}, \pg{types}, it is rather certainly re-synchronization that seems to be the least desirable, as the one most substantially affecting the relation of temporal measurements with spatial ones.  
One could therefore ask: When a pair of mutually URMoving RFs is adjustable without re-synchronization to a generalized Lorentzian pair? 
A characterization of such pairs in terms of adjustment without re-synchronization to reciprocal pairs of RFs was given by Theorem \ref{RECIP-cor}, \pg{RECIP-cor}; 
once again, by Remark \ref{polar}, \pg{polar}, generalized Lorentzian adjustment without re-synchronization means the same as generalized Lorentzian rescaling.
 
In this subsection, it is shown that pairs of RFs that can be rescaled to generalized Lorentzian pairs constitute, in a certain sense, a majority of pairs of mutually URMoving RFs.

Moreover, it is possible to give a necessary and sufficient condition for the existence of a generalized Lorentzian rescaling of a pair $(f,g)$ of mutually URMoving RFs in terms of the RFCT matrix $A^{g,f}$. That condition is rather cumbersome if given with the utmost generality, accounting for a number of exceptions of purely mathematical character, which cannot even be experimentally detected.
However, if the consideration is restricted to the strictly proper pairs, defined in Subsection \ref{proper}, \pg{proper}, then the necessary and sufficient condition can be expressed quite simply.





\paragraphh{Theorem:\ 
A majority of pairs of RFs admit a generalized Lorentzian rescaling}
\label{SYNCHRO-FREE}
\ \\
Let $C$ be any nonzero real number. 
Let $(f,g)$ be a strictly proper pair of mutually URMoving RFs
and let $A:=A^{g,f}$.
Then $(f,g)$ can be rescaled (or, equivalently, adjusted without re-synchronization) to a $C$-Lorentzian pair of RFs
if and only if $\mu<1$ and $C\mu>0$, where 
\beq{mu}
\mu:=\mu^{g,f}:=\frac{A_{01}A_{11}^{-1}A_{10}}{A_{00}}.
\eeq
\qedd

Thus indeed, a generalized Lorentzian rescaling exists for a ``majority" of pairs of \\
URMoving RFs: if $\mu\not<1$, then one can fix this violation e.g. by merely replacing any one of the four blocks, $A_{00}$, $A_{01}$, $A_{10}$, or $A_{11}$ by its opposite $(-A_{00})$, $(-A_{01})$, $(-A_{10})$, or $(-A_{11})$ so that to switch the sign of $\mu$ and thus get $\mu\le-1<1$; 
then, however, one would need to switch the sign of $C$ as well, to satisfy the condition $C\mu>0$.  

One now sees that
$\mu$ is an important characteristic of a pair of RFs.
It is dimensionless, invariant with respect to any adjustment without re-synchronization and with respect to the interchange of the roles of $f$ and $g$: $\mu^{f,g}=\mu^{g,f}=\mu^{\tilde g,\tilde f}$, where $\tilde g$ and $\tilde f$ are any adjustments of $g$ and $f$ without re-synchronization, and has the following expressions: 
$$\mu=-\dfrac{\dfrac{\partial t}{\partial\xx'} \vv^{f,g} }
		{\dfrac{\partial t}{\partial t'} }
=-\dfrac{\dfrac{\partial t'}{\partial\xx} \vv^{g,f} }
		{\dfrac{\partial t'}{\partial t} }
=-\dfrac{\dfrac{\partial t}{\partial\xx'} \dfrac{\partial\xx'}{\partial t} }
	{\dfrac{\partial t'}{\partial t} \dfrac{\partial t}{\partial t'} }
=-\dfrac{\dfrac{\partial t'}{\partial\xx} \dfrac{\partial\xx}{\partial t'} }
	{\dfrac{\partial t'}{\partial t} \dfrac{\partial t}{\partial t'} }
=-\dfrac1 2
\dfrac{\dfrac{\partial t}{\partial\xx'} \dfrac{\partial\xx'}{\partial t} 
	+\dfrac{\partial t'}{\partial\xx} \dfrac{\partial\xx}{\partial t'} }
	{\dfrac{\partial t'}{\partial t} \dfrac{\partial t}{\partial t'} }$$
in terms of Subsection \ref{proper}, \pg{proper}.

Note that the pair $(f,g)$ is strictly proper if and only if the relative speeds $|\vv^{g,f}|$ and $|\vv^{f,g}|$ are both finite and 
$\mu\ne0$. 

Note also that if the pair $(f,g)$ can be rescaled (or, equivalently, adjusted without re-synchronization) to a $C$-Lorentzian pair of RFs $(\tilde f,\tilde g)$, then
$\mu=Cv^2=1-\g_v^{-2}$ -- cf. \req{mu1}, page \pageref{mu1} -- and so, $C\mu=C^2 v^2$, where $\vv:=\vv^{\tilde g,\tilde f}$.

Proof of Theorem \ref{SYNCHRO-FREE} is given in 
Appendix 
\ref{Details on Theorem SYNCHRO-FREE}, \pg{Details on Theorem SYNCHRO-FREE}.




\paragraphh{Example: 
Non-transitivity of generalized Lorentzian rescaling}
\label{LOR-nontransitivity}
\ \\
There are three RFs $f$, $g$, and $h$ such that each of the pairs $(f,g)$ and $(g,h)$ is a generalized Lorentzian pair, while the pair $(f,h)$ cannot be rescaled, and hence
cannot be adjusted without re-synchronization, to a generalized Lorentzian pair. 
Indeed, let, e.g., 
$$g:=\diag\left(2^{-3/2}\pmatrix{4&-8\cr1&-4\cr},I_2\right)h\quad
{\rm and}\quad 
f:=\diag\left(2^{-1/2}\pmatrix{2&-1\cr2&-2\cr},I_2\right)g,$$
for an arbitrary RF $h$, so that
$f=\diag\left(\dfrac1 4\pmatrix{7&-12\cr6&-8\cr},I_2\right)h$.
Then the pairs $(f,g)$ and $(g,h)$ are (1/2)-Lorentzian and 8-Lorentzian, respectively, while according to Theorem \ref{SYNCHRO-FREE}, the pair $(f,h)$ cannot be rescaled to a generalized Lorentzian pair.
\qedd 




\paragraphh{Remark:\ A universal generalized Lorentzian rescaling of a pairwise reciprocal and proper family of RFs need not exist}
\label{UNIVERSAL-SYNCHRO-FREE}
\ \\
Now consider the problem of the existence of a generalized Lorentzian universal rescaling of a family $\F$ of mutually URMoving RFs. As Example \ref{LOR-nontransitivity} shows, as a minimum, one should impose here the condition that {\em each} pair of RFs in $\F$ can be rescaled to a generalized Lorentzian pair. 
However, we shall see that this condition will not suffice, even if every pair of RFs in $\F$ is known to be proper and reciprocal (and thus, by Theorem \ref{RECIP}, \pg{RECIP}, can be rescaled to a generalized Lorentzian pair) and even if $\F$ is known to consist of only three RFs;
elaboration on this statement is given in Appendix \ref{EX-UNIVERSAL-SYNCHRO-FREE}, \pg{EX-UNIVERSAL-SYNCHRO-FREE}. 

This shows that in Theorem \ref{universal rescaling}, \pg{universal rescaling}, the isotropy stipulation cannot be dropped and that, moreover, it could not be dropped even if the conclusion of Theorem \ref{universal rescaling} were weakened in the following two aspects at once: (i) isotropic rescalability were replaced by mere rescalability or, equivalently, by adjustability without re-synchronization and (ii) a $C$-Lorentzian family with a universal constant $C=C_\F$ were replaced by a generalized Lorentzian family, with $C$ depending on the choice of a pair of RFs in $\F$.
\qedd

\section{Testing reciprocity and/or isotropy and executing an appropiate generalized Lorentzian adjustment}
\label{experiment}
In this section, we summarize developed in the previous sections {\em special} theories of relativity in order to consider relevant problems of testing of reciprocity and isotropy assumptions and the corresponding problems of execution of adjustment. 

Let $(f,g)$ be a strictly proper pair of mutually URMoving RFs, so that the relative velocity $\vv:=\vv^{g,f}$ is finite and nonzero.
Physically, as explained in Introduction and Section \ref{definitions}, 
the notion of such a pair may have many different kinds of physical realization.
However, of foremost interest to us here is the 
standard autonomous construction for inertial RFs, described in Subsection \ref{recip-iso-natur}, which will be assumed in this section. 

Our main objective in this section is to propose a method to test the
hypothesis that all the pairs of inertial RFs obtained via a standard autonomous construction are adequately modeled by the notion of natural pairs.
Recall that a pair of mutually URMoving RFs $(f,g)$ is defined as natural if it can be adjusted via re-orientation and isotropic rescaling to a reciprocal and isotropic pair of RFs. 
Thus, to test whether a pair of RFs is natural means to test properties of reciprocity and isotropy. 
We approach this task at the three main levels described in Introduction, page \pageref{levels}.

But before we proceed towards that end, we shall indicate how to {\em test} whether the two so constructed physical RFs can be adequately described as a a pair of mutually URMoving RFs $(f,g)$. 
That will be the case whenever the RFCT $\A^{g,f}$ is affine.
Any affine RFCT can be completely determined by the measurement of the time-space coordinates $X_i:=\pmatrix{t_i\cr\xx_i}:=f(e_i)$ and $X'_i:=\pmatrix{t'_i\cr\xx'_i}:=g(e_i)$, $i=0,\dots,4$, of any 5 events $e_0,\dots, e_4$ in RFs $f$ and $g$, 
assuming that 
the $X'_i\,$'s are affine-independent. 
Taking more events: $e_5,e_6,\dots$, the ``observers" can test whether the RFCT is indeed affine, that is, whether the two physical RFs under consideration may be described, with an appropriate degree of accuracy, as mutually URMoving. 
(To exchange the information on the identification of the events and on their time-space measurements, the ``observers" in $f$ and $g$ must each possess a signal with the relative speed greater than the relative speed of the other RF.)

\subsection{Level 0: Executing an appropiate generalized Lorentzian adjustment with no assumptions on a pair of mutually URMoving RFs}
\label{level-0}

By Theorem \ref{TH1-cor}, \pg{TH1-cor}, for any real $C$, any pair of mutually URMoving RFs can be adjusted to a $C$-Lorentzian pair.
Therefore, if all the types of adjustment listed in Subsection \ref{adjustment} are permitted, then {\em the only testing needed here} is the described above testing whether the two RFs in question are mutually URMoving.  

Hence, at Level 0, it only remains to show how to execute an appropiate generalized Lorentzian adjustment.

If $Cv^2<1$ (recall, $\vv:=\vv^{g,f}$ was supposed to be finite and nonzero) then, 
by Theorem \ref{BOOST-ADJUST}, \pg{BOOST-ADJUST}, there exists a unique adjustment 
\beq{tilde-g}
\tilde g:=\pmatrix{\tau&\bb^T\cr \0&S\cr}g
\eeq 
of RF $g$ 
such that the pair $(f,\tilde g)$ is $C$-boost. 
Even when the condition $Cv^2<1$ is not satisfied for the given pair $(f,g)$ and a given $C$, it is satisfied if $f$ is replaced by an appropriate (say temporal) adjustment $\tilde f$ of $f$ $\left(\mbox{note that if }\tilde f=\diag(\tau,I_3)f \mbox{, then }\vv^{g,\tilde f}=\dfrac{\vv^{g,f}}\tau \right)$. 

Thus, without loss of generality, the condition $Cv^2<1$ may be assumed to take place. 
Then all the parameters of the needed here adjustment \req{tilde-g} can be uniquely determined using relations \req{vv}, \req{S}, \req{bb}, and \req{tau}, established below in 
Appendix \ref{proof of TH BOOST-ADJUST}, \pg{proof of TH BOOST-ADJUST}: 
\begin{eqnarray*}
S&=&S^{\vv}\left({A_{10}A_{01}\over A_{00}}-A_{11}\right),\\
\bb^T&=&C\vv^T S+\g_v^{-1}A_{01},\\
\tau&=&{A_{00}\over\g_v},
\end{eqnarray*}
where $\vv:=\vv^{g,f}$ and $\g_v$ are computed according \req{vv-matrix}, page \pageref{vv-matrix}, and \req{gamma}, page \pageref{gamma}, respectively.

It is seen that neither the value nor the sign of $C$ is determined by the mere fact that a pair $(f,g)$ can be adjusted to a $C$-Lorentzian pair.

\subsection{Level 1: Testing reciprocity only and 
executing an appropiate generalized Lorentzian adjustment}
\label{level-1}

According to our hypothesis in its strongest form, all the pairs of inertial RFs obtained via a standard autonomous construction are adequately modeled by the notion of natural pairs and thus can be adjusted via re-orientation and isotropic rescaling to a reciprocal and isotropic pair of RFs. 

However, in this subsection we want to describe a method of testing of reciprocity only, rather than of both reciprocity and isotropy, and describe how to execute a corresponding generalized Lorentzian adjustment.

Because our construction is {\em autonomous}, there is no reason to expect that the given pair of RFs $(f,g)$ will be reciprocal by itself, without any adjustment. 
At the same time, re-synchronization is not needed here.
Moreover, if re-synchronization were allowed here as well, than in view of Theorem \ref{TH1}, \pg{TH1}, reciprocity could not be possibly tested. 

By Theorem \ref{RECIP-cor}, \pg{RECIP-cor}, $(f,g)$ can be can be adjusted without re-synchronization to a reciprocal pair of RFs
if and only if it can be 
rescaled (or, equivalently, adjusted without re-synchronization) to a generalized Lorentzian pair of RFs;
in turn, by Theorem \ref{SYNCHRO-FREE}, \pg{SYNCHRO-FREE}, this is equivalent to the system of two inequalities 
\beq{ineq}
\mu<1\quad\mbox{and}\quad C\mu>0,
\eeq 
where 
$\mu:=\dfrac{A_{01}A_{11}^{-1}A_{10}}{A_{00}}$.

Thus, this system of inequalities constitutes a definitive {\em test} of the reciprocity or, more exactly, a test of the adjustability without re-synchronization to a reciprocal pair.

In case the results of this test are positive, appropriate (but not unique at that) adjustments 
$\tilde f:=\diag(1,N^{-1})f$ and $\tilde g:=\diag(\tau,M)g$ of RFs $f$ and $g$, such that the pair $(\tilde f,\tilde g)$ is $C$-boost, are described by the formulas  
\begin{eqnarray*}
N&=&((\aaa^T\bb)^{-1}\bb\bb^T+\bb_2\bb_2^T+\bb_3\bb_3^T)^{1/2},\\
\tau&=&A_{00}/\g_v,\\
M&=&-(NS^{\vv})^{-1}A_{11},
\end{eqnarray*}
where $\aaa:=A_{00}(A_{11}^T)^{-1}A_{01}^T$, $\bb:=CA_{10}$, $\bb_2:=\aaa\times\bb$, $\bb_3:=\aaa\times\bb_2$, $\vv:=C^{-1}N(A_{11}^T)^{-1}A_{01}^T$, and $\g_v$ is given by \req{gamma}, page \pageref{gamma} [cf. \req{N-r}--\req{vv-r}, \req{00rf}, \req{11rf}, and the paragraph that precedes \req{N-r}]. 

By the second inequality in \req{ineq}, the sign of $C$ is uniquely determined by the pair $(f,g)$. 
However, in view of Remark \ref{UNIVERSAL-SYNCHRO-FREE}, \pg{UNIVERSAL-SYNCHRO-FREE}, with the reciprocity property only, the sign of $C$ can hardly be considered a local property of the physical spacetime, since a universal generalized Lorentzian rescaling of a pairwise reciprocal family need not exist. 
Moreover, the sign of $C$ may depend on the choice of the pair of RFs $(f,g)$ in such a family.
E.g., if RFs $f$, $g$, and $h$ are such that $A^{g,f}=B^{C_1,v}$ and $A^{h,g}=B^{C_2,u}$ are scalar boost matrices with, say, $C_1=1$, $v=0.2$, $C_2=3$, and $u=0.1$, then one has $\mu^{h,f}<0$, and so, by \req{ineq}, pair of RFs $(f,h)$ can be rescaled to a $C$-Lorentzian pair only with $C<0$, while $C_1>0$ and $C_2>0$.

\subsection{Another Level 1: Testing isotropy only and 
executing an appropiate generalized Lorentzian adjustment}
\label{another-level-1}

In this subsection we want to describe methods of testing of the condition of weak isotropy. 
Again, although the spacetime may be adequately described as isotropic in a domain containing the given pair of RFs $(f,g)$,  
there is no reason to expect that $(f,g)$ will be isotropic or weakly isotropic as it is, without any adjustment -- because the physical construction is {\em autonomous} for each of the two RFs under consideration. 

According to Theorem \ref{skew-free-charact}, \pg{skew-free-charact}, $(f,g)$ can be adjusted via spatial re-orientation to a weakly-isotropic pair of RFs $(\hat f,\hat g)$ if and only if it can be adjusted via isotropic rescaling and re-synchronization to a proper $C$-Lorentzian pair of RFs $(\tilde f,\tilde g)$.

It can be seen from the proof of Theorem \ref{skew-free-charact} [cf. \req{!}, page \pageref{!}]
that $(f,g)$ can be adjusted via spatial re-orientation to a weakly-isotropic pair of RFs if and only if 
\beq{iso-test}
(I_3-P^\vv)A_{11}A_{11}^T(I_3-P^\vv)=\xi^2(I_3-P^\vv)
\eeq
for some $\xi>0$, where $\vv:=\vv^{g,f}$.

This is a definitive {\em test} of the weak isotropy. 

In case the result of this test is positive, appropriate adjustments 
$
\tilde f:=\pmatrix{\tau_1&\bb^T_1\cr \0&I_3\cr}f
$
and 
$\tilde g:=\pmatrix{\tau&\bb^T\cr \0&\xi Q\cr}g$ of RFs $f$ and $g$, such that the pair $(\tilde f,\tilde g)$ is $C$-boost, are uniquely described -- given $C$ and $\tau_1$ and given that $\tau>0$ -- 
\begin{eqnarray}
\xi&=&\dfrac{|A_{11}^T \xx^\perp|}{|\xx^\perp|},\\
Q&=&Q_\vp:=\vp\dfrac{P^\uu(A_{11}^T)^{-1}} {|A_{11}^{-1}\uu^\circ|}
-\dfrac1\xi(I_3-P^\uu)A_{11},\label{Q-ttest}\\
\tau&=&\tau_1 \dfrac{|A_{10}|} {\xi\vp|A_{11}^{-1}A_{10}|}
\,(A_{01}A_{11}^{-1}A_{10}-A_{00}),\label{tau-test}\\
\bb_1^T&=&u^{-2}(\uu^T A_{11}-\tau_1 u^2 A_{01}+\g^{-1}\xi\uu^T Q) A_{11}^{-1},\label{bb1-test}\\
\bb^T&=&\g^{-1}(\tau_1 A_{01}+\bb_1^T A_{11}+\xi C\g\uu^T Q),\label{bb-test}
\end{eqnarray}
where $\vp:=\mbox{sign}\,\left[(A_{01}A_{11}^{-1}A_{10}-A_{00})\tau_1 \right]$, 
\beq{uu-test}
\uu:=\dfrac{A_{10}}{\sqrt{\tau^2+C|A_{10}|^2}},
\eeq and $\g:=\g_u$ (recall \req{gamma}, page \pageref{gamma}) 
[cf. \req{xi}, \req{Q-vp}, \req{tau-final}, \req{b1}, \req{b-final}, \req{vp-final}, \req{skew-free-v}, and \req{u-final}]; 
here, $\tau_1$ is any nonzero real number with the large enough absolute value 
so that $\tau$ in \req{tau-test}
is large enough
so that $\uu$ can be defined by \req{uu-test}; in particular, if $C\ge0$, then $\tau_1$ may be taken to be any nonzero real number. 

It is seen that neither the value nor even the sign of $C$ is uniquely determined by the mere fact that a pair $(f,g)$ can be adjusted via isotropic rescaling and re-synchronization to a $C$-Lorentzian pair.

\subsection{Level 2: Testing reciprocity and isotropy and 
executing an appropiate generalized Lorentzian adjustment}
\label{level-2}

In this subsection, we shall describe how to test the full content of our main hypothesis that all the pairs of inertial RFs obtained via a standard autonomous construction are adequately modeled by the notion of natural pairs and thus can be adjusted via re-orientation and isotropic rescaling to a reciprocal and isotropic pair of RFs. 
We shall also describe how to execute an appropriate generalized Lorentzian adjustment, which will be shown to be unique in a certain sense.
What is even more important, it will be shown that the constant $C$ can also be uniquely determined here.  

By Theorem \ref{RECIP-AND-ISO-cor}, \pg{RECIP-AND-ISO-cor}, pair $(f,g)$ is natural if and only if it can be isotropically rescaled to a generalized Lorentzian pair.
Therefore, in view of Proposition \ref{LOR}, \pg{LOR}, pair $(f,g)$ is natural if and only if if the RFCT matrix $A:=A^{g,f}$ admits a representation of the form \req{synchro-free}, page \pageref{synchro-free}, with $N=\xi_1 I_3$ and $M=\xi Q$ for some positive real $\xi$ and $\xi_1$, and some orthogonal $3\times3$ matrix $Q$.
Note that 
$\diag(1,\xi_1 I_3)B^{C,\vv}=B^{C\xi_1^{-2},\,\xi_1\vv}\,\diag(1,\xi_1 I_3)$. 
Hence, without loss of generality, we shall assume that $\xi_1=1$, so that the condition that $(f,g)$ is natural may be rewritten as 
\beq{repr-test}
A=B^{C,\vv}\diag(\tau,\xi Q)
\eeq
or, equivalently, as the system of equations
\bea
A_{00}&=&\g_v\tau, \label{00test}\\
A_{01}&=&-\g_v C\xi\vv^T Q, \label{01test}\\
A_{10}&=&\g_v\tau\vv, \label{10test}\\
A_{11}&=&-\xi S^{\vv}Q. \label{11test}
\eea
Eqs. \req{10test} and \req{00test} uniquely determine $\vv=\dfrac{A_{10}}{A_{00}}(=\vv^{g,f})$. 
Note next that the existence of an orthogonal matrix $Q$ satisfying \req{11test} is equivalent to the condition 
\beq{test1}
A_{11}A_{11}^T=\xi^2 (S^\vv)^2
\eeq
for some $\xi>0$.


Thus, \req{test1} uniquely determines $\xi>0$; alternatively and equivalently, $\xi$ may be uniquely determined by \req{xi}, \pg{xi}, which follows from \req{!}, which follows from \req{11test}.
Also, \req{11test} implies
\beq{Q-test}
Q=-\xi^{-1}(S^\vv)^{-1}A_{11}.
\eeq
This implies $\g_v \xi \vv^T Q=-\vv^T A_{11}$. 
Hence, given \req{11test}, equation \req{01test} can be rewriten as
\beq{test2}
A_{01}=C\vv^T A_{11}.
\eeq
This uniquely determines the value of $C$, say by the formula
\beq{C-test}
C=\dfrac{A_{01}A_{01}^T}{\vv^T A_{11}A_{01}^T}.
\eeq
Hence, $\tau$ is uniquely determined by \req{00test}, and 
$Q$ is uniquely determined by \req{Q-test}, taking into account \req{S-v} and \req{gamma}, page \pageref{S-v}.

Note that representation \req{repr-test} means that the pair $(f,\tilde g)$ is $C$-boost, where $\tilde g:=\diag(\tau,\xi Q)g$ is an adjustment of $g$ obtained via re-orientation and isotropic rescaling. 

Thus, all the elements of representation \req{repr-test} -- $C$, $\vv$, $\tau$, $\xi$, and $Q$ -- are uniquely determined. 
In particular, the adjustment $(f,\tilde g)$ of pair $(f,g)$ is uniquely determined.
But the most important fact here is that the value of $C$ is uniquely determined.

Moreover, in view of Theorem \ref{scale-ortho} (or Theorem \ref{universal rescaling}) and Proposition \ref{C-F}, \pg{C-F}, the sign of $C$ can be considered truly a local property of the physical spacetime provided that the main hypothesis is true in its full form, as stated in the beginning of this subsection.  

At the same time, one has a definitive {\em test} as to whether $(f,g)$ is natural, i.e., can be adjusted via re-orientation and isotropic rescaling to a reciprocal and isotropic pair of RFs. 
This test consists of the following two conditions [cf. \req{test2} and \req{test1}]:
\begin{enumerate}
\item
vectors $A_{01}^T$ and $A_{11}^T A_{10}$ are collinear with each other and 
\item
$A_{11}A_{11}^T=\xi^2 (S^\vv)^2$ for some $\xi>0$,
where $S^\vv=S^{C,\vv}$ is defined by \req{S-v}, page \pageref{S-v}, $\vv=\vv^{g,f}$, and $C$ is determined by \req{C-test}.
\end{enumerate}

\section{Waves of transformation of spacetime}
\label{waves}

\subsection{Equations of waves of transformation of spacetime, wave duality, and wave interpretation of $C$}
\label{wave eqs.}

For the local, or special, theory of relativity the notion of the RF introduced in Subsection \ref{RF-RFCT} as a 1-to-1 mapping of the event space $\E$ onto $\R^4$ is adequate. 
In the general theory, $\E$ and $\R^4$ should be replaced by subsets of theirs. 
Respectively, an RFCT in the general theory is a mapping of a subset of $\R^4$ onto some, perhaps other, subset of $\R^4$. 

Let $\A$ be such an RFCT, which is defined and differentiable on some open set $\D$ in $\R^4$ and whose Jacobian matrix at point $X$ is $A:=A(X)$, for any $X$ in $\D$. 
Matrix $A$ can be considered as the matrix $A^{g,f}$ of the RFCT from an RF $g$ to another RF $f$, URMoving relative to $g$, where both RFs $f$ and $g$ can be considered as located in an infinitesimally small neighborhood of the point $X$ of the domain $\D$ in $\R^4$. 
By Theorem \ref{TH1-cor} (\pg{TH1-cor}), Proposition \ref{LOR} (\pg{LOR}), and equations \req{scalar} and \req{scalar-infty} (\pg{scalar-infty}), the pair $(f,g)$ can be adjusted to a $C$-boost pair $(\tilde f,\tilde g)$, for every given $C$. 
Thus, the matrix $\tilde A:=\tilde A(X):=A^{\tilde g,\tilde f}$ is $C$-boost, at every point $X$ in $\D$.   

Suppose that such local adjustments can be done in a consistent fashion, so that the resulting $C$-boost matrices $\tilde A(X)$, $X\in\D$, constitute a family of the Jacobian matrices of a differentiable mapping defined on domain $\D$. 

The question is, What are characteristic properties of the family of the $C$-boost matrices $\tilde A(X)$, $X\in\D$? 

To simplify the notation and without loss of generality, we shall assume that $A(X)=\tilde A(X)$ for all $X$ in $\D$, so that the original family $A(X)$, $X\in\D$, already consists of $C$-boost matrices, where the local value of $C=C(X)$ at point $X$ in $\D$ may of course depend on $X$. 
Likewise, the speed parameter $v=v(X)$ in \req{scalar-boost}, \pg{scalar-boost},
may depend on the point $X=:(t,x,y,z)^T$ in $\D$.

Let the four-dimensional vector $(\tau,\xi,\eta,\zeta)^T$ in $\R^4$ denote the image of a point $X=(t,x,y,z)^T$ in $\D$ under the mapping $\A$, i.e.,  $(\tau,\xi,\eta,\zeta)^T=\A((t,x,y,z)^T)$, so that here $\tau$ is the ``new", transformed temporal coordinate, while $\xi$, $\eta$, and $\zeta$ are the ``new" spatial coordinates of an event with the ``old" temporal coordinate  $t$ and ``old" spatial coordinates $x$, $y$, and $z$.

Thus, the scalar $C$-boost Jacobian matrix $A=A(X)=A((t,x,y,z)^T)$ has the form
\beq{A-wave}
A=\diag\left(J,-I_2\right),
\eeq
where $J:=\dfrac{\partial(\tau,\xi)}{\partial(t,x)}:=
\pmatrix{\tau_t&\tau_x\cr\xi_t&\xi_x\cr}$ is a $2\times2$ Jacobian matrix;
the subscripts ${}_t$ and ${}_x$ stand for the partial derivatives with respect to $t$ and $x$. 
Let us disregard such experimentally non-detectable degeneracies as some of the elements of $J$ being zero at some point. 

Then one can see that the scalar-boost property of $A$ is completely characterized by the system of equations $\mbox{trace}\,J=0$ and $\mbox{det}\,J=-1$, that is, 
\bea
\tau_t+\xi_x&=&0,\label{waves1}\\
\tau_t\xi_x-\tau_x\xi_t&=&-1.\label{waves2}
\eea
By \req{scalar-boost}, \pg{scalar-boost}, one has
\beq{C-wave}
C=-\frac{\tau_x}{\xi_t}.
\eeq

Rewrite system \req{waves1}--\req{waves2} as 
\bea
\xi_x&=&-\tau_t,\label{w1}\\
\xi_t&=&\dfrac{1-\tau_t^2}{\tau_x}.\label{w2}
\eea
The latter two equations, together with $\xi_{xt}=\xi_{tx}$, yield
\beq{u}
\tau_x^2\tau_{tt}-2\tau_t\tau_x\tau_{tx}+(\tau_t^2-1)\tau_{xx}=0.
\eeq
Conversely, if $\tau$ is a solution of equation \req{u}, then there exists a solution $\xi$ of system \req{w1}--\req{w2}, and so, system \req{waves1}--\req{waves2} is solved, in principle. 


\paragraphh{Remark:\ 
Wave duality between time and space}
\label{duality}
\ \\ 
System \req{waves1}--\req{waves2} is self-dual in the sense that it remains invariant when the ``new" temporal coordinate 
$\tau$ is interchanged with the ``new" spatial coordinate $\xi$ and, simultaneously, the ``old" temporal coordinate $t$ is interchanged with the ``old" spatial coordinate $x$. 
Therefore, given a family of solutions $\tau=\tau(t,x)$ and $\xi=\xi(t,x)$ 
of system \req{waves1}--\req{waves2}, one can obtain another, {\em dual}, family of solutions $\hat\tau(t,x):=\xi(x,t)$ and $\hat\xi(t,x):=\tau(x,t)$ 
by such interchanging of variables. 
Obviously, if a family of solutions of \req{waves1}--\req{waves2} is dual to another family, then vice versa is also true, so that one can refer in this case to the two families as to a {\em dual pair}. 
If a family of solutions of \req{waves1}--\req{waves2} is dual to itself, let us call it {\em self-dual}.
To avoid misunderstanding, note that in a self-dual family of solutions of \req{waves1}--\req{waves2}, every member of the family is dual to a possibly different member of the same family, not necessarily to itself. 
Note also that any family of solutions of \req{waves1}--\req{waves2} can be (at least formally) extended to a self-dual family, namely, to the union of the given family with its dual.

In the next two subsections, we shall present, as two models, 
two dual pairs of explicitly described families of non-linear solutions of \req{waves1}--\req{waves2}.
The two families of the first dual pair are identical to each other, so that in fact one has one self-dual family. 
In contrast, the two families of the second dual pair are different from each other. 
\qedd

As an immediate consequence to Remark \ref{duality}, one has the following, dual to \req{u}, equation:
\beq{v}
\xi_t^2\xi_{xx}-2\xi_t\xi_x\xi_{tx}+(\xi_x^2-1)\xi_{tt}=0.
\eeq

Equations \req{u} and \req{v} are non-linear wave equations, since they are of the hyperbolic type; indeed, their discriminants are everywhere positive, equal to $(2\tau_t\tau_x)^2-4\tau_x^2(\tau_t^2-1)=4\tau_x^2$ for \req{u} and $4\xi_t^2$ for \req{v}.


\paragraphh{Remark:\ 
Wave interpretation of $C$}
\label{wave-C}
\ \\ 
Recall that any equation of the form $\psi=\psi(\al t+\beta x)$, with $\beta\ne0$, represents a wave propagating along the $x$-axis with constant velocity
\beq{v-psi}
v^\psi=-\dfrac\al\beta=-\dfrac{\psi_t}{\psi_x}.
\eeq 

Hence, $\tau$ and $\xi$, the solutions to the wave equations \req{u} and \req{v}, may be considered as the time wave and the space wave, respectively, propagating along the $x$-axis with not necessarily constant velocities
\beq{v-tau}
v^\tau=-\dfrac{\tau_t}{\tau_x}
\eeq 
and
\beq{v-xi}
v^\xi=-\dfrac{\xi_t}{\xi_x}.
\eeq 
It follows from \req{C-wave}, \req{v-tau}, \req{v-xi}, and \req{waves1} that $\dfrac1C$ is the product of the velocities of the time and space waves along the $x$-axis:
\beq{C-interpret}
\dfrac1C=v^\tau v^\xi.
\eeq
In particular, it follows, once again, that $C$ has the dimension of (velocity)${}^{-2}$.
\qedd

In view of the C\'auchy--Kowalevsky Theorem, one can impose arbitrary analytical initial conditions on $\tau_t$, $\tau_x$, $\xi_t$, and $\xi_x$ in problem \req{waves1}--\req{waves2}. 
Thus, there exist solutions of \req{waves1}--\req{waves2} with local values of $C$ of both signs, depending on the point in the spacetime.

We shall present three explicitly given families of nonlinear solutions of \req{waves1}--\req{waves2}.
For each solution belonging to the first of these families, $C$ may take on values of both signs, depending on $t$ and $x$. 
For each solution belonging to either of the other two families, $C$ is everywhere positive.

\subsection{Self-dual sum-of-two-waves family of solutions}
\label{linear+log} 


In search of an interesting family of explicit solutions of system \req{waves1}--\req{waves2} or, equivalently, \req{u} or \req{v}, one could first try a single wave -- say $\tau=\tau(x-ut)$ as a solution to \req{u} -- propagating with a constant velocity $u$ along the $x$-axis. 
However, as it is easy to see, that would lead only to the trivial family of linear solutions of \req{waves1}--\req{waves2} that correspond to the scalar $C$-boost matrices \req{A-wave} independent of $X=(t,x,y,z)^T$, with $\tau_t$, $\tau_x$, $\xi_t$, and $\xi_x$ being arbitrary constants satisfying \req{waves1}--\req{waves2}.

Any such trivial solution is a member (corresponding to $\psi=0$ below) of the following much richer and more interesting family of explicit solutions of \req{waves1}--\req{waves2}, described by the formulae
\bea
\tau&=&\g(t-C^{\rm lin}vx)+\psi(x-ut)+\tau_0,\label{tau-2-wave}\\
\xi&=&\g(vt-x)+u\psi(x-ut)+\xi_0,\label{xi-2-wave}
\eea
where
\beq{g-wave}
\g:=\dfrac{\vp_1}{\sqrt{1-C^{\rm lin}v^2}},
\eeq
\beq{u-wave}
u:=
\dfrac{\g+\vp_2}{C^{\rm lin}\g v},
\eeq
$\vp_1=\pm1$, $\vp_2=\pm1$, while $C^{\rm lin}$, $v$, $\tau_0$, and $\xi_0$ are arbitrary real parameters, except that $C^{\rm lin}$ and $v$ are assumed to be nonzero and such that the definition of $\g$ by \req{g-wave} makes sense; 
here, the function $\psi$ can be considered as an arbitrary infinite-dimensional, functional parameter.  

This family was derived assuming that $\tau$ or, equivalently, $\xi$ is the sum of two waves each with a constant velocity; it is then necessary that at least one of the two waves be linear, as in \req{tau-2-wave} and in \req{xi-2-wave}. 
We omit the derivation. 
Let us only indicate that it is straightforward to check that indeed the functions $\tau$ and $\xi$ given by \req{tau-2-wave}--\req{xi-2-wave} satisfy the system \req{waves1}--\req{waves2} at all points $X=(t,x,y,z)^T$ where $\psi(x-ut)$ is differentiable. 

Let us emphasize that $u$ in \req{tau-2-wave}--\req{xi-2-wave} is not arbitrary but is determined by $C^{\rm lin}$, $v$, and $\vp_2$ according to \req{u-wave}.

The family \req{tau-2-wave}--\req{xi-2-wave}
is especially interesting when the $\psi$-terms are small as compared to the linear terms, and so, may be considered as non-linear perturbation waves. 

Notice that equations \req{tau-2-wave} and \req{xi-2-wave} have the same functional form with respect to the arguments $t$ and $x$.
  
What is more interesting is that family \req{tau-2-wave}--\req{xi-2-wave} is self-dual in the sense of Remark \ref{duality}:
when $\tau$ is interchanged with $\xi$ and, simultaneously, $t$ is interchanged with $x$, any member of the family \req{tau-2-wave}--\req{xi-2-wave} turns into another member of the same family, with certain ``dual" values of the numerical parameters $\vp_1$, $\vp_2$, $C^{\rm lin}$, $v$, $\tau_0$, and $\xi_0$, and the functional parameter $\psi$; namely,
$\hat\vp_1:=-\vp_1$, $\hat\vp_2:=\vp_2$, $\widehat{C^{\rm lin}}:=1/C^{\rm lin}$, $\hat v:=C^{\rm lin}v$, $\hat\tau_0:=\xi_0$, $\hat\xi_0:=\tau_0$, and $\hat\psi(\la):=u\psi(-u\la)$ for all $\la$, where $u$ is defined by \req{u-wave};
note that the ``dual" value of $u$, that is,  
$\hat u:=(\hat\g+\hat\vp_2)/(\widehat{C^{\rm lin} }\hat\g \hat v)$ -- 
is reciprocal to $u$, $\hat u=1/u$.

Each of equations \req{tau-2-wave}--\req{xi-2-wave} describes a linear 
superposition of two waves, a linear wave and an 
arbitrary wave with a constant (but not arbitrary) velocity; let us refer to 
the latter wave as to the $\psi$-wave.

The linear wave components of $\tau$ and $\xi$ in \req{tau-2-wave}--\req{xi-2-wave}, i.e. $\tau^{\rm lin}:=
\g(t-C^{\rm lin}vx)+\tau_0$ and $\xi^{\rm lin}:=\g(vt-x)+\xi_0$, jointly describe the mutual URMotion of a scalar $(C^{\rm lin})$-boost pair of RFs, with constant 
relative velocity 
$v$ along the $x$-axis. 

Recall that $C^{\rm lin}$ can take on values of either sign.
Therefore, in the case when the derivative of $\psi$ is uniformly small enough, the true local value of $C$ obtained according to \req{C-wave} will have the same sign as $C^{\rm lin}$ everywhere in spacetime, and so, it can be everywhere positive or everywhere negative. 
On the other hand, taking e.g. $\psi(\la):=\ln|\la|$, it is easy to see that for every solution of \req{tau-2-wave}--\req{xi-2-wave}, the sign of $C$ can vary depending on $t$ and $x$. 

The $\psi$-wave components of $\tau$ and $\xi$ in \req{tau-2-wave}--\req{xi-2-wave}, i.e. $\tau^\psi:=\psi(x-ut)$ and 
$\xi^\psi:=u\psi(x-ut)$ describe waves moving with constant velocity $u$.

Note that in the domains of the spacetime where $\psi$-wave components of $\tau$ and $\xi$ are much larger than the linear ones, e.g. in a neighborhood of the plane of singularity $x-ut=0$ in $\R^4$ in the case $\psi(\la)\equiv\ln|\la|$, the true local value of $C$ will be close to [cf. \req{C-wave}] 
\beq{C-psi}
C^{\psi}:=-\frac{\tau^\psi_x}{\xi^\psi_t}=\dfrac1{u^2},
\eeq
which is always positive.

Let us also note that for every member of the family of solutions \req{tau-2-wave}--\req{xi-2-wave}, the velocities $v$ of the linear wave and $u$ of the $\psi$-wave are different from each other.

\subsection{Dual sum-and-product wave families of solutions}
\label{sum-and-product} 

Another family of explicit solutions of \req{waves1}--\req{waves2} is described by the formulae
\bea
\tau&=&\dfrac1{2\al}\ln\dfrac{(e^{2\al t+\beta_2}-1)^2} {e^{2\al t+\beta_2}}
-\dfrac1\al \ln|\al x+\beta_1|+\tau_0,\label{tau-sum}\\
\xi&=&-x\dfrac{e^{2\al t+\beta_2}+1}{e^{2\al t+\beta_2}-1}
-\dfrac{2\beta_1/\al}{e^{2\al t+\beta_2}-1}+\xi_0.\label{xi-prod}
\eea
Here, $\al\ne0$, $\beta_1$, $\beta_2$, $\tau_0$, and $\xi_0$ are arbitrary real parameters.

Note that $\tau$ in \req{tau-sum} is the {\em sum} of two functions, one of which depends only on $t$ and the other, only on $x$. 
It is a wave propagating along the $x$-axis with variable velocity [see \req{v-tau}]
\beq{v-tau-sum}
v^\tau=-\dfrac{\tau_t}{\tau_x}
=(\al x+\beta_1)\dfrac{e^{2\al t+\beta_2}+1}{e^{2\al t+\beta_2}-1}.
\eeq

Next, in the case $\beta_1=\xi_0=0$, $\xi$ in \req{xi-prod} is the {\em product} of two functions, one of which depends only on $t$ and the other, only on $x$. 
It is a wave propagating along the $x$-axis with variable velocity [see \req{v-xi}]
\beq{v-xi-prod}
v^\xi=-\dfrac{\xi_t}{\xi_x}
=(\al x+\beta_1)\dfrac{4e^{2\al t+\beta_2}}{e^{4\al t+2\beta_2}-1}.
\eeq

It follows from \req{C-wave}, \req{tau-sum}, and \req{xi-prod} or, alternatively, from 
\req{C-interpret}, \req{v-tau-sum}, and \req{v-xi-prod} that
$$C=\dfrac{(e^{2\al t+\beta_2}-1)^2} {4e^{2\al t+\beta_2}(\al x+\beta_1)^2}.$$
Thus, $C$ is positive everywhere. 

Asymptotic behavior of $\tau$ and $\xi$ for large $t$ is given by 
\[\begin{array}{lll}
\tau &\approx& 
	\left\{ \begin{array}{ll}
			&t+\dfrac{\beta_2}{2\al}-\dfrac1\al \ln|\al
 			x+\beta_1|+\tau_0,\quad \al t\to\infty,\\
			\vspace*{-6pt} \nonumber\\
			-&t-\dfrac{\beta_2}{2\al}-\dfrac1\al \ln|\al
 			x+\beta_1|+\tau_0,\quad \al t\to-\infty,
		\end{array}
	\right. \\
\vspace*{-6pt} \\
\xi &\approx& 
	\left\{ \begin{array}{ll}
			-&x+\xi_0,\quad \al t\to\infty,\\
			\vspace*{-6pt} \nonumber\\
			&x+2\beta_1/\al+\xi_0,\quad \al t\to-\infty
		\end{array}
	\right. 
\end{array}
\]

It follows that the direction of ``time" $\tau$ is the same as that of $t$ when $\al t\to\infty$ and is opposite when $\al t\to-\infty$. 
A similar relation takes place between $\xi$ and $x$ when $\al t\to\pm\infty$. 

The family dual to the one given by \req{tau-sum}--\req{xi-prod} is described by the formulae
\bea
\hat\tau&=&-t\dfrac{e^{2\al x+\beta_2}+1}{e^{2\al x+\beta_2}-1}
-\dfrac{2\beta_1/\al}{e^{2\al x+\beta_2}-1}+\xi_0,\label{tau-prod}\\
\vspace*{-6pt} \nonumber\\
\hat\xi&=&\dfrac1{2\al}\ln\dfrac{(e^{2\al x+\beta_2}-1)^2} {e^{2\al x+\beta_2}}
-\dfrac1\al \ln|\al t+\beta_1|+\tau_0.\label{xi-sum}
\eea

\appendix
\section*{}
\label{proofs}



\aparagraphh{Proof of Proposition \ref{re-label}}
\label{re-label-proof}
\ \\
Assume that $\A^{g_1,f_1}=\A^{g,f}$, i.e. $f_1\circUp g_1^{-1}=f\circUp g^{-1}$. 
Then the 1-to-1 mapping $f_1^{-1}\circUp f$ is identical to $g_1^{-1}\circUp g$; let us denote this mapping by $\ell$, so that $\ell=f_1^{-1}\circUp f=g_1^{-1}\circUp g$. 
Thus, $\ell$ is a re-labeling of the event space.
Moreover, one obviously has $f_1=f\circUp \ell^{-1}$ and $g_1=g\circUp \ell^{-1}$, so that $f_1=f^\ell$ and $g_1=g\,^\ell$. 

Conversely, suppose that if $f_1=f^\ell=f\circUp \ell^{-1}$ and $g_1=g\,^\ell=g\circUp \ell^{-1}$ are the re-labeled versions of $f$ and $g$ under any re-labeling $\ell$. 
Then it is easy to see that $f_1\circUp g_1^{-1}=f\circUp g^{-1}$, i.e., $\A^{g_1,f_1}=\A^{g,f}$. 
\qedd

\aparagraphh{Proof of Proposition \ref{LOR}}
\label{Proof of Theorem LOR}
\ \\
Let matrix $A$ be partitioned as in \req{block}, \pg{block}.
It is $C$-Lorentzian if and only if \req{lor-matrix}, \pg{lor-matrix}, takes place 
or, equivalently, 
\bea
A_{00}^2-CA_{10}^T A_{10}&=&1, \label{L1}\\
A_{00}A_{01}-CA_{10}^T A_{11}&=&\0^T, \label{L2}\\
A_{01}^T A_{01}-CA_{11}^T A_{11}&=&-CI_3. \label{L3}
\eea
First, it is straightforward to check that either \req{L} or \req{L-infty} implies that $A$ is $C$-Lorentzian.
On the other hand, \req{L} can be rewritten as the system of equations 
\bea
A_{00}&=&\vp\g_v, \label{1a}\\
A_{01}&=&-C\g_v\vv^T Q, \label{1b}\\
A_{10}&=&\vp\g_v\vv, \label{1c}\\
A_{11}&=&-S^{\vv}Q. \label{1d}
\eea
Note that \req{1a} and \req{gamma} imply
\beq{*}
A_{00}\ne0,
\eeq
which excludes case \req{L-infty}, in particular.
Moreover, \req{1a}, \req{1c}, and \req{1d} yield
\bea
\vp&=&\sign A_{00}, \label{1'}\\
\vv&=&{A_{10}\over A_{00}}, \label{1''}\\
Q&=&-(S^{\vv})^{-1}A_{11}. \label{1'''}
\eea
Hence, $\vp$ and $\vv$ are uniquely determined by $A$, and $Q$ is uniquely determined by $A$ and $C$. 
Note also that \req{L1} and \req{1''} imply $Cv^2<1$, so that \req{gamma} makes sense.

Thus, in case \req{*}, it suffices to show that equations \req{L1}--\req{L3} together with \req{1'}--\req{1'''} imply that $Q$ is orthogonal and that equations \req{1a}--\req{1d} hold. 
It is easy to check that
\beq{**}
(S^{\vv})^{-1}=I_3+(\g_v^{-1}-1)P^\vv.
\eeq
Hence, using \req{1'''}, \req{P}, \req{gamma}, \req{1''}, \req{L2}, and \req{L3}, one has 
$$Q^T Q 
=A_{11}^T(I_3+(\g_v^{-2}-1)P^\vv) A_{11}
=A_{11}^T A_{11}-C^{-1}A_{01}^T A_{01}=I_3,$$ 
i.e., $Q$ is indeed orthogonal.
Next, \req{L1}, \req{1''}, \req{1'}, and \req{gamma} imply \req{1a}. 
Further, \req{1a} and \req{1''} yield \req{1c}, and \req{1'''} is equivalent to \req{1d}. Finally, \req{1'''}, \req{**}, \req{P}, and \req{L2} imply \req{1b}. 

It remains to consider the case when \req{*} is false. In this case, \req{L1} implies $C<0$. Hence, \req{L-infty} may be rewritten as 
\bea
A_{01}&=&\sqrt{-C}\ee^T Q, \label{2a}\\
A_{10}&=&{\ee\over\sqrt{-C}}, \label{2b}\\
A_{11}&=&(P^\ee-I_3)Q, \label{2c}
\eea
plus $A_{00}=0$.
Then \req{2b} yields 
\beq{2'}
\ee=\sqrt{-C} A_{10},
\eeq
whence, using \req{2a}, one has 
$P^\ee Q=\ee\ee^T Q=\sqrt{-C}A_{10}\ee^T Q=A_{10}A_{01}$, and so,
by \req{2c}, 
\beq{2''}
Q=A_{10}A_{01}-A_{11}.
\eeq
Thus, $\ee$ and $Q$ are uniquely determined by $A$ and $C$. 
It remains to show that $\ee^T\ee=1$ and that $Q$ is orthogonal.
But $\ee^T\ee=1$ follows from \req{2'}, \req{L1}, and $A_{00}=0$, while 
$Q^T Q=I_3$ follows from \req{2''}, \req{L1}, \req{L2}, \req{L3}, and $A_{00}=0$.  
\qedd

\aparagraphh{Proof of Theorem \ref{TH1}}
\label{proof of TH1}
\ \\
To prove Theorem \ref{TH1}, we shall need

\aparagraphh{\quad\quad Lemma: Nonsingularity of the ``determinant" matrix}
\label{non-singular}
\ \\
If a $4\times4$ real matrix $A$ is non-singular and $A_{00}\ne0$, then the matrix 
$A_{00}A_{11}-A_{10}A_{01}$ is also non-singular [recall the convention \req{block}].
\qedd
{\em Proof}\quad
If $(A_{00}A_{11}-A_{10}A_{01})\xx=\0$
for some $\xx\in\R^3$, then 
$A\pmatrix{\la\cr\xx\cr}=\pmatrix{0\cr\0\cr}$ for $\la=-A_{01}\xx/A_{00}$, and so, by the non-singularity of $A$, one has $\xx=\0$.
\qedd

Let us now return to
{\em Proof of Theorem \ref{TH1}}. \quad  
That Condition 1 of Theorem \ref{TH1} implies Condition 2 therein follows 
immediately from the definition of adjustment in terms of being relatively at 
rest, which implies transitivity: if an RF $h$ is an adjustment of 
(i.e., is at rest relative to) an RF $g$ and RF $g$ is an adjustment of 
(i.e., is at rest relative to) an RF $f$, then RF $h$ is an adjustment 
of 
(i.e., is at rest relative to) RF $f$.  

It remains to prove that Condition 2 of Theorem \ref{TH1} implies Condition 1.

Let 
$A:=A^{g,f}$ and $B:=A^{g_1,f_1}$, where
the pairs $(f,g)$ and $(f_1,g_1)$ satisfy Condition 2 of Theorem \ref{TH1}. 
According to \req{rest}, page \pageref{rest}, it remains to show that there are two nonsingular matrices of the form
$$\pmatrix{\tau&\bb^T\cr \0&S\cr}\quad\mbox{and}\quad
\pmatrix{\tau_1&\bb^T_1\cr \0&S_1\cr},$$
where $S$ and $S_1$ are $3\times3$, such that
\beq{adjust}
B\pmatrix{\tau&\bb^T\cr \0&S\cr}
=\pmatrix{\tau_1&\bb^T_1\cr \0&S_1\cr}A,
\eeq
that is, 
\begin{eqnarray}
\tau B_{00}&=&\tau_1A_{00}+\bb_1^T A_{10},\label{00}\\
B_{00}\bb^T+B_{01}S&=&\tau_1 A_{01}+\bb_1^T A_{11},\label{01}\\
\tau B_{10}&=&S_1 A_{10},\label{10}\\
B_{10}\bb^T+B_{11}S&=&S_1A_{11}.\label{11}
\end{eqnarray}
Without loss of generality, $B_{00}\ne0$. Indeed, otherwise, $B_{10}\ne\0$, since $B$ is non-singular. Then, one can replace $B$ by, e.g., 
$$\tilde B=\pmatrix{\tilde B_{00}&\tilde B_{01}\cr \tilde B_{10}&\tilde B_{11}\cr}
:=\pmatrix{1&\tilde\bb^T\cr \0&I_3\cr}B$$
with some $\tilde\bb$ such that $\tilde B_{00}=\tilde\bb^T B_{10}\ne0$. 

Hence, \req{00} and \req{01} may be rewritten, respectively, as
\beq{00'}
\tau =\frac1{B_{00}}(\tau_1A_{00}+\bb_1^T A_{10})
\eeq
and
\beq{01'}
\bb^T=\frac1{B_{00}}(\tau_1 A_{01}+\bb_1^T A_{11}-B_{01}S).
\eeq
Using the last expression, one can rewrite \req{11} as
\beq{11'}
S=(B_{00}B_{11}-B_{10}B_{01})^{-1}
[B_{00} S_1 A_{11}-B_{10}(\tau_1 A_{01}+\bb_1^T A_{11})];
\eeq
by Lemma \ref{non-singular}, the matrix $B_{00}B_{11}-B_{10}B_{01}$ is non-singular. 
Since $A$ is non-singular, one can always choose a nonzero real number $\tau_1$ and a vector $\bb_1\in\R^3$ so that in \req{00'}, $\tau\ne0$. 
In fact, $A_{10}$ and $B_{10}$ are either both nonzero or both zero, because of the condition that $f$ and $g$ are not at rest relative to each other and $f_1$ and $g_1$ are not at rest relative to each other or, alternatively, $f$ and $g$ are at rest relative to each other and $f_1$ and $g_1$ are at rest relative to each other. 
Hence, one can always find a non-singular matrix $S_1$ to satisfy \req{10}. 
Then, all the relations \req{00}--\req{11} will take place if $\tau$, $\bb$, and $S$ are given by \req{00'}--\req{11'}. 
Note finally that in view of \req{adjust}, the matrix $\pmatrix{\tau&\bb^T\cr \0&S\cr}$ will be non-singular; this follows because $\tau_1\ne0$ and $S_1$ is non-singular, and so, $\pmatrix{\tau_1&\bb^T_1\cr \0&S_1\cr}$ is non-singular.
\qedd


\aparagraphh{Detais on Remark \ref{HYPERB}}
\label{Detais on Remark HYPERB}
\ \\
This remark is immediate from Theorem \ref{univ} and the following observation.
Let $\tilde f$ be an adjustment of some RF $f$ in $\F$, so that
$$\tilde f=\pmatrix{\tau&\bb^T\cr \0&S\cr}f.$$
If $g$ is in $\F$, $\uu=\vv^{g,\tilde f}$, and $\vv=\vv^{g,f}$, 
then $u=|S\vv|/|\tau+\bb^T\vv|$; if $\vv$ is infinite, 
this formula still works ``in the limit"; thus, $u=|S\ee|/|\bb^T\ee|$ if $\vv$ is infinite and has the direction of the line carrying unit vectors $\pm\ee$. 
It remains to notice the following:

(i) for any small enough $C>0$, the set of the terminal points of the vectors $\vv$ satisfying the inequality $|S\vv|/|\tau+\bb^T\vv|<1/\sqrt C$ is the inside of a two-sheet hyperboloid if $\bb\ne\0$; moreover, the inside of any two-sheet hyperboloid in $\R^3$ is contained in the set $\{\vv\in\R^3\colon\ |S\vv|/|\tau+\bb^T\vv|<1/\sqrt C\}$, for appropriate $S$, $\tau$, and $C$; the same inequality describes the inside of an ellipsoid if $\bb=\0$;

(ii) the relation $|S\vv|/|\tau+\bb^T\vv|<\infty$ describes the complement to $\R^3$ of the plane defined by the equation $\tau+\bb^T\vv=0$ if $\bb\ne\0$; otherwise, it describes the set of all finite velocities $\vv$.
\qedd

\aparagraphh{Proof of Theorem \ref{RECIP} and Remark \ref{IMPROPER} }
\label{Proof of Theorem RECIP}
\ \\
Let $A:=A^{g,f}$. The reciprocity means $A^2=I_4$, or 
\bea
A_{00}^2+A_{01}A_{10}&=&1, \label{r00}\\
A_{00}A_{01}+A_{01}A_{11}&=&\0^T, \label{r01}\\
A_{00}A_{10}+A_{11}A_{10}&=&\0, \label{r10}\\
A_{10}A_{01}+A_{11}^2&=&I_3. \label{r11}
\eea

Multiplying \req{r11} by $A_{11}$ on the right and then using \req{r01} to
replace $A_{01}A_{11}$ by $-A_{00}A_{01}$, one has 
$-A_{00}A_{10}A_{01}+A_{11}^3=A_{11}$. 
Again using \req{r11}, now to replace $A_{10}A_{01}$ by $I_3-A_{11}^2$, one obtains 
\beq{A11}
A_{11}^3+A_{00}A_{11}^2-A_{11}-A_{00}I_3=0.
\eeq
Hence the eigenvalues of $A_{11}$ satisfy the equation 
\beq{lambda}
\la^3+A_{00}\la^2-\la-A_{00}\equiv(\la+A_{00})(\la^2-1)=0,
\eeq
and so, may equal only to 1, $(-1)$, or $(-A_{00})$. 
In particular, now we see that all the eigenvalues of $A_{11}$ must be real.
Therefore, there exists a non-singular $3\times3$ {\em real} matrix
$S$ such that the matrix $S^{-1}AS$ is in a Jordan
canonical form.  

But, in view of \req{similar}, $A$ may be replaced by
$\diag(1,S^{-1})\,A\,\diag(1,S)$, for any non-singular $3\times3$ real matrix
$S$; then $A_{11}$, $A_{01}$, and $A_{10}$ become replaced by
$S^{-1}A_{11}S$,
$A_{01}S$, and
$S^{-1}A_{10}$, respectively.  Therefore, $A_{11}$ may be
assumed to be in a Jordan canonical form.
Thus, only the following three cases are possible.

{\em Case 1} \quad $A_{11}=\diag(\la_1,\la_2,\la_3)$, where
$\{\la_1,\la_2,\la_3\}\subseteq\{1,-1,-A_{00}\}$.\\
Then \req{r11} implies that $A_{10}A_{01}=I_3-A_{11}^2=\diag(1-\la_1^2,1-\la_2^2,1-\la_3^2)$ is a diagonal matrix of rank $\le\mbox{rank}\,(A_{10})\le1$, and so, for some permutation matrix $P$, 
$P^{-1}A_{10}A_{01}P$ equals to either $\diag(1-A_{00}^2,0,0)$ or the zero matrix. 
Hence, by \req{r00}, one always has $P^{-1}A_{10}A_{01}P=\diag(1-A_{00}^2,0,0)$, 
and so, by \req{r11}, $(P^{-1}A_{11}P)^2=\diag(A_{00}^2,1,1)$.
Replacing now $A_{01}$, $A_{10}$, and $A_{11}$ by 
$A_{01}P$, $P^{-1}A_{10}$, and
$P^{-1}A_{11}P$, respectively, that is, replacing $A$ by $\diag(1,P^{-1})\,A\,\diag(1,P)$, one has 
\beq{1001}
A_{10}A_{01}=\diag(1-A_{00}^2,0,0)
\eeq
and 
\beq{a11}
A_{11}=\diag(\vp_1A_{00},\vp_2,\vp_3)
\eeq
for some $\vp_1$, $\vp_2$, and $\vp_3$ in $\{1,-1\}$. 

\hskip3cm 
{\em Subcase 1.1}\quad $A_{00}=\vp_0$ for some $\vp_0\in\{1,-1\}$.\\
	Then, by \req{a11}, $A_{11}=\diag(\vp_1,\vp_2,\vp_3)$ for some $\vp_1$, $\vp_2$, and $\vp_3$ in $\{1,-1\}$, and, by \req{1001}, either $A_{10}=\0$ or $A_{01}=\0^T$. 
Therefore, letting $\alpha\to0$ or $\alpha\to\infty$ depending on whether $A_{10}=\0$ or $A_{01}=\0^T$, one sees that
$\diag(1,\alpha^{-1}I_3)A\diag(1,\alpha I_3)$ converges to the matrix $\diag(\vp_0,\vp_1,\vp_2,\vp_3)$, which is $C$-Lorentzian for all real $C$; in particular, it is 0-Lorentzian.
Thus, $A$ is asymptotically spatially similar to a $0$-Lorentzian pair of RFs, in the sense of Remark \ref{IMPROPER}.

\hskip3cm 
{\em Subcase 1.2}\quad $A_{00}\not\in\{1,-1\}$.\\
In this subcase, \req{1001} implies that for some nonzero real $a$, $A_{01}=(a^{-1},0,0)$  and $A_{10}=((1-A_{00}^2)a,0,0)^T$.
Hence, \req{r01} implies that in \req{a11}, $\vp_1=-1$.
Thus, 
$$A=\pmatrix{A_{00}&a^{-1}&0&0\cr
 		(1-A_{00}^2)a&-A_{00}&0&0\cr
		0&0&\vp_2&0\cr
		0&0&0&\vp_3},$$
whence $A$ is $C$-Lorentzian with $C:=-a^{-2}/(1-A_{00}^2)$.  

{\em Case 2, in which 
$$A_{11}=\pmatrix{\la&1&0\cr
		0&\la&0\cr
		0&0&\mu\cr},$$
where $\la,\mu\in\{1,-1,-A_{00}\}$.}

In this case, in view of \req{A11}, $\la$ must be a double root of \req{lambda}, wherefore $\la=-A_{00}=-\delta$ for some $\delta\in\{1,-1\}$, and so, $\mu=\vp$ for some $\vp\in\{1,-1,A_{00}\}=\{1,-1\}$. 
Now \req{r11} yields
$$A_{10}A_{01}=
\pmatrix{0&2\delta&0\cr
0&0&0\cr
0&0&0\cr},$$
whence, for some nonzero real $b$, one has $A_{10}=(b,0,0)^T$ and $A_{01}=(0,2\delta/b,0)$. 
Therefore,
\beq{improper}
A=\pmatrix{\delta&0&2\delta/b&0\cr
		b&-\delta&1&0\cr
		0&0&-\delta&0\cr
		0&0&0&\vp\cr}
\eeq
for some $\delta$ and $\vp$ in $\{1,-1\}$, and so, 
$$\diag(1,1,\alpha^{-1},1)\,A\,\diag(1,1,\alpha,1)
\mathop{\longrightarrow}\limits_{\alpha\to0}
\pmatrix{\delta&0&0&0\cr
		b&-\delta&0&0\cr
		0&0&-\delta&0\cr
		0&0&0&\vp\cr},$$
the latter being a 0-Lorentzian matrix.
Thus, $A$ is asymptotically spatially similar to a $0$-Lorentzian pair of RFs.

{\em Case 3, in which 
$$A_{11}=\pmatrix{\la&1&0\cr
		0&\la&1\cr
		0&0&\la\cr},$$
where $\la\in\{1,-1,-A_{00}\}$.}
This case is in fact impossible, since it would imply, in view of \req{A11}, that $\la$ is a triple root of equation \req{lambda}, which cannot have a triple root for any value of $A_{00}$. 
\qedd

\aparagraphh{Proof of Theorem \ref{skew-free-charact} and Remark \ref{skew-free-uniqueness} }
\label{Details on skew-free}
\ \\
Note that for any $\xi>0$ and any orthogonal $3\times3$ matrix $Q$, one has 
$\diag(1,\xi Q) B^{C,\vv}=B^{C\xi^{-2},\,\xi Q\vv}\diag(1,\xi Q)$; recall \req{boost}, \pg{boost}, for the definition of $B^{C,\vv}$.
Hence, in view of Proposition \ref{LOR}, \pg{LOR}, pair $(f,g)$ can be adjusted via isotropic rescaling and re-synchronization to a proper $C$-Lorentzian pair if and only if the matrix $A:=A^{g,f}$ satisfies the equation 
\beq{skew-free1}
B^{C,\uu}\pmatrix{\tau&\bb^T\cr \0&\xi Q\cr}
=\pmatrix{\tau_1&\bb^T_1\cr \0&I_3\cr}A,
\eeq
for some $\uu\ne\0$, $\bb_1$, and $\bb$ in $\R^3$, $\xi>0$, orthogonal matrix $Q$, and real $\tau\ne0$ and $\tau_1\ne0$.

Equation \req{skew-free1} is a special case of \req{adjust}, with 
\beq{special}
S_1=I_3,\quad S=\xi Q,\quad \mbox{and}\ B=B^{C,\uu}.
\eeq
Let us define $\g$ by
\beq{skew-free-g}
\g=\g_u;
\eeq
recall \req{gamma}, \pg{gamma}.
Note that \req{10} can be rewritten here as 
\beq{skew-free-v}
\uu=\dfrac{A_{10}}{\g\tau}.
\eeq
Hence, in view of \req{vv-matrix}, \pg{vv-matrix}, vectors $\uu$, $\vv:=\vv^{g,f}$, and $A_{10}$ have the same direction, and so, 
\beq{Pu=Pv}
P^\uu=P^\vv.
\eeq
Given \req{skew-free-v}, equation
\req{skew-free-g} is equivalent to
\beq{u-final}
u=\dfrac{|A_{10}|} {\sqrt{\tau^2+C|A_{10}|^2}}.
\eeq

Let $\vv^\circ$ stand for the unit vector along $\vv$ (or, equivalently, along $\uu$):
\beq{v0}
\vv^\circ:=\dfrac\vv v=\dfrac{A_{10}} {|A_{10}|}=\dfrac\uu u.
\eeq

In view of \req{special} and \req{boost}, \req{11} now implies 
\beq{!}
(I_3-P^\vv)A_{11}Q^T=-\xi(I_3-P^\vv),
\eeq
which in turn implies that $(f,g)$ can be adjusted via spatial re-orientation to a weakly-isotropic pair of RFs $(\hat f,\hat g)$, where $\hat f:=f$ and $\hat g:=\diag(1,Q)g$. 
This demonstrates the ``only if" part of Theorem \ref{skew-free-charact}.
Also, this verifies the last of the three statements of Remark \ref{skew-free-uniqueness}.

Let us now verify the second, ``uniqueness" statement of Remark \ref{skew-free-uniqueness}.
This amounts to showing that $\tau$, $\bb$, $\xi$, $Q$, and $\bb_1$ are uniquely determined in \req{skew-free1} given $\tau_1$ and given that $\tau>0$. 


Rewrite \req{!} as $Q^T(I_3-P^\vv)\xx=-\xi^{-1} A_{11}^T (I_3-P^\vv)\xx$ for all $\xx$ in $\R^3$ or, equivalently, as
\beq{***}
Q^T \xx^\perp=-\xi^{-1} A_{11}^T \xx^\perp
\eeq
for all $\xx^\perp$ in $\R^3$ that are orthogonal to $\vv$.
This implies
\beq{xi}
\xi=\dfrac{|A_{11}^T \xx^\perp|}{|\xx^\perp|},
\eeq
for any $\xx$ with $\xx^\perp\ne\0$.

Note also that 
if \req{***} takes place for some orthogonal matrix $Q$, then 
there exist exactly two orthogonal matrices $Q$ satisfying 
\req{***}.
Using \req{!}, it is straightforward to check that in such a case 
those two matrices $Q$ are 
\beq{Q-vp}
Q_\vp:=\vp\dfrac{P^\uu(A_{11}^T)^{-1}} {|A_{11}^{-1}\uu^\circ|}
-\dfrac1\xi(I_3-P^\uu)A_{11},\quad \vp=\pm1;
\eeq
note that $(A_{11}^T)^{-1}$ exists since the pair $(f,g)$ is strictly proper.

For $B=B^{C,\uu}$, as in \req{special}, one has $B_{00}B_{11}-B_{10}B_{01}=
-\g I_3+(\g-1)P^\uu$, and so, equation \req{11'} can be rewritten as 
\beq{skew-free11}
\xi(-\g I_3+(\g-1)P^\uu)Q=\g(A_{11}-\tau_1 \uu A_{01}-\uu\bb_1^T A_{11}).
\eeq
Multiplying both sides of this equation by $\uu^T$ on the left and by $A_{11}^{-1}$ on the right, one obtains
\beq{b1}
\bb_1^T=u^{-2}(\uu^T A_{11}-\tau_1 u^2 A_{01}+\g^{-1}\xi\uu^T Q) A_{11}^{-1}.
\eeq
Replacing here $Q$ by $Q_\vp$ from \req{Q-vp}, multiplying by $\uu$ on the right, and then using \req{skew-free-v}, one has 
\beq{1-b1 v}
\tau\g(1-\bb_1^T\uu)=\tau_1 A_{01}A_{11}^{-1}A_{10}
		-\tau\xi\vp \dfrac{|A_{11}^{-1}A_{10}|} {|A_{10}|}.
\eeq
On the other hand, in view of \req{skew-free-v}, 
equation \req{00} can be rewritten here as 
$$\tau\g(1-\bb_1^T\uu)=\tau_1 A_{00}.$$
This, together with \req{1-b1 v} and the condition $\tau>0$, implies
\beq{vp-final}
\vp=\mbox{sign}\,\left[(A_{01}A_{11}^{-1}A_{10}-A_{00})\tau_1 \right]
\eeq
and
\beq{tau-final}
\tau=\tau_1 \dfrac{|A_{10}|} {\xi\vp|A_{11}^{-1}A_{10}|}
\,(A_{01}A_{11}^{-1}A_{10}-A_{00}).
\eeq

Equation \req{01} can be rewritten here as
\beq{b-final}
\bb^T=\g^{-1}(\tau_1 A_{01}+\bb_1^T A_{11}+\xi C\g\uu^T Q).
\eeq

Now we can demonstrate the second, uniqueness statement of Remark \ref{skew-free-uniqueness}. 
We see that $\vp$ is uniquely determined by \req{vp-final}. 
Also, since $\vv=\vv^{g,f}$ is uniquely determined by the pair $(f,g)$, the value of $\xi$ is uniquely determined by \req{xi}.
Therefore, $\tau$ is uniquely determined by \req{tau-final}.
Next, the direction and the length of $\uu$ are uniquely determined by 
\req{skew-free-v} and \req{u-final}, respectively. 
Now one can compute also $\g$ using \req{skew-free-g}.
Then $Q=Q_\vp$ is uniquely determined by \req{Q-vp}, and so, $\bb_1$ is uniquely determined by \req{b1}, and finally, $\bb$ is uniquely determined by \req{b-final}. 

It remains to prove the ``if" part of Theorem \ref{skew-free-charact} and the first part of Remark \ref{skew-free-uniqueness}. 
Thus, suppose that the pair $(f,g)$ can be adjusted via spatial re-orientation to a weakly-isotropic pair $(\hat f,\hat g)$.
Without loss of generality, one may assume that $(f,g)$ itself is weakly-isotropic. 
This means that for some $\xi>0$ and for 
all $\xx$ in $\R^3$, one has $|(I_3-P^\vv)A_{11}\xx|=\xi|(I_3-P^\vv)\xx|$.
Hence,
\req{!} takes place for some orthogonal matrix $Q$. 
Note that $A_{01}A_{11}^{-1}A_{10}-A_{00}\ne0$; indeed, otherwise, one would have $A\pmatrix{1\cr\xx\cr}=\pmatrix{0\cr\0\cr}$ if $\xx=-A_{11}^{-1}A_{10}$, and so, $A$ would be singular.
Next, define $\vp$ by \req{vp-final} and then $\tau$ by \req{tau-final}, choosing $\tau_1$ to be any nonzero real number with the large enough absolute value 
so that $\tau$ is large enough
so that $u$ can be defined by \req{u-final}, and thus $\g$ can be defined by \req{skew-free-g}; then define $\uu$ by \req{skew-free-v}.
Now define $Q:=Q_\vp$ by \req{Q-vp},
$\bb_1$ by \req{b1}, and finally $\bb$ by \req{b-final}. 

Then it is straightforward to check that equation \req{skew-free1} is satisfied. 
This proves the ``if" part of Theorem \ref{skew-free-charact} and the first part of Remark \ref{skew-free-uniqueness}; 
it is obvious that if $C\ge0$, then $\tau_1$ can be taken to be equal to 1 (or to any other nonzero real) in order for the R.H.S. of \req{u-final} to be defined. 
\qedd

\aparagraphh{Counterexample for Remark \ref{weak iso vs iso} }
\label{weak-iso-and-iso}
\ \\
Let $g$ be any RF and let $f:=Ag$, where 
$$A:=B^{C,v}+\diag(0,0,2,2)
+\left(\dfrac{1-\sqrt{1-Cv^2}}v,1,0,0\right)^T(0,0,a,b),$$ 
$C$, $v$, $a$, and $b$ are nonzero reals, and $Cv^2<1$; recall here definition \req{scalar-boost}, page \pageref{scalar-boost}. 
Then the pair $(f,g)$ is reciprocal and weakly isotropic;
at the same time, pair $(f,g)$ 
cannot be isotropically rescaled to a generalized Lorentzian pair.
Indeed, otherwise, one could find a real number $\tilde C$ and positive real numbers $\tau^f$, $\xi^f$, $\tau^g$, and $\xi^g$ such that the matrix 
$$\diag(\tau^f,\xi^f I_3)^{-1}A\,\diag(\tau^g,\xi^g I_3)
=\pmatrix{(\tau^f)^{-1}\tau^g A_{00} & (\tau^f)^{-1}\xi^g A_{01}\cr   
	(\xi^f)^{-1}\tau^g A_{10} & (\xi^f)^{-1}\xi^g A_{11}\cr}$$
is $\tilde C$-Lorentzian, which would imply, in particular, (cf. \req{L2}, \pg{L2})
that the vectors \\
$A_{00}A_{01}=v^{-1}(\g_v-1)(-1-\g_v,a,b)$ 
and $A_{10}^T A_{11}=\g_v v(-\g_v,a,b)$ are collinear, which is obviously not true.
\qedd

\aparagraphh{Proof of Proposition \ref{one-angle} and Theorem \ref{RECIP-AND-ISO}}
\label{Proof of Theorem RECIP-AND-ISO and of Remark 1Q}
\ \\
Let us first consider the case $\vv:=\vv^{g,f}\ne\0$.
Obviously, Condition 3 of Proposition \ref{one-angle} implies Condition 1 implies Condition 2. 
To complete the proof of Proposition \ref{one-angle}, it remains to show that Condition 2 of Proposition \ref{one-angle} implies Condition 3.
Let $B$ be the matrix 
of the RFCT $\A^{g,f}$ in an orthonormal basis of the form 
$\pmatrix{1\cr\0\cr},\pmatrix{0\cr\vv/v\cr},\pmatrix{0\cr\bb_2\cr},\pmatrix{0\cr\bb_3\cr}$. 
Then Condition 2 implies $B\diag(I_2,R)=\diag(I_2,R)B$ for some $2\times2$ matrix $R$ of rotation through not a multiple of $180^\circ$.
Writing $R$ as $\pmatrix{\cos\theta&-\sin\theta\cr \sin\theta&\cos\theta\cr}$
with $\sin\theta\ne0$,
it is now easy to obtain Condition 3. 
Thus, Proposition \ref{one-angle} is proved. 

Let us now prove Theorem \ref{RECIP-AND-ISO}.
Since the reciprocity means $B^2=I_4$, one has the system of the equations 
(i) $B_0^2=I_2$ and (ii) $\la^2 P^2=I_2$, whence $\la=1$. 
Since $v\ne0$ and in view of \req{vv-matrix}, one can represent $B_0$ as $\g\pmatrix{1&-Cv\cr v&-a\cr}$, for some real numbers $C$, $\g$, and $a$, provided $v<\infty$.
Now (i) implies $a=1$ and $\g=\pm\g_v=\pm(1-Cv^2)^{-1/2}$. 
It is thus shown that $B=\diag(B_0,P)$, where $B_0=\pm\g_v\pmatrix{1&-Cv\cr v&-1\cr}$ and $P$ is a $2\times2$ rotation matrix. 
Hence, $B^T\diag(1,-CI_3)B=\diag(1,-CI_3)$, and so, 
$B$ is $C$-Lorentzian provided that $0<v<\infty$. 
The possibility $v=\infty$ is treated in the same manner. 
Here, one can write $B_0=\pmatrix{0&-Cu\cr u&-a\cr}$, for some real numbers $C$, $u\ne0$, and $a$. 
Using now (i) $B_0^2=I_2$, one has $a=0$ and $Cu^2=-1$, wherefore $B$ is again $C$-Lorentzian.

The case $\vv=\0$ is only easier. Here, the matrix $Q$ in the isotropy condition is any orthogonal $3\times3$ matrix, and so,  $A:=A^{g,f}=\diag(\mu,\nu I_3)$, for some real numbers $\mu$ and $\nu$. Now the reciprocity $A^2=I_4$ yields $\mu=\pm1$ and $\nu=\pm1$, whence $A$ is $C$-Lorentzian for any real $C$. 
\qedd

\aparagraphh{Proof of Theorem \ref{scale-ortho} }
\label{proof of [universal rescaling]}
\ \\
The ``if" part is immediate from Theorem \ref{RECIP-AND-ISO-cor}. 
To prove the ``only if" part, let us assume that $\F$ is a natural family of RFs. 

By Theorem \ref{RECIP-AND-ISO-cor}, for each pair of RFs $(f,g)$ in $\F$ there exist a real number $C^{f,g}$ and positive real numbers $\tau^{f,g}$, $\xi^{f,g}$, $\tau^{g,f}$, and $\xi^{g,f}$ such that the isotropically rescaled pair $(\tilde f^g,\tilde g^f)$ is $C^{f,g}$-Lorentzian, where $\tilde f^g:=\diag(\tau^{f,g},\xi^{f,g}I_3)f$ and $\tilde g^f:=\diag(\tau^{g,f},\xi^{g,f}I_3)g$.  

For any $f$ and $g$ in $\F$, since the pair $(\tilde f^g,\tilde g^f)$ is $C^{f,g}$-Lorentzian, the pair $(f,\tilde{\tilde g}^f)$ is $\tilde C^{f,g}$-Lorentzian, where 
$\tilde{\tilde g}^f:=\diag\left(\dfrac{\tau^{g,f}}{\tau^{f,g}},
	\dfrac{\xi^{g,f}}{\xi^{f,g}} I_3\right)g$, which is an isotropic rescaling of $g$, and 
$\tilde C^{f,g}:=C^{f,g}(\xi^{f,g}/\tau^{f,g})^2$.

Therefore, without loss of generality, one may assume 
that $\F$ has the property that all the relative velocities within $\F$ are nonzero; otherwise, consider first, in place of $\F$,
any maximal subfamily $\F_0$ of $\F$ with this property; then, by the last part of Remark \ref{C-uniq}, $(f,\tilde{\tilde g}^f)$ will be $C$-Lorentzian for {\em any} real $C$, if $f$ is any RF in $\F_0$ and $g$ is any RF in $\F$ with $\vv^{g,f}=\0$ (since for such $f$ and $g$, one will have $\vv^{\tilde{\tilde g}^f,f}=\0$). 

Now, let us first consider the case of {\em non-collinearity} when there are three RFs $f$, $g_1$, and $g_2$ in $\F$ such that the relative velocities $\vv^{g_1,f}$ and $\vv^{g_2,f}$ are non-collinear with each other. 
Let us fix any such $f$, $g_1$, and $g_2$.
Note that $\vv^{\tilde{\tilde g}_1^f,f}$ and $\vv^{\tilde{\tilde g}_2^f,f}$ are non-collinear, since $\vv^{g_1,f}$ and $\vv^{g_2,f}$ are so. 

Hence, without loss of generality one may assume that (i) for every $g$ in $\F$ the pair $(f,g)$ is $C^{f,g}$-Lorentzian for some real $C^{f,g}$, (ii) $\vv^{g_1,f}$ and $\vv^{g_2,f}$ are non-collinear for some $g_1$ and $g_2$ in $\F$, and (iii) $\vv^{g,h}\ne\0$ for any two RFs $g$ and $h$ in $\F$.  

Conditions (ii) and (iii) imply that for every $g$ in $\F$, either 
$\vv^{g,f}$ and $\vv^{g_1,f}$ are non-collinear or $\vv^{g,f}$ and $\vv^{g_2,f}$ are non-collinear.
Thus, in the non-collinearity case it remains to prove the following.

Suppose that $f$, $g$, and $h$ are three RFs such that (i) $\vv^{g,f}$ and $\vv^{h,f}$ are linearly independent and (ii)
the pairs $(f,g)$, $(f,h)$, and $(\tilde g,h)$ are $C_1$-, $C_2$-, and $C_3$-Lorentzian, respectively, for some real $C_1$, $C_2$, and $C_3$ and for some isotropic rescaling $\tilde g=\diag(\tau,\xi I_3)g$ of $g$, where $\tau$ and $\xi$ are some positive reals. 
Then $C_1=C_2$; note that, because of the group property, $C_1=C_2=C$ for some $C$ would imply that the pair $(g,h)$ is $C$-Lorentzian, as well as $(f,g)$ and $(f,h)$ are.  

Let $A:=A^{g,f}$ and $B:=A^{h,f}$.
In view of Proposition \ref{LOR}, \pg{LOR}, and because re-orientation preserves $C$-Lorentzian pairs, one may assume without loss of generality that $A=B^{C_1,\vv}$ or $A=B^{C_1,\ee_1}_\infty$
and $B=B^{C_2,\uu}$ or $B=B^{C_2,\ee_2}_\infty$ for some $\vv$, $\uu$, and unit $\ee_1$ and $\ee_2$ in $\R^3$ such that $\vv$ or $\ee_1$ is linearly independent of $\uu$ or $\ee_2$, as applicable.
Then $A^{g,h}=B^{-1}A=BA$, and $A^{\tilde g,h}=BA\diag(\tau,\xi I_3)^{-1}$. 

Consider first the case of the finite relative velocities, when $A=B^{C_1,\vv}$ and $B=B^{C_2,\uu}$.
Since $(\tilde g,h)$ is $C_3$-Lorentzian and in view of definition \req{lor-matrix}, \pg{lor-matrix}, and identity $(B^{C,\vv})^{-1}=B^{C,\vv}$, one has
$$\diag(1,-C_3 I_3)B^{C_2,\uu}B^{C_1,\vv}=
(B^{C_1,\vv}B^{C_2,\uu})^T \diag(1,-C_3 I_3) \diag(\tau^2,\xi^2 I_3),$$
which can be rewritten as the system of equations 
\bea
\g_u \g_v(1-C_2\uu^T\vv)&=&\g_u \g_v(1-C_1\uu^T\vv)\tau^2,\label{00-uni}\\
-\g_u \g_v C_3(\uu-\g_u^{-1}S^\uu \vv)&=&
\g_u \g_v\tau^2(-C_2\uu+C_1\g_u^{-1}S^\uu \vv),\label{10-uni}\\
\g_u \g_v(-C_1\vv+C_2\g_v^{-1}S^\vv \uu)&=&
-\g_u \g_v C_3 \xi^2(\vv-\g_v^{-1}S^\vv \uu),\label{01-uni}\\
-C_3(-\g_u \g_v C_1 \uu\vv^T+S^\uu S^\vv)&=&
-C_3 \xi^2(-\g_u \g_v C_2 \uu\vv^T+S^\uu S^\vv).\label{11-uni}
\eea
Since $\vv$ and $\uu$ are linearly independent, \req{10-uni} implies $C_3=C_1\tau^2$ and $C_3=C_2\tau^2$, whence $C_1=C_2$.

If one or both of the two relative velocities is infinite, that is, if 
$A=B^{C_1,\ee_1}_\infty$ and/or $B=B^{C_2,\ee_2}_\infty$, then the corresponding equations may be obtained from \req{00-uni}--\req{11-uni} by the limit transition(s) with $\vv=v\ee_1$ as $v\to\infty$ and/or $\uu=u\ee_2$ as $u\to\infty$, so that
$\g_v\vv\to\ee_1/\sqrt{-C_1}$ and $S^\vv\to I_3-P^{\ee_1}$ and/or
$\g_u\uu\to\ee_2/\sqrt{-C_2}$ and $S^\uu\to I_3-P^{\ee_2}$. 

The case when only one of the two relative velocities is infinite, i.e. $A=B^{C_1,\ee_1}_\infty$ or $B=B^{C_2,\ee_2}_\infty$, is similar to the the case of finite relative velocities; one uses here the limit version of \req{01-uni} if $A=B^{C_1,\ee_1}_\infty$ and that of \req{10-uni} if $B=B^{C_2,\ee_2}_\infty$.

If now both of the two relative velocities are infinite, i.e. $A=B^{C_1,\ee_1}_\infty$ and $B=B^{C_2,\ee_2}_\infty$, then the limiting versions of \req{10-uni}--\req{11-uni} may be written as 
\bea
C_3 (I_3-P^{\ee_2})\ee_1&=&
C_1 \tau^2(I_3-P^{\ee_2})\ee_1,\label{10-uni-1}\\
C_2 (I_3-P^{\ee_1})\ee_2&=&
C_3 \xi^2(I_3-P^{\ee_1})\ee_2,\label{01-uni-1}
\eea
\beq{11-uni-1}
C_3\left(-\dfrac{C_1}{\sqrt{C_1C_2}}\ee_2\ee_1^T+
(I_3-P^{\ee_2})(I_3-P^{\ee_1})\right)
=
C_3\xi^2 \left(-\dfrac{C_2}{\sqrt{C_1C_2}}\ee_2\ee_1^T+
(I_3-P^{\ee_2})(I_3-P^{\ee_1})\right).
\eeq
Note that $(I_3-P^{\ee_1})\ee_2\ne\0$ and $(I_3-P^{\ee_2})\ee_1\ne\0$ since
$\ee_1$ and $\ee_2$ are linearly independent.
Hence, if $C_3=0$, then \req{10-uni-1} and \req{01-uni-1} imply $C_1=0=C_2$.
Note also that the matrix $I_3$ is linearly independent of $P^{\ee_1}$, $P^{\ee_2}$, $\ee_2\ee_1^T$, and $P^{\ee_2}P^{\ee_1}$, since 
$P^{\ee_2}P^{\ee_1}=(\ee_2^T\ee_1)\ee_2\ee_1^T$ and
${\rm rank}(aP^{\ee_1}+bP^{\ee_2}+c\ee_2\ee_1^T)=
{\rm rank}(aP^{\ee_1}+\ee_2(b\ee_2+c\ee_1)^T)\le{\rm rank}(P^{\ee_1})+{\rm rank}(\ee_2)=2$ for all real $a$, $b$, and $c$, while ${\rm rank}(I_3)=3$. 
Hence, in the case $C_3\ne0$, \req{11-uni-1} implies $\xi^2=1$, and so, $C_1=C_2$. 

Thus, $C_1=C_2$ whenever the case of {\em non-collinearity} obtains. 

Otherwise, one may assume that $\vv=(v,0,0)^T$, $\uu=(u,0,0)^T$, and $\ee_1=\ee_2=(1,0,0)^T$. 
Then -- for the finite relative velocities -- equations \req{00-uni}--\req{11-uni} assume the form
\bea
\g_u \g_v(1-C_2 uv)&=&\g_u \g_v(1-C_1 uv)\tau^2,\label{00-uni-2}\\
-\g_u \g_v C_3(u-v)&=&
\g_u \g_v\tau^2(-C_2 u+C_1 v),\label{10-uni-2}\\
\g_u \g_v(-C_1 v+C_2 u)&=&
-\g_u \g_v C_3 \xi^2(v-u),\label{01-uni-2}\\
-\g_u \g_v C_3(-C_1 uv+1)&=&
-\g_u \g_v C_3 \xi^2(-C_2 uv+1),\label{11-uni-2}\\
-C_3 I_2&=&-C_3\xi^2 I_2.\label{11-uni-2'}
\eea
Here, the two eqs. \req{11-uni-2} and \req{11-uni-2'} correspond to the single eq. \req{11-uni}.

If $C_3\ne0$, then \req{11-uni-2'} implies $\xi^2=1$, and so, $C_1=C_2$ by \req{11-uni-2}, since $uv\ne0$.

If $C_3=0$, then the matrix
\beq{C3=0}
A^{\tilde g,h}=
\diag\left(\g_u \g_v \pmatrix{1-C_2 uv & 0\cr u-v & 1-C_1 uv},I_2\right)
\diag(\tau,\xi I_3)^{-1}
\eeq
is $0$-Lorentzian. 
Hence, by \req{0-L}, \pg{0-L}, the matrix $\diag(\g_u \g_v (1-C_1 uv),1,1)$ is orthogonal.
This means that $(\g_u \g_v (1-C_1 uv))^2=1$, or 
\beq{algebra}
2C_1 uv=C_1 v^2+C_2 u^2.
\eeq 
But \req{10-uni-2} and $C_3=0$ imply $C_2 u=C_1 v$. 
The latter eq. together with \req{algebra} imply $C_1(u-v)=0$, since $v\ne0$. 
If $u=v$, then $C_2 u=C_1 v$ yields $C_1=C_2$. 
If $u\ne v$, then $C_1=0$, and again $C_2 u=C_1 v$ yields $C_2=0=C_1$.

It remains to consider the limiting versions of eqs. \req{00-uni-2}--\req{C3=0} with $v\to\infty$ and/or $u\to\infty$. 

For instance, the limiting versions of eqs. \req{00-uni-2}, \req{10-uni-2}, and \req{01-uni-2} with only $u\to\infty$ imply $C_2=C_1 \tau^2$, 
$C_3=C_2 \tau^2$, and $C_2=C_3 \xi^2$, respectively. 
Hence, if $C_3=0$, then $C_2=0=C_1$.
If $C_3\ne0$, then \req{11-uni-2'} implies $\xi^2=1$, and so, $C_2=C_3$; hence,
$\tau^2=C_3/C_2=1$; thus, $C_2=C_1 \tau^2=C_1$. 

The limiting case with only $v\to\infty$ is completely similar to the latter one.

Consider finally the limiting versions of eqs. \req{00-uni-2}--\req{C3=0} with both $v\to\infty$ and $u\to\infty$. 
If $C_3\ne0$, then $C_1=C_2$ follows from \req{11-uni-2'} and the limiting version of \req{11-uni-2}. 
If $C_3=0$, then the limiting version of \req{C3=0} is
$$
A^{\tilde g,h}=
\diag\left(-\dfrac{C_2}{\sqrt{C_1C_2}},-\dfrac{C_1}{\sqrt{C_1C_2}},1,1\right)
\diag(\tau,\xi I_3)^{-1}.
$$
By \req{0-L}, \pg{0-L}, the matrix $\xi^{-1}\diag\left(-\dfrac{C_1}{\sqrt{C_1C_2}},1,1\right)$ must be orthogonal. 
This implies $C_1=C_2$.
\qedd

\aparagraphh{Proof of Proposition \ref{C-F}}
\label{proof of C-F}
\ \\
It is easy to check that if a family $\F$ of RFs is $C$-Lorentzian, then its isotropic rescaling $\tilde\F$ defined by $\tilde f:=\diag(\tau,\xi I_3)f$ for all $f$ in $\F$ with $\tau$ and $\xi$ independent of $f$ is 
$\tilde C$-Lorentzian with 
$\tilde C:=C\tau^2/\xi^2$. 
This implies Part 1 of the proposition. 

To verify the rest of the proposition, take any two RFs $f$ and $g$ in $\F$. 
Then there are isotropic rescalings $\tilde f:=\diag(\tau^f,\xi^f I_3)f$  and 
$\tilde g:=\diag(\tau^g,\xi^g I_3)g$ of $f$ and $g$ such that the pair 
$(\tilde f,\tilde g)$ is $C$-Lorentzian for some real $C=:C_\F$;
here, $\tau^f$, $\xi^f$, $\tau^g$, and $\xi^g$ are positive reals. 
Let $A:=A^{g,f}$.

Consider first the case when the relative velocity $\vv^{g,f}$ is finite. 
Then, by Proposition \ref{LOR},
\beq{LOR-CF}
\diag(\tau^f,\xi^f I_3)A=B^{C,\vv}\diag(\vp,Q)\diag(\tau^g,\xi^g I_3)
\eeq 
for some $\vv$ in $\R^3$, $\vp=\pm1$, and orthogonal matrix $Q$.
Equivalently, 
\bea
\tau^f A_{00}&=&\vp\tau^g\g_v, \label{00CF}\\
\tau^f A_{01}&=&-\xi^g C\g_v\vv^T Q, \label{01CF}\\
\xi^f A_{10}&=&\vp\tau^g\g_v\vv, \label{10CF}\\
\xi^f A_{11}&=&-\xi^g S^{\vv}Q. \label{11CF}
\eea
Eq. \req{00CF} implies
\beq{vp-CF}
\vp=\sign\,A_{00}
\eeq
and $A_{00}\ne0$.
Also, \req{11CF} implies that $A_{11}$ is non-singular.
Next, \req{10CF}  and \req{00CF} yield 
\beq{vv-CF}
\vv=\dfrac{\xi^f}{\tau^f} \dfrac{A_{10}}{A_{00}}.
\eeq
It follows from \req{11CF} that
\beq{xi-CF}
\xi^g I_3=\xi^f(A_{11}A_{11}^T)^{1/2} (S^\vv)^{-1}
\eeq
and 
\beq{Q-CF}
Q=-(A_{11}A_{11}^T)^{1/2} A_{11}.
\eeq
It also follows from \req{11CF} that $\xi^f\vv^T A_{11}=-\xi^g\g_v\vv^T Q$.
Comparing this with \req{01CF}, one has $\tau^f A_{01}=C\xi^f\vv^T A_{11}$.
This can be rewritten, in view of \req{vv-CF}, as 
\beq{C-CF}
C\dfrac{A_{10}^T A_{11}}{A_{00}}=\left(\dfrac{\tau^f}{\xi^f}\right)^2 A_{01}.
\eeq
Also, \req{00CF} implies
\beq{tau-CF}
\tau^g=\tau^f \dfrac{A_{00}}{\vp\g_v}.
\eeq

Now, Parts 2 and 4 follow from \req{C-CF}. 
Note that the positive real numbers $\tau^f$ and $\xi^f$, determining the isotropic rescaling $\tilde f=\diag(\tau^f,\xi^f I_3)f$ of $f$, can be chosen completely arbitrarily; 
then one can compute $\vv$ by \req{vv-CF} and $C$ by \req{C-CF}; after that, $\g_v$ by \req{gamma} (\pg{gamma}), and finally uniquely determine the isotropic rescaling 
$\tilde g=\diag(\tau^g,\xi^g I_3)g$ of $g$ using \req{xi-CF} and \req{tau-CF}, 
for any $g$ in $\F$ with a finite relative velocity $\vv^{g,f}$. 
This partially proves Part 3 of the proposition.

It remains to treat the case when the relative velocity $\vv^{g,f}$ is infinite. 
Here, we need to consider the limiting versions of eqs. \req{00CF}--\req{11CF} when $\vv=v\ee$ with $v\to\infty$ and $\ee$ being a unit vector in $\R^3$:
\bea
\tau^f A_{00}&=&0, \label{00CF-infty}\\
\tau^f A_{01}&=&\xi^g\sqrt{-C}\ee^T Q, \label{01CF-infty}\\
\xi^f A_{10}&=&\tau^g \dfrac{\ee}{\sqrt{-C}}, \label{10CF-infty}\\
\xi^f A_{11}&=&-\xi^g (I_3-P^\ee)Q; \label{11CF-infty}
\eea
here, one can always choose $\vp=1$; cf. \req{boost-infty}, \pg{boost-infty}.
The treatment of this case is similar.

First, \req{10CF-infty} yields
\beq{ee-CFinfty}
\ee=\dfrac{A_{10}}{|A_{10}|}
\eeq
and 
\beq{tau-CFinfty}
\tau^g=\xi^f\sqrt{-C}|A_{10}|.
\eeq
It follows from \req{11CF-infty} that
\beq{xi-CFinfty}
(\xi^g)^2(I_3-P^\ee)=(\xi^f)^2 A_{11}A_{11}^T.
\eeq
Next, \req{01CF-infty} implies
\beq{C-CFinfty}
C=- \left(\dfrac{\tau^f}{\xi^g}\right)^2 A_{01}A_{01}^T.
\eeq

Here, given any positive reals $\tau^f$ and $\xi^f$, one uniquely determines $\ee$ by \req{ee-CFinfty}, then $\xi^g>0$ by \req{xi-CFinfty}, next $C$ by \req{C-CFinfty}, and finally $\tau^g$ by \req{tau-CFinfty}.

This completes the proof of the proposition.

Note that $Q$ here is also uniquely determined. 
Indeed, \req{01CF-infty} and \req{10CF-infty}
imply $\dfrac{\tau^f \xi^f}{\tau^g \xi^g}A_{10}A_{01}=P^\ee Q$. 
This and \req{11CF-infty} now imply
\beq{Q-CFinfty}
Q=\dfrac{\tau^f \xi^f}{\tau^g \xi^g}A_{10}A_{01}
	-\dfrac{\xi^f}{\xi^g}A_{11}.
\eeq
Hence, in any case, 
all the parameters $\vp$, $\vv$, $\ee$, and $Q$ are uniquely determined -- by \req{vp-CF}, \req{vv-CF}, \req{ee-CFinfty}, and \req{Q-CF} or \req{Q-CFinfty}.
\qedd

\aparagraphh{Details on Remark \ref{dimension}}
\label{details on [dimension]}
\ \\ 
Consider first the case of $\R^1$ in place of $\R^3$.
Here, let $f$ be any RF and let then e.g. 
$g:=\dfrac1{2\sqrt2}\pmatrix{3&-1\cr1&-3\cr}f$ and 
$h:=\dfrac1{\sqrt2}\pmatrix{2&-2\cr1&-2\cr}f$.
Let $\tilde g:=\diag(2/\sqrt5,\sqrt5/2)\,g$. 
Then the pairs $(f,g)$, $(f,h)$, and $(\tilde g,h)$ are 
$C_1$-, $C_2$-, and $C_3$-Lorentzian, respectively,
with $C_1:=1$, $C_2:=2$, and $C_3:=16/5$. 

Consider second $\R^2$ in place of $\R^3$.
One counterexample for this case is as follows. 
Let $C_2$ be any negative real. Let $\xi$ be any positive real except 1. 
Let $C_1:=C_2\xi^2$, $C_3:=C_2\xi^{-2}$, and $\tau:=\xi^{-2}$. 
Let $\ee_1$ and $\ee_2$ be any two orthogonal unit vectors in $\R^2$.
For any negative real $C$ and any unit vector $\ee$ in $\R^2$, let us define here $B^{C,\ee}_\infty$ as in \req{boost-infty}, \pg{boost-infty}, but with $I_2$ in place of $I_3$.
Let now $f$ be any RF, and define RFs $g$, $h$, and $\tilde g$ by $g:=B^{C_1,\ee_1}_\infty f$, $h:=B^{C_2,\ee_2}_\infty f$, and 
$\tilde g:=\diag(\tau,\xi I_2)g$. 
Then the pairs $(f,g)$ and $(f,h)$ are obviously $C_1$- and $C_2$-Lorentzian.
Also, the pair $(\tilde g,h)$ is $C_3$-Lorentzian, since 
$$A^{\tilde g,h}=B^{C_3,-\ee_1}_\infty \diag(1,Q),$$
where the matrix $Q:=\ee_1\ee_2^T-\ee_2\ee_1^T$ is orthogonal (since $\ee_1$ and $\ee_2$ are orthogonal).

Let now $\G$ stand for any one of the two above triples $(f,g,h)$, constructed with $\R^1$ or $\R^2$ in place of $\R^3$. 
In either case, we have seen that every pair of RFs in $\G$ can be isotropically rescaled to a generalized Lorentzian pair. 
Moreover, the pairs $(f,g)$ and $(f,h)$ are already $C_1$- and $C_2$-Lorentzian.

Let us show now that there is no generalized Lorentzian isotropic rescaling 
$\hat\G:=(\hat f,\hat g,\hat h)$ of $\G$, where 
$\hat f:=\diag(\tau^f,\xi^f I)f$, 
$\hat g:=\diag(\tau^g,\xi^g I)g$, and $\hat h:=\diag(\tau^h,\xi^h I)f$, 
for 
any positive reals $\tau^f$, $\xi^f$, $\tau^g$, $\xi^g$, $\tau^h$, and $\xi^h$; 
here, $I$ stands either for $I_1=1$ or $I_2$, according to the number of the spatial dimensions. 
Assume that, to the contrary, there is a generalized Lorentzian isotropic rescaling 
$\hat\G:=(\hat f,\hat g,\hat h)$ of $\G$.

It follows from the second statement of Part 3 of Proposition \ref{C-F}, \pg{C-F}, applied to $(f,g)$ or $(f,h)$ in place of $\F$, that there exists at most one choice of $\tau^g$, $\xi^g$, $\tau^h$, and $\xi^h$ given $\tau^f$ and $\xi^f$ and given that the pairs $(\hat f,\hat g)$ and $(\hat f,\hat h)$ are generalized Lorentzian. 
But it is easy to check that the choice
$\tau^g:=\tau^h:=\tau^f$ and $\xi^g:=\xi^h:=\xi^f$ 
makes 
the pairs $(\hat f,\hat g)$ and $(\hat f,\hat h)$ generalized Lorentzian, namely,
$\hat C_1$- and 
$\hat C_2$-Lorentzian with $\hat C_1:=C_2(\tau^f/\xi^f)^2$ and 
$\hat C_2:=C_2(\tau^f/\xi^f)^2$.
Hence, this is the only choice of $\tau^g$, $\xi^g$, $\tau^h$, and $\xi^h$. 

With such a choice, the pair $(\hat g,\hat h)$ being generalized Lorentzian implies, in the same manner, that  
$(g,h)$ is so.
But $(\tilde g,h)$ is generalized Lorentzian as well, and $\tilde g=\diag(\tau,\xi I)\,g$ for some positive $\tau$ and $\xi$.
It follows again from the second statement of Part 3 of Proposition \ref{C-F} -- applied now to $(g,h)$ in place of $\F$, $h$ in place of $f$, and 
$(\tilde g,h)$ in place of $\tilde\F$ --
that $\tau=\xi=1$,
which contradicts the above constructions, in which $\xi$ (as well as $\tau$) differs from 1.
\qedd

\aparagraphh{Proof of Theorem \ref{BOOST-ADJUST}}
\label{proof of TH BOOST-ADJUST}
\ \\
Let 
$A:=A^{g,f}$. 
To prove Part I of Theorem \ref{BOOST-ADJUST}, \pg{BOOST-ADJUST}, it suffices to show that 
the representation
\beq{Cnot0Anot0}
A=B^{C,\uu}\pmatrix{\tau&\bb^T\cr \0&S\cr}
\eeq
takes place for a nonzero real number $\tau$, a non-singular real $3\times3$ matrix $S$, and vectors $\bb\in\R^3$ and $\uu\in\R^3$ if and only if $v<\infty$ and $Cv^2<1$, and then necessarily $\uu=A_{10}/A_{00}$ [cf. \req{vv-matrix}]. 

Toward that end, rewrite \req{Cnot0Anot0} as the system of the equations
\bea
A_{00}&=&\g_u\tau, \label{00a}\\
A_{01}&=&\g_u(\bb^T-C\uu^T S), \label{01a}\\
A_{10}&=&\g_u\tau\uu, \label{10a}\\
A_{11}&=&\g_u\uu\bb^T-S^{\uu}S; \label{11a}
\eea
here, $S^{\uu}$ is defined as in \req{S-v} or \req{S-0}, \pg{S-0}.
Then \req{10a} and \req{00a} imply $A_{00}\ne0$ and 
\beq{vv}
\uu={A_{10}\over A_{00}}=\vv;
\eeq
hence, the condition that $v<\infty$ and $Cv^2<1$ simply means that one can define $\g_u=\g_v$ as in \req{gamma}.
Next, \req{01a} is equivalent to 
\beq{bb}
\bb^T=C\uu^T S+\g_u^{-1}A_{01}.
\eeq
Now, \req{11a} together with \req{vv}, \req{bb}, and \req{**} yield 
\beq{S}
S=S^{\uu}\left({A_{10}A_{01}\over A_{00}}-A_{11}\right).
\eeq
By \req{00a}, 
\beq{tau}
\tau={A_{00}\over\g_u}.
\eeq

Vice versa, \req{vv}, \req{bb}, \req{S}, and \req{tau} imply \req{00a}--\req{11a} or, equivalently, \req{Cnot0Anot0}. 
In turn, \req{Cnot0Anot0} implies that $S$ is nonsingular, as well as $A$ is.
Hence, Part I of Theorem \ref{BOOST-ADJUST} is proved.

To prove Part II of Theorem \ref{BOOST-ADJUST}, it suffices to show that (i)
the representation
\beq{Cnot0Ais0}
A=B_\infty^{C,\ee}\pmatrix{\tau&\bb^T\cr \0&S\cr}
\eeq
takes place for a nonzero real number $\tau$, a non-singular real $3\times3$ matrix $S$, and a unit vector $\ee\in\R^3$ if and only if $v=\infty$ and $C<0$, and then necessarily either $\ee$ or $-\ee$ has the direction of $A_{10}$, and (ii) given the sign of $\tau$, the matrices $\bb$ and $S$ are uniquely determined.   

Here, the necessity of the conditions $v=\infty$ and $C<0$ is obvious.
Also, by definition, $v=\infty$ implies $A_{00}=0$. 
Rewrite now \req{Cnot0Ais0} as the system of the equations
\bea
A_{00}&=&0, \label{00b}\\
A_{01}&=&\sqrt{-C}\ee^T S, \label{01b}\\
A_{10}&=&{\tau\over\sqrt{-C}}\ee, \label{10b}\\
A_{11}&=&{1\over\sqrt{-C}}\ee\bb^T+(P^\ee-I_3) S. \label{11b}
\eea
Then, \req{10b} implies
\beq{ee}
\ee=\vp{A_{10}\over|A_{10}|}
\eeq
where $\vp:=\mbox{sign}\,\tau=\pm1$,
and
\beq{tau-infty}
\tau=\vp\sqrt{-C}|A_{10}|.
\eeq
By \req{P}, \pg{P}, $P^\ee=\ee\ee^T$; 
next, \req{01b} means $\ee^T S=A_{01}/\sqrt{-C}$; 
hence, \req{11b} can be rewritten as 
\beq{S-b}
S=-A_{11}+{\ee\over\sqrt{-C}}(\bb^T+A_{01}).
\eeq
Substituting this expression for $S$ into \req{01b}, one has
\beq{b-b}
\bb^T=\sqrt{-C}\ee^T A_{11}.
\eeq
Substituting this expression for $\bb^T$ into \req{S-b}, one obtains
\beq{S-c}
S=(P^\ee-I_3)A_{11}+
{\ee A_{01}\over\sqrt{-C}},
\eeq
with $\ee$ given by \req{ee}.

Vice versa, \req{ee}, \req{tau-infty}, \req{b-b}, and \req{S-c} imply \req{01b}--\req{11b}.
Hence, Part II of Theorem \ref{BOOST-ADJUST} is proved as well.
\qedd


\aparagraphh{Proof of Theorem \ref{SYNCHRO-FREE} }
\label{Details on Theorem SYNCHRO-FREE}
\ \\
The theorem can be restated as follows: Let $C$ be any nonzero real number and let $A:=A^{g,f}$ for a strictly proper pair of RFs $(f,g)$. 
Then there exist some $\vv\in\R^3$, real $\tau\ne0$, and non-singular $3\times3$ matrices $M$ and $N$ such that 
\beq{synchro-free}
A=\diag(1,N)B^{C,\vv}\diag(\tau,M)
\eeq
if and only if $\mu<1$ and $C\mu>0$, where $\mu$ is given by \req{mu}, \pg{mu}.
It is easy to see that using here the matrix $\diag(1,N)$ of the form less general than that of $\diag(\tau,M)$ in fact does not diminish generality.


Rewrite \req{synchro-free} as the system of equations
\bea
A_{00}&=&\g_v\tau, \label{00rf}\\
A_{01}&=&-\g_v C\vv^T M, \label{01rf}\\
A_{10}&=&\g_v\tau N\vv, \label{10rf}\\
A_{11}&=&-NS^{\vv}M. \label{11rf}
\eea
Substituting these expressions into \req{mu}, one has
\beq{mu1}
\mu=Cv^2,
\eeq
which implies $\mu<1$, in order for $\g_v$ to exist.
Also, \req{mu1}, together with \req{mu} and with $(f,g)$ being strictly proper, implies $C\mu>0$. 
This demonstrates the ``only if" part of the theorem.

To prove the ``if" part, observe first that for any two vectors $\aaa$ and $\bb$ in $\R^3$ such that $\aaa^T\bb>0$, there exists a symmetric positive-definite matrix $P$ such that $P\aaa=\bb$; 
for instance, choose $P=(\aaa^T\bb)^{-1}\bb\bb^T+\bb_2\bb_2^T+\bb_3\bb_3^T$, where $\bb_2$ and $\bb_3$ are any vectors in $\R^3$, which are orthogonal to $\aaa$ and, together with $\bb$, form a basis in $\R^3$ (e.g., one can take $\bb_2:=\aaa\times\bb$ and then $\bb_3:=\aaa\times\bb_2$).

Hence, whenever $\aaa^T\bb>0$, 
there exists a non-singular $3\times3$ matrix $N$ such that
\beq{N-r}
NN^T\aaa=\bb.
\eeq
Now, apply this observation to the vectors 
\beq{aaa}
\aaa:=A_{00}(A_{11}^T)^{-1}A_{01}^T
\eeq
and 
\beq{bbb}
\bb:=CA_{10},
\eeq
which satisfy the inequality $\aaa^T\bb>0$, because $\aaa^T\bb=C\mu A_{00}^2$ and $C\mu>0$. 
Next, let
\beq{vv-r}
\vv:=C^{-1}N^T(A_{11}^T)^{-1}A_{01}^T.
\eeq
Then \req{vv-r}, \req{N-r}, \req{aaa}, and \req{bbb} imply
$Cv^2=C\vv^T\vv=\mu<1$, and so, $\g_v$ can be determined by \req{gamma}, \pg{gamma}.
Solving now \req{00rf} for $\tau$ and \req{11rf} for $M$, one can easily check that all equations \req{00rf}--\req{11rf} are thus satisfied.
\qedd


\aparagraphh{Counterexample for Remark \ref{UNIVERSAL-SYNCHRO-FREE} }
\label{EX-UNIVERSAL-SYNCHRO-FREE}
\ \\
Let $f$, $g$, and $h$ be RFs such that $g=Af$ and $h=Bf$, where 
$$A=\pmatrix{3&-8/3& 0& 0\cr
	3&-3&0&0\cr
	0&0&1&0\cr
	0&0&0&-1\cr}
\quad\mbox{and}\quad
B=\pmatrix{2& -16/3& 4& 140/9\cr 
	6& -10& 3& 70/3\cr 
  	-3/2& 1& 2& 0\cr 
	9/4& -3& 0& 6\cr}.$$
Note that
$A^2=B^2=(BA^{-1})^2=I_4$, so that
every pair of RFs among $f$, $g$, and $h$ is reciprocal; moreover, every such pair is strictly proper 
since $A_{11}$ and $B_{11}$ are non-singular, $\aaa_1^T\bb_1\ne0$, and $\aaa_2^T\bb_2\ne0$, where $\aaa_1:=(A_{11}^T)^{-1}A_{01}^T= (8/9, 0, 0)^T$, 
$\bb_1:=A_{10}=(3,0,0)^T$,
$\aaa_2:=(B_{11}^T)^{-1}B_{01}^T=(8/3, -2, -70/9)^T$,
and $\bb_2:=B_{10}=(6, -3/2, 9/4)^T$.

Note that, provided $\aaa$ and $\bb$ are given by 
\req{aaa} and \req{bbb}, condition \req{N-r} is not only sufficient but necessary for \req{synchro-free}, since \req{N-r} follows from \req{00rf}--\req{11rf}.

Therefore, if the triple $(f,g,h)$ is can be adjusted without re-synchronization to a 
triple $(\tilde f,\tilde g,\tilde h)$ such that the pairs $(\tilde f,\tilde g)$ and $(\tilde f,\tilde h)$ are generalized Lorentzian, then
there exists a non-singular $3\times3$ matrix $N$ such that
$$NN^T\aaa_1=\la_1\bb_1\quad\mbox{and}\quad NN^T\aaa_2=\la_2\bb_2$$
for $\aaa_1$, $\bb_1$, $\aaa_2$, $\bb_2$ defined above and 
for some real numbers $\la_1$ and $\la_2$, which must be then nonzero.

Hence, for all real $\al_1$ and $\al_2$, one has 
$$\sum_{i,j=1}^2\al_i\al_j\aaa_i^T \la_j\bb_j
=\left(N^T\sum_{i=1}^2\al_i\aaa_i\right)^T \left(N^T\sum_{i=1}^2\al_i\aaa_i\right)\ge0;$$
this implies
$4\la_1\la_2(\aaa_1^T\bb_1)(\aaa_2^T\bb_2)
\ge\la_1^2(\aaa_2^T\bb_1)^2+\la_2^2(\aaa_1^T\bb_2)^2
+2\la_1\la_2(\aaa_1^T\bb_2)(\aaa_2^T\bb_1)$ for some real nonzero $\la_1$ and $\la_2$, which further implies
$(\aaa_1^T\bb_1)^2(\aaa_2^T\bb_2)^2\ge
(\aaa_1^T\bb_1)(\aaa_2^T\bb_2)(\aaa_1^T\bb_2)(\aaa_1^T\bb_2)$; however, the latter inequality is false for the above $\aaa_1$, $\bb_1$, $\aaa_2$, and $\bb_2$.
Thus, our triple $(f,g,h)$ is not adjustable without re-synchronization to a 
triple $(\tilde f,\tilde g,\tilde h)$ such that the pairs $(\tilde f,\tilde g)$ and $(\tilde f,\tilde h)$ are generalized Lorentzian, even though every pair of RFs among $f$, $g$, and $h$ is reciprocal and proper (and therefore can be rescaled to a generalized Lorentzian pair).
\qedd

\end{document}